\documentclass[preprint,authoryear,12pt]{elsarticle}

\usepackage{amssymb}
\usepackage{graphics}

\journal{Astronomy and Computing (now published)   }

\begin{document}

\begin{frontmatter}

\title{N-Body Simulations of Collective Effects
in Spiral and Barred Galaxies}

\author{Xiaolei Zhang}

\address{School of Physics, Astronomy, and Computational
Sciences, George Mason University, 
4400 University Drive, Fairfax, VA 22030, USA, xzhang5@gmu.edu}

\begin{abstract}
We present gravitational N-body simulations of the
secular morphological evolution of disk galaxies induced
by density wave modes.  In particular, we address the 
demands collective effects place on the choice of simulation 
parameters, and show that the common practice of the use
of a large gravity softening parameter was responsible for
the failure of past simulations to correctly model
the secular evolution process in galaxies, even for those
simulations where the choice of basic state allows an
unstable mode to emerge, a prerequisite for obtaining
the coordinated radial mass flow pattern needed for
secular evolution of galaxies along the Hubble sequence.  We also 
demonstrate that the secular evolution rates measured in
our improved simulations agree to an impressive degree with
the corresponding rates predicted by the recently-advanced 
theories of dynamically-driven secular evolution of galaxies.  
The results of the current work, besides having direct 
implications on the cosmological evolution of galaxies, 
also shed light on the general question of how irreversibility 
emerges from a nominally reversible physical system.

\end{abstract}


\end{frontmatter}

\section{INTRODUCTION}

\subsection{Self-Organization in Nonequilibrium Systems} 

Understanding the self-organization behavior of nonlinear,
far from equilibrium systems is among the most challenging
of problems of contemporary research.  This multi-disciplinary
study is relevant to answering many key questions in 
condensed matter physics, biological and life sciences, social sciences, 
as well as the processes of structure formation in the universe.  

Several leading scientists of the 20th century considered the 
exploration  of collective behaviors, or the
processes of morphogenesis, to be of central 
importance to the future advancement of science.  More than half a 
century ago, Richard Feynman closed volume II of 
``The Feynman Lectures on Physics" by declaring: ``The next great era of 
awakening of human intellect may well produce a method of understanding 
the qualitative content of equations.  Today we cannot.  Today
we cannot see that the water flow equations contain such
things as the barber pole structure of turbulence that one
sees between rotating cylinders. Today we cannot see whether
Schrodinger's equation contains frogs, musical composers, or
morality -- or whether it does not'' (Feynman et al. 1964).
P.W. Anderson, a noted condensed matter physicist, penned an influential 
article for {\em Science} titled ``More is different'' (Anderson 1972), 
in which he called attention to the hierarchical organization
of physical systems, and the ability of many-degrees-of-freedom
systems to form emergent structures and dynamics, which break the 
symmetries of equations describing the underlying micro-dynamics.  

Among the studies on morphogenesis, one of the most prophetic is that of 
Ilya Prigogine.  In his theory of ``dissipative structures'', Prigogine 
emphasized the entropy-production-enhancing function of self-organized global 
patterns in far-from-equilibrium systems, as well as the constructive 
role of dissipation in maintaining these patterns  (Prigogine 1980).  
To paraphrase this theory, most of the self-organized structures in nature 
come in roughly two types: (1) equilibrium
structures, which are formed through equilibrium phase transitions.
The examples of equilibrium structure formation include the phase transition
of water to form ice when the temperature is lowered to zero degree 
Celsius, as well as the formation of minerals of a specific crystal structure
when the environmental temperature and pressure satisfy a range of conditions 
(i.e. the formation of diamond crystal from carbon in a high pressure
environment).  (2) nonequilibrium structures, which are formed in 
systems far from equilibrium, and which (in addition to sharing
some common features with structures formed in equilibrium
phase transitions) have the added feature 
that these self-organized structures are generally in a dynamical equilibrium 
state, meaning they are sustained through the competition of growth and decay 
tendencies (as highlighted by the well-known fluctuation-dissipation theorem,
proved mostly at close-to-equilibrium regimes but is also valid at 
far-from-equilibrium situations for self-organized dissipative structures). 
There is in general also a continuous flux of energy and entropy through the
system to maintain the nonequilibrium fluctuation.  The formation of
self-organized dissipative structures in nonequilibrium systems often 
serves the important function of greatly accelerating the speed of 
entropy evolution of the parent systems (Prigogine 1980)\footnote{The
transportation of entropy to its environment is the chief means
a dissipative structure can maintain a low or constant entropy
state despite it being a very efficient engine at the local
production of entropy.  This aspect also resolves the paradox of
how nature can generate complex biological entities such as
human being as a result of nonequilibrium evolution, despite being governed
by the second law of thermodynamics universally.  The study of
nonequilibrium dissipative structures tells us that entropy
increasing evolution does not always mean the rush towards homogeneity
everywhere, at least not for open, many degree-of-freedom, 
far-from-equilibrium systems.}.

The hierarchy of self-organization processes 
in a nonequilibrium system often leads to a series of effective 
singularities in the dynamics, which allow {\em emergent} new dynamics 
to form that cannot be {\em deductively} derived from the 
differential formulation one starts the analysis with.  A synthetic
approach uniting the various local aspects to achieve global
self-consistency will need to be adopted. 
The correlations among the fluctuations of the individual degrees of freedom 
of the many-body system are shown to play a crucial constructive
role at the juncture of nonequilibrium phase transitions.

Although the mechanism for the generation of new dynamics through 
so-called ``spontaneous breaking of gauge symmetry'' had been routinely 
proposed in high energy physics, it was mostly used in a model context, 
rather than derived from first principles in a self-consistent fashion.  
Quoting once again condensed matter physicist P.W. Anderson: ``... the 
concept of broken symmetry has been borrowed by the elementary particle 
physicists, but their use of the term is strictly an analogy, whether 
a deep or a specious one remaining to be understood'' (Anderson 1972).  
On the other hand, the fact that, as of now, there have been very few examples 
of self-organized dissipative systems being analyzed from first principles 
(in Prigogine's work, chemical clock was used as one prominent example)
is partly due to the intrinsic complexities of such problems, i.e., the many
degrees of freedom of the components, and the correlations among the
components, which invalidate many basic assumptions underlying the usual
kinetic theory approach for treating many-particle systems.  
One of the most important
assumptions used in the derivation of the Boltzmann kinetic equation is 
the so-called ``molecular chaos'' assumption (see, e.g., Kreuzer [1981]
for a detailed description of the BBGKY procedure for the derivation
of the collisional Boltzmann equation), or the assumption that particle 
collisions are uncorrelated.  This assumption is crucial to Boltzmann's
arriving at his famous H-theorem, or that entropy never decreases in
nonequilibrium processes.  The inter-particle correlations, however,
are the necessary ingredient for obtaining 
self-organized behavior, and their
re-introduction into the study of nonequilibrium dynamics
allow local-entropy-decreasing processes to be admitted into the analyses. 

N-body simulations of self-gravitating systems, in our case the
simulation of disk galaxies containing self-organized density wave patterns,
offer a rare chance to observe at close range the modification 
of differential dynamics to arrive at new (i.e. emergent)
meta-laws, thus offer clues to the common features
of spontaneous symmetry breaking processes in many-degrees-of-freedom 
dynamics.  In this case, an added advantage is that a parallel
theoretical development has also been accomplished in the past
few decades, which can serve as standards of comparison with the
simulation results.

\subsection{Secular Evolution of Galaxies in the Context of
Nonequilibrium Phase Transition}

The striking coherence of spiral and bar patterns in disk galaxies
has long captured our awe and fascination, but it was only since the
advent of density wave theory (Lindblad 1963; Lin \& Shu 1964;
Kalnajs 1965) that these patterns were understood as propagating waves of 
over-density in differentially rotating galaxy disks\footnote{The
publication of the first round of density wave papers followed shortly  
after the appearance of the first volume of ``The Feynman Lectures on 
Physics" in which Feynman suggested to the freshman and sophomore
physics students in his class: ``Incidentally, if you are looking for a good problem,
the exact details of how the arms are formed and what determines
the shapes of these galaxies has not been worked out" (Feynman,
Leighton, \& Sands 1963).}. 

As the study of density waves in galaxies progressed over the past few 
decades, the propagating wave picture of the initial studies 
further evolved into a {\em modal} view, in which the oppositely-propagating
trains of density wave in the radial direction superpose to
form growing density wave {\em modes} (Lin \& Lau 1979 and the 
references therein; Bertin et al. 1989a,b).  In this formulation the so-called
``grand-design'' spirals and bars observed in nearby galaxies
were regarded as spontaneously growing modes 
in a galactic resonant cavity whose properties are characterized by 
the axisymmetric distributions of disk-mass surface density, 
stellar and gaseous velocity dispersions, as well as the overall 
gravitational potential field (which include contributions not only
from the disk mass, but also from the more spherically distributed luminous 
and dark halos, as well as the galactic bulge), which together form the 
so-called ``basic state'' of the galactic disk, upon which the wave modes 
grow as unstable harmonic perturbations\footnote{The ``basic state'' of
the galactic disk is formally equivalent to the so-called ``boundary
condition'' in an electromagnetic resonant cavity.}. 

During the early decades of density wave study, the basic state
of the galactic disk was treated as a stationary background
from which the unstable trains and modes
of density waves were calculated to varying orders of approximation
(according to the orderings of either the degree of non-linearity, or else
the degree of locality in the successive WKBJ approximations). 
The prospect of the secular evolution of the mass distribution
of the basic state itself was never seriously considered, apart
from phenomenological inferences of the possible role of gas
accretion (Kormendy 1979).  As we know, gaseous mass in galaxies forms only 
a small percentage of the total disk mass, and the disk mass in most
intermediate- and early-type galaxies was dominated by stellar mass.
Therefore, a significant transformation of the Hubble type of a
galaxy during its lifetime will necessarily involve the secular
redistribution of the {\em stellar} mass, in conjunction with the
redistribution of the gas mass.

The long-held view that the stellar disks of galaxies remain mostly
unchanged throughout a galaxy's lifetime is partly a result of the belief
that stars behave ``adiabatically'' during their orbital motion,
and do not dissipate their orbital energy when interacting
with a stationary density wave except at the wave/particle
resonances (Lynden-Bell \& kalnajs 1972) -- a behavior summarized by the 
well-known ``conservation of the Jacobi integral'' of a single
star's orbit in the rotating frame of a stationary wave perturbation 
(Binney \& Tremaine 2008).  Another often-used phrase to describe
this quasi-stability of the stellar orbit is the so-called
``angular momentum barrier'' to secular redistribution of
stellar mass in a disk galaxy.

Observationally, there is growing evidence that galaxy morphology 
does evolve significantly throughout the cosmic history, in general following 
the trend from a disk-dominated late-Hubble-type to a bulge-dominated 
earlier-Hubble-type (Zhang 2003; Kormendy \& Kennicutt 2004; as well as the
references therein).  Though galaxy mergers had previously been proposed 
as responsible for a large fraction of galaxy Hubble-type evolution, there
are strong dynamical reasons why mergers cannot be responsible for most
of the observed morphological evolution of disky galaxies (Zhang 2003, 2008;
as well as the references therein).
Boxy early-type disk galaxies, and especially the massive cD-type galaxies 
in the central regions of dense clusters, are the only prime candidates 
for the merger mechanism of disk-galaxy morphological evolution.

How then could stars in disk galaxies overcome the
angular momentum barrier, break the constraint of the
conservation of the Jacobi, and initiate significant redistribution
of stellar mass during the lifetime of a galaxy?  As it turned out, the
answer had much to do with the dynamics of the self-organization process
which formed the global density wave patterns in the first place,
as well as with the theory of dissipative structures.  Zhang (1996, 1998, 1999,
hereafter Z96, Z98, Z99, respectively)
showed that stars in galaxy disks possessing spontaneously-formed
nonaxisymmetric density wave modes, such as spirals and bars,
are able to display dissipation-like behavior just like their
gaseous counterpart\footnote{In fact, as it turned out, gas also achieves
efficient secular mass redistribution through {\em the same} gravitational
torque mechanism as stars (Z98), since its participation
in the interaction with the density wave is mostly through the
scattering of clouds.  The microscopic viscosity in the cloud
medium can be shown to be entirely insignificant (compared to cloud
scattering which has much greater mean-free-path) to the secular 
redistribution of gas mass in a galaxy disk environment. See further
the discussions in Appendix G.}, 
as long as the {\em collective mutual interactions}
of the stars are taken into account.  The conservation of
the Jacobi integral for a single star's orbit is shown to be
purely a consequence of having treated such an orbit as {\em passively}
responding to an {\em applied} potential field, which ignored the {\em mutual
interactions} of stars in a {\em self-sustained} density wave mode.  
The passive treatment does not incorporate the graininess effect of the 
spiral and bar potential which reflects the inherent correlations 
among the orbits of the stars (as well as gas clouds).

The detailed analyses of the workings of collective effects in spiral and 
barred galaxies were previously presented in Z96 and Z98.  For a quick rehash, 
we look into the dynamical processes in galaxies from both the global 
and the local points of view -- and ultimately, these two views need to give 
consistent results for spontaneously-formed and self-sustained density wave 
modes.  Globally, modal formation is due to the fact a galaxy disk
containing a density wave mode is generally more energetically 
favorable than the axisymmetric basic state from which the mode emerges. 
Therefore, the marginally axisymmetrically stable galactic disk, 
formed from the primordial collapse and dissipation of the gas clouds, 
is unstable to the formation of nonaxisymmetric spiral or bar modes.  
From another perspective, modal growth is due to the fact that
the galactic resonant cavity for propagating density waves has 
a positive gain for the wave amplitude
during each round trip of the wave-train propagation between
the corotation circle and the inner galaxy.  The wave amplification 
happens mostly near the corotation region through the
so-called ``over-reflection'' of the outward propagating wave train to
become an inward propagating wave train of higher amplitude,
and the accompanying transmission of angular momentum across the 
corotation radius to a third wave branch which propagates
to the outer disk to dissipate the energy and angular 
momentum to the environment\footnote{Note that an idealized density wave 
mode rotates in the azimuthal
direction with a fixed pattern speed, and since the galaxy disk itself
possesses differential rotation, the disk matter rotates faster than
the density wave inside the corotation radius, and thus overtakes
the density wave; and vice versa outside corotation.  {\em The wave
thus have negative angular momentum density inside
corotation relative to the basic state, and positive angular momentum
density relative to the basic state outside corotation.}}. 

\begin{figure}
\vspace{6cm}
\includegraphics{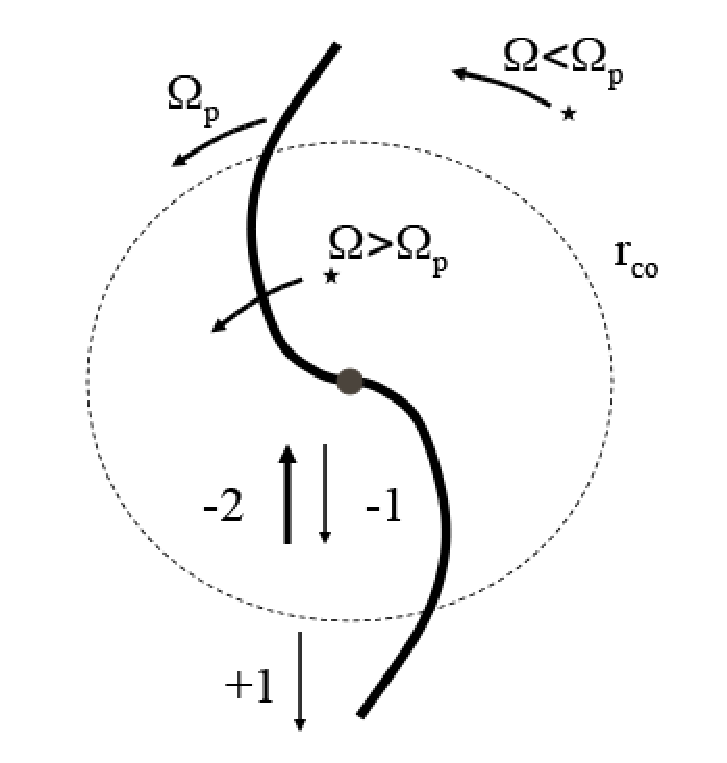}
\caption{Schematic of the wave propagation and amplification
in the galactic resonant cavity to form infinitely-growing
density wave modes.  The symbols $\Omega$ and $\Omega_p$
denote the galaxy (or disk matter) circular/angular speed
and density wave pattern speed, respectively.  Inside the corotation
radius the stars overtake the wave during their circulation
around the galaxy, and vice versa outside corotation.
Here the circle with radius $r_{co}$ is the corotation circle.  
The numbers next to the arrows indicate the 
normalized amount of wave angular momentum relative to the basic state
for the respective segment of the outward propagating (incoming) and
inward propagating (over-reflected) wave trains inside corotation,
as well as the outward propagating (transmitted) wave train outside
the corotation radius.  Note that the over-reflection
process conserves angular momentum, which we can check
by summing the two numbers after the over-reflection process
(-2 and 1) and compare the result with the number before
the over-reflection process (-1).
The feedback loop is completed by
an inner Q-barrier (usually identified with the galactic
bulge region), where the inward-propagating
wave train gets turned around to become the outward propagating
wave train for the next cycle of over-reflection at corotation.
This illustration is for the milder version
of the over-reflection mechanism WASER (Mark 1976).  
For the SWING mechanism (Toomre 1981),
the over-reflection factor can be significantly bigger
than the factor of 2 illustrated here. The SWING and WASER
mechanisms differ by whether the incoming wave train towards
corotation is that of the leading or the trailing type, respectively,
produced in turn by either the inward-propagating trailing wave
tunneling through the central region of the galaxy
when no Q-barrier exists, and emerging as a leading wave; 
or else by the inward-propagating trailing wave being reflected by the Q-barrier
when one exists, which produces an outward-propagating
trailing wave. The SWING mechanism tends
to produce bar-like modes, and the WASER mechanism spiral
modes.}
\label{Figure1}
\end{figure}

Figure \ref{Figure1} shows a schematic
of the wave amplification process (Mark 1976; Toomre 1981).
The over-reflection mechanism at corotation removes
angular momentum from the waves inside corotation, and
delivers it to outside corotation to be dissipated
in the outer disk.  Due to the change of sign of wave
angular momentum density across the corotation radius,
this angular momentum redistribution process through the
wave propagation and over-reflection at corotation causes the wave branches
both inside and outside corotation to grow indefinitely,
if no nonlinear and dissipative mechanism is present
to counter the growth tendency.

The {\em global} spontaneous growth tendency of the unstable mode in 
realistic galaxy disks is counter-acted by the {\em local} dissipation 
at the spiral (or bar) arm locations, since each time as disk material 
crosses the arm density wave crest during its circum-galactic rotation, 
it loses angular momentum to the wave potential field through the mediation 
of a collisionless shock (Z96), for matter inside the corotation 
radius.  The sense of angular momentum transmission is reversed, i.e. 
from the wave to the disk matter, for the orbiting matter outside 
corotation.  This angular momentum exchange between the wave and the 
basic state disk matter is of the correct sense both inside and outside 
corotation to lead to the damping of the growing wave amplitude 
(since the wave has negative angular momentum density with respect 
to the basic state, upon receiving angular momentum from the disk matter 
its amplitude decreases.  Similarly the sense of angular momentum exchange,
and the sign of angular momentum density of the wave, outside corotation lead to
the damping of the wave outside corotation as well).  Eventually,
at sufficiently nonlinear amplitude, the wave mode achieves a 
quasi-steady amplitude and the galaxy settles into a
dynamical equilibrium state (Z98).

This equilibrium is called {\em dynamical} because it is maintained
through the competition of opposing tendencies of wave-mode growth and 
damping, and is maintained also at the expense of dissipative
secular evolution of the basic state mass distribution: the
disk matter inside corotation, upon giving angular momentum
to the wave at each arm crossing, spirals slightly inward; and
the matter outside corotation spirals slightly outward after
each arm crossing.  The dissipative secular evolution of the
basic state mediated by the density wave mode thus allows the 
accomplishment of another mission of
dissipative structures, that of enhanced entropy
production and export by these structures.  As
is well known the direction of entropy evolution for a self-gravitating
system is towards ever increasing central concentration, together
with the buildup of an extended outer envelope (Antonov 1962;
Lynden-Bell \& Wood 1968).  This direction of entropy evolution 
for self-gravitating systems is the same as that of the galaxy
morphological evolution along the Hubble sequence from 
the late to the early Hubble types.

The angular momentum exchange between the wave and the basic state is made 
possible through a collisionless gravitational shock at the density wave crest 
(Z96), similar in nature to the well-known plasma 
collisionless shocks (Balogh \& Treumann 2013 and the references therein).  
The correlated interactions of the orbiting stars (which can also be 
viewed as mediated by the correlation between the grainy potential field 
of the wave and the individual particles' motion), scatter the
stars mildly each time they cross an arm, to allow just the right amount 
of angular momentum exchange between that contributing to the
stellar orbital motion and that contributing to the wave motion, in order 
to counter the growth tendency of the wave mode as well as to initiate the 
secular orbital decay (inside corotation) or increase (outside corotation), 
at the quasi-steady state of the wave mode.  

The meeting point between the local and global points of view is the global 
self-consistency requirement, i.e., since the dissipative density wave mode 
is self-sustaining and self-limiting, the combined effects of
various dynamical processes on the wave and the basic state should be 
globally balanced and quasi-steady.  The fact that
this global self-consistency is achievable in galaxies
is due both to the long-range nature of gravitational interaction,
as well as to the dependence of wave amplification and damping
efficiencies on the nonlinearity of the density wave pattern
(Z96, Z98).

\begin{figure} 
\vspace{6cm}
\includegraphics{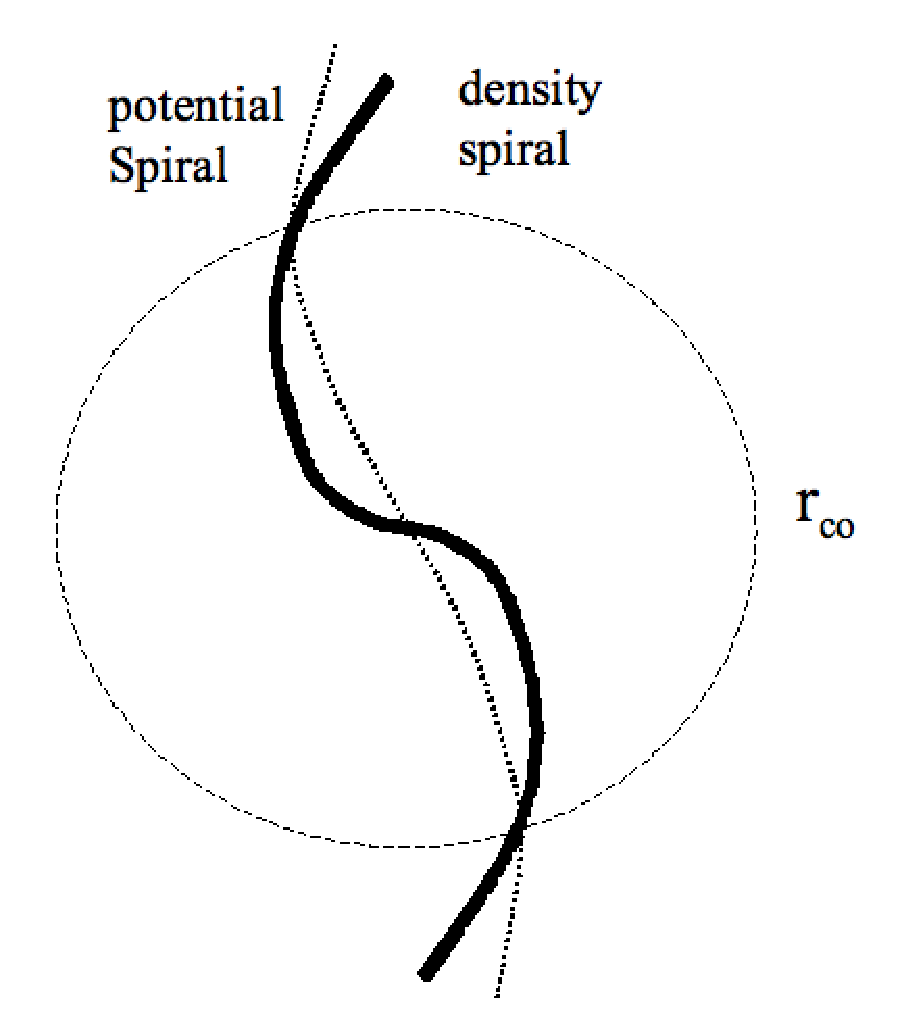}
\caption{Schematic of the radial distribution
of the azimuthal potential-density phase shift
for a self-sustained density wave mode in a galaxy disk.
The solid line indicates the locus of the
peaks of the density spiral, and the dotted line indicates
the locus of the troughs of the potential spiral.  The circle
indicates the corotation circle where the density
wave pattern speed and the basic state matter's circular
speed are the same. It is also
where the potential-density phase shift is zero for a spontaneously
formed density wave mode (Z96; Zhang \& Buta [2007];
Buta \& Zhang [2009]).}
\label{Figure2}
\end{figure}

The long-range nature of gravitational
interaction (as manifested through the form of the Poisson integral)
means that the density and its associated potential field do not
always coincide in space.  In the case of a skewed density wave mode,
this nonlocality of potential manifests as a characteristic
radial distribution of the azimuthal potential-density phase shift.  
As shown schematically in Figure \ref{Figure2},
the radial distribution of the potential-density phase shift,
for an unstable spiral or bar mode, is such that the spiral perturbation
potential lags the perturbation density inside the corotation, and vice
versa outside the corotation.  This characteristic distribution
of the potential-density phase shift is admitted by both
the Poisson integral and by the equations-of-motion for
the particular set of solutions of disk-mass and particle-velocity
distributions that support a spontaneously-formed density 
wave mode (Z96, see especially the Appendices there).  

At the quasi-steady state of the wave mode, the characteristic phase shift 
distribution represented in Figure \ref{Figure2} implies a secular torque 
action by the wave potential on the basic state disk matter, which leads to 
secular angular momentum exchange between the wave and the basic state 
both inside and outside corotation. This secular torque action 
can be shown to be responsible both for the 
spontaneous emergence of the wave mode in the linear regime, 
as well as for the maintenance of the mode to a quasi-steady amplitude 
at the expense of dissipative basic state evolution (Z98).

The characteristic phase shift distribution represented in 
Figure \ref{Figure2} is both the symptom, as well
as the driving dynamical mechanism, for the underlying secular 
dissipative process.  For example, the existence of this phase shift 
is one of reasons that collisionless shock at the spiral arm crossing
can form to initiate dissipation, since the streaming disk matter, while 
marginally gravitationally stable before arm crossing, experiences
added compression due to the nonlocal gravitational
potential contributed by the rest of the disk matter, which drives
the streaming matter below their instability threshold
and thus enabling the collisionless shock (Z96).
On the other hand, secular angular momentum exchange
between the density wave potential and the underlying
disk mass distribution (as needed for wave damping and
basic state evolution) necessarily leads to an azimuthal 
potential-density phase shift, a fact
already hinted at in Kalnajs (1972), though there
it was the interaction between a non-self-consistent
{\em gaseous} density wave and a driving {\em stellar} potential wave
that was the subject of study (i.e., the exploration of using gaseous wave 
as a candidate for the damping of stellar wave), rather than the
self-interaction of spontaneously-formed density wave mode.

As valid for all collective or cooperative effects, here
we witness in spiral and barred galaxies a seeming miracle, of all the
diverse components of the system cooperate to accomplish
a central goal: to accelerate the entropy evolution of the
nonequilibrium system.  These coordinated interactions
are accomplished through a global instability pattern (here the
density wave modal pattern), which spontaneously emerges in 
far-from-equilibrium systems.   Their longevity is one of the
reasons we observe them (in nature there are no lack of systems
that self-destruct not long after emergence, and thus have short lifetime.  
Even for spiral galaxies themselves there are constant appearance 
of transient noise, as well as slow-growing modal components 
that die out in the competition with the most unstable modes -- which
is one reason why dissipation is so important in the operation
of these systems: In sufficiently nonlinear regime all fluctuations
receive damping, but only the dominant modal components receive
sufficient amplification to offset the damping).  So the seeming
miracle of cooperation among the components of self-organized
systems is partly a result of a natural selection
process, though this selection happens on much shorter
timescale (as fraction of the lifetime of a single dynamical
system) than biological natural selection, which happens through
many generations of individual members.  We can instead call
this rapid natural selection process in self-organized systems
a ``winner takes all'' process, i.e., in the case of a galaxy,
the dominant mode rapidly solicits all resources of a disk galaxy's
mass distribution to modal activity, and after a few dynamical
timescales little is left for the slower-growing modes and noise,
which all receive additional damping through the local
dissipation mechanism at the density wave crest.
Other examples of such dissipative structures in nature include
the hexagonal convection cells in the well-known B\'enard problem of atmospheric
heat flow, as well as human beings who are capable of the most efficient
entropy production and export.  In some sense, dissipative
structures are the tools nature manufactures to accomplish the task of
accelerated evolution in nonequilibrium systems.

\subsection{N-body Simulations of The Secular Evolution of
Disk Galaxies}

N-body simulation of self-gravitating systems has had a long history
(see, for example, Sellwood [1987] and Hockney \& Eastwood [1988] 
for reviews of the early decades' work).  For the modeling of disk galaxies, 
N-body simulations were often carried out in a two-dimensional (2D) 
configuration, emulating the thin disks in observed galaxies, with the 
spherical bulge and halo modeled as inert and time-independent.  
To save computation time, 2D galaxy-disk simulations 
often employed a particle-mesh 
approach, with the force calculation performed on a regularly-spaced mesh and 
then interpolated to the individual disk particles, rather than being
performed directly. 

After assigning (usually constant) mass to the simulation particles 
(each has many solar masses depending on the total number of 
particles used in a given simulation and the total disk mass), 
with the particles distributed according to a given galaxy's 
disk surface density, the particle masses are further 
grouped/assigned onto the corresponding grid points.  Grid
potential (or force) is calculated using the Poisson integral,
aided by Fourier transform techniques whenever possible,
and then interpolated back onto individual super-particles
(usually through the same algorithm as the mass assignment algorithm
in order to conserve momentum).  These forces are used to calculate
accelerations on the individual particles, and the particle
trajectories are advanced through a standard finite difference
algorithm such as leap frog.  Then the cycle starts anew.  Time
step and spatial grid spacings are selected to ensure stability
and accuracy, and special treatments at the inner and outer boundaries
of the disks are needed to take care of in-spilled and out-spilled
particles leaving the computation grid.

Earlier N-body simulations of disk galaxies containing density wave
patterns focused on the emergence phase of the patterns, and checked
the growth rates against linear density wave theory predictions, 
and derived pattern speeds and other related density wave parameters
(Sellwood 1987 and the references therein).  Donner \& Thomasson
(1994) first explored the longevity of density wave modes in
N-body disk simulations. They found initial evidence of the
mass inflow/outflow behavior for stars inside and outside of corotation,
respectively, but did not attribute a dynamical mechanism to it.

Starting from her Ph.D. dissertation work (Zhang 1992), the current
author had carried out a systematic study of the dynamical mechanisms
underlying the secular morphological evolution of disk galaxies enabled
by the self-organized density wave modes.  Building on the numerical algorithms
for N-body simulation of Thomasson (1989), which was a direct descendant of 
polar-grid code originally developed by Miller (1976), Z96 \& Z98 explored the
signature of collisionless shock in spiral galaxies, the potential-density
phase shift distribution, as well as the secular evolution of the 
basic state mass distribution in the simulated galactic disks, and 
compared these with the theoretical predictions.  

Despite the confirmation of theoretically-derived mass flow rates 
in the accompanying N-body simulations (Z98), which established 
the viability of the analytical approach, the simulated
mass flow rates were found to be small, far from being adequate to lead to 
significant morphological transformation of a galaxy within
a Hubble time.  This small mass flow rate was traced back
to the small amplitudes of the spiral patterns formed in the
simulations, which are far short of the extremely-nonlinear
wave amplitudes observed in physical galaxies.  This feature of the
small wave amplitude applies to all the disk galaxy simulations performed
in the past few decades, and its cause was barely addressed.

In order to circumvent the difficulty with simulations, 
Zhang \& Buta (2007, 2015) used near-and-mid-infrared images 
of galaxies and applied the analytical mass flow rate equation 
of Z96 \& Z98 (see also equation \ref{eq:accre} in Appendix A of
the current paper) directly to the derived mass surface density
of physical galaxies.  In these studies, 
it was found that mass flow rates from a few $M_{\odot}$ per year 
to over one hundred $M_{\odot}$ per year were typical, depending
on the density wave amplitudes and pitch angles of galaxies.  
These derived mass flow rates for physical galaxies far exceeded 
those obtained from past N-body simulations, mainly as a result 
of the fact that mass flow rate is proportional to the wave amplitude 
squared (Appendix A), and the amplitudes in physical galaxies are oftentimes 
a factor of 3-10 times higher than those obtained in N-body simulations, 
implying a difference in mass flow rates of a factor of 10 - 100 
between physical and previously-simulated galaxies.

Given the importance of understanding secular evolution processes in 
physical galaxies, one could not help but wonder what exactly 
were the factors that prohibited the simulated disk galaxies from 
achieving the range of density wave amplitudes and the level 
of mass inflow rates in observed galaxies.  It is this question 
that the current paper addresses.  We show that an
artificial parameter inserted into the numerical equation for
calculating gravitational potential, i.e. the so-called {\em 
softening parameter}, underlies much of the discrepancy between 
the simulated and observed wave amplitudes.  Softening was introduced 
to represent both the finite thickness effect of realistic galaxy disks
in 2D simulations, and to restrain the artificial relaxation effect 
that is more pronounced in simulated disks which have a much smaller number
of particles compared to observed galaxies.  When properly chosen,
it was found in the past that the exact values of softening
parameter do not impact significantly the morphology of the 
simulated density wave patterns.  On the other hand, as we will 
show in this paper, when our goal is to model the quantitative
characteristics of the self-organized density wave modes, 
especially the nonlinear equilibrium amplitude of the
quasi-steady wave modes which in turn determines the
rate of the secular evolution of the basic state mass 
distribution, the role of softening turns out to be more subtle, 
due to the fact that the spontaneously formed modes are stabilized 
through the dynamical balancing act of global amplification 
and local dissipation.  Both the modal amplification process (which is 
sensitively dependent on the self-gravity of the disk, especially near the 
corotation region) and the local dissipation of the wave (through the 
collisionless shock enabled by the mutual interactions of local matter 
within the arm instability region) depend on the exact form of the force law, 
or the departure of this law from the original Newtonian form as
a result of softening.  The net effect is that smaller softening
leads to equilibrium wave-mode amplitude and the resulting
secular mass flow rates much closer to what we had measured
from observed galaxies (Zhang \& Buta 2007; 2015).

Besides impacting the equilibrium wave amplitude, we show that small softening
leads to easier attainment of quasi-steady modes. 
The wave patterns obtained in small softening runs, though locally often 
appearing more noisy, are in fact globally closer to pure modal 
forms, as borne out from power spectrum analyses.  
Even the noisier local mass 
distribution itself in the small softening simulations is in fact closer 
to how observed galaxy mass is really distributed, i.e., an average galaxy 
disk is often populated by stellar clusters, globular clusters,
giant molecular cloud complexes, and other spiral arm inhomogeneity.
In the face of all the inhomogeneities in mass distribution,
the grand-design density wave modal patterns are nonetheless often
obtained among observed galaxies.  It is thus reassuring that the 
robustness of observed density wave modes\footnote{See also 
Zhang \& Buta (2015).  In the cases of M51 and NGC 3627 analyzed there, 
the intrinsic modal features of 
these galaxies were shown to be preserved even after strong tidal 
interactions with companion galaxies.} can be reproduced in the simulations
using small softening parameters.  This shows that self-organized
nonequilibrium patterns possess so-called ``asymptotic
stability'' and are unaffected by the incidentals of the
noisy background.  If anything, the noisy mass distribution
provides the necessary initial seeds for the rapid spontaneous
growth of the instability pattern.

\section{THE EFFECTS OF
SOFTENING IN N-BODY SIMULATIONS OF DISK GALAXIES}

In this section, we motivate the study in the main body 
of this paper (sections \S3-\S4) by giving the historic context and 
rationales for the introduction of the softening parameter in N-body 
simulations of disk galaxies, as well as the potential limitations when 
softening is used in modeling the long-term (secular) evolution behavior of
disk galaxies containing self-organized density wave modes.  We
also present a first set of simulations showing the dependence of the rate
of disk mass redistribution on the choice of the softening parameter,
as well as on other parameters of the simulation.

\subsection{Background}

In essentially all of the gravitational N-body simulations of galaxies,
a so-called softening parameter is used to control 
the artificial relaxation effects due to the smaller number of particles 
used in these simulations compared to that in physical galaxies.
The softened force law can take various analytical forms.
In the often employed Plummer-sphere softening scheme, the gravitational 
potential due to a point mass $m$ at a distance $r$ is calculated as

\begin{equation}
\Phi (r)
= {{-Gm} \over {\sqrt{(r^2 + {a_{soft}}^2)}}}
\label{eq:soften}
\end{equation}
where $a_{soft}$  is the softening parameter, and G is the gravitational
constant.  A choice of $a_{soft}=0$ corresponds to the original Newtonian
force law.  A finite softening parameter reduces the amount of unrealistic 
close encounters which occur more frequently for small N (N here denotes
the number of particles used in N-body simulations) systems,
and thus reduces the level of artificial relaxation in such systems.
Furthermore, in the particle-mesh approach commonly adopted in
simulating disk galaxies, the mesh size itself provides additional
softening effect.  The hope is that a {\em collisionless} configuration 
would result from a well-matched N and an {\em effective} softening 
parameter which takes into account both particle and 
grid softening effects.  
In the context of 2D disk galaxy simulations, both particle and mesh
softening also emulate the realistic physical configuration of
finite disk thickness.  2D simulation incorporating softening
has been shown to be able to model the realistic macroscopic properties 
of disk galaxies containing density wave patterns with $10^3 - 10^7$ 
less number of particles than that present in physical galaxies.

One of the rationales for introducing the softening parameter
is to achieve an effectively collisionless environment for
the duration of the simulation run, so that only long-range
interactions determine the formation of density wave patterns in their
parent disks.  However, with the realization that galaxies which possess
global density wave modes {\em always} involve partly-local
collective dissipation processes, one soon faces the fact that it is 
impossible to completely avoid collision-like behavior in the
simulation of unstable density wave modes, no matter how large 
a particle number N is used (whether in real galaxies or in simulations): 
After all, collective effects in these systems {\em depend} on the 
near-collision or small-angle scattering of particles in the global 
instabilities (i.e. density wave modes) to set up
long-range correlations, in order to achieve self-organization
and to induce secular evolution of the basic state of the disk (Z96).  
In some sense, these galactic systems are in a {\em forced relaxation 
configuration}, with the forcing accomplished by the density wave 
collisionless shocks.  The {\em effective} local instability parameter 
Q$_{eff}$ in the spiral arms is thus always less than one to enable 
interparticle correlation.  {\em A true collisionless configuration
will never be able to support self-organization behavior}. 

The custom choice of large softening in the past N-body simulations
of disk galaxies, however, reduces the very interparticle interaction 
that is the backbone support of collective effects, and thus can be 
responsible for reducing density wave amplitudes in simulated galaxies, 
as had already been noticed early on (Sellwood 1987).  Despite perhaps
subconsciously knowing the correlation between the choice of
softening parameter and the simulated wave amplitude by some researchers, 
the practice of using large softening (i.e. using a value comparable
to the smallest grid size) continued almost universally, partly because 
these earlier N-body simulations were mainly interested in obtaining
a comparable appearance of density wave morphology as that in
observed galaxies.  They are not geared towards the study of secular
evolution effects induced by the collective density wave modes, which
will require the accurate simulation both of the morphology
as well as the amplitude of these density wave modes.

In a series of papers, Romeo (1994a, 1997, 1998) conducted systematic
studies of softening on the quality of N-body simulations, and advocated
against using large particle softening in N-body simulation.
Furthermore, he pointed out that ``A delicate aspect of the
relaxation problem that has not been considered in the previous
discussion concerns the effects of collective interactions
between particles and self-consistent fluctuations on the
dynamical evolution of the system (e.g., Romeo 1990 and the references
therein; Weinberg 1993; Zhang 1996). A thorough treatment
of collective effects would demand titanic efforts even in
simpler models (cf. Weinberg 1993)'' (Romeo 1997)\footnote{Weinberg
(1993) treated the effects of collective interaction on
the relaxation processes in symmetric and periodic systems.
These systems did not contain self-organized density wave patterns,
thus are passive systems.}.

It is indeed to these aspects of the collective interactions of
particles, as well as the self-consistent fluctuations in galaxy dynamics,
that the work of Z96, Z98, Z99 and Zhang \& Buta 
(2007, 2015) were devoted. The current paper is a continuation 
of this line of work, focusing in particular on factors that
produced the difference between the simulated and observed
galaxy characteristics.  We highlight below that overly-softened gravity 
modifies the Newtonian force into a more sluggishly-interacting 
one that hampers the proper operation of collective effects.  
This might not be a serious concern if one's interest is in obtaining a 
coherent modal morphology that mimics the observed galaxy morphology 
(e.g. Donner \& Thomasson 1994, hereafter DT94).  However, 
it can be detrimental to the determination of realistic secular mass flow 
rates that are relevant to physical galaxies.

\subsection{Softening and Its Effect on Secular Mass Flow Rates}

\begin{figure}
\vspace{170pt}
\centerline{
\includegraphics{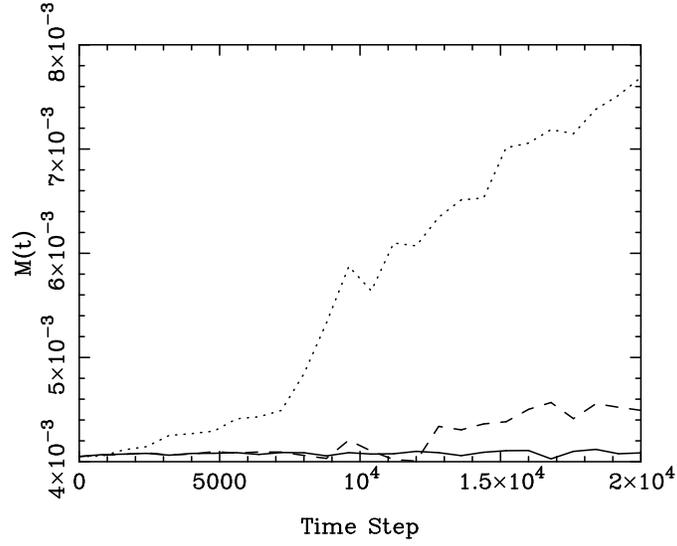}
}
\caption{Evolution of enclosed disk mass within central r=3.5 radius
of a set of 2D N-body simulations
with different softening parameter $a_{soft}$. Solid: $a_{soft}=1.5$,
Dashed: $a_{soft}=0.75$, Dotted:i $a_{soft}=0.25$. N=1 million
particles are used to represent the active disk.}
\label{Figure3}
\end{figure}

In Figure \ref{Figure3}, we present a set of N-body 
simulations of the density-wave-induced
secular evolution of disk galaxy mass distribution,
using the basic state specification similar to that 
first explored in DT94, and subsequently used for the study of collective 
effects and secular evolution in Z96, Z98, Z99.  
The details of the basic state properties (mass distribution,
rotation curve, and velocity dispersion distribution) 
are given in Appendix B of the current paper, while the
dynamical mechanism which induced the secular evolution of the basic
state mass distribution is summarized in Appendix A.  In the current
section we will only present the signatures of radial
mass flow, while leaving the quantitative comparison to previously
derived analytical mass flow rates to \S3 (see especially \S3.6).

All the simulations in the current paper are performed in a two-dimension 
(2D) configuration, using a particle-mesh approach on a polar grid 
(Miller 1976).  The polar codes used in the current paper were written 
by the author based on algorithms described in Thomasson (1989). 
These codes have been used extensively in the development of
the secular evolution theory in Z96, Z98, Z99, and many
results were cross-checked with that 
presented in DT94.  More details of the simulation grid and simulation
procedure are given in Appendix C of the current paper.  
N=1 million active disk particles are used in this first set of simulations 
presented in Figure \ref{Figure3}.

The polar grid used has 220 radial rings and 256 azimuthal
spokes.  The distribution of the radial spacing is exponential,
and the azimuthal spokes have equal spacings.
The time step is chosen such that 1256 time steps represent 
one rotation period at r=20.  The gravitational constant G is 
renormalized to accommodate this choice of spatial and temporal 
normalization and resolution (see further details in Appendix C).  
{\em These grid and time step resolutions are used in the rest of 
the simulations presented in this paper as well (except for
the grid-resolution tests presented in Appendix D3), 
and only the particle numbers N and the softening parameters $a_{soft}$
are changed and will be noted in each case.}

Incidentally, if one does not feel comfortable with the
normalized units used here (which was originally introduced
in Thomasson [1989] and Donner \& Thomasson [1994]),
one can attach a ``kpc" to the normalized length unit, 
a ``$10^{11} M_{\odot}$'' to the normalized mass unit, 
and can scale the time steps to the unit of 1256 steps 
for one rotation period at r=20, for the results 
presented in the main body of the paper (and half as
many steps per rotation period for results presented
in Appendix D3 for a coarser grid and a larger effective
time step choice) to aid in the intuitive comparison 
with physical galaxy properties.

The curves in Figure \ref{Figure3} 
are for the evolution of enclosed disk mass within the
central $r=3.5$ region (in the scale of this disk which has
corotation radius in the approximate range of 18-30 depending
on the softening length used and on the epoch of secular evolution, 
the disk region within $r=3.5$ corresponds roughly to
the central bulge region).  The three curves, 
from bottom to top, correspond to softening parameter values of 1.5 
(as used in DT94 and Z98), 0.75, and 0.25 in the unit of the
radius of the innermost grid ring (or inner computation boundary)
size of 1.  As can be seen from the figure, the choice of softening
of 1.5 (which is a common choice by N-body simulators) produced 
a barely noticeable mass inflow in the duration of the simulation run, 
which results in a central mass growth of 6.5\% over roughly 25
rotation periods, which is obviously quite inadequate
in transforming galaxy morphology in a Hubble time.
Gradually reducing softening as shown in Figure \ref{Figure3} is seen to
systematically increase the mass inflow rate.  

The price one pays for reduced softening in small-N simulations
is the increase in relaxation (heating) rate and the formation of
local instability clumps (see later among the morphological plots in
\S3.1).  To compensate for the unrealistic degree of
relaxation in small-softening simulations, one needs
to use a correspondingly increased number of particles. 
How much the particle number needs to be increased for a given
choice of softening parameter can in fact be obtained empirically.
For 2D simulations, we found that
for a factor of $f$ reduction in the particle softening parameter,
an increase of $f^2$ is needed in the number of particles
to keep the heating effect in check.  This scaling
behavior had also been found in previous studies
by other authors (e.g. Thomasson, Donner \& Elmegreen 1991).

\begin{figure}
\vspace{170pt}
\centerline{
\includegraphics{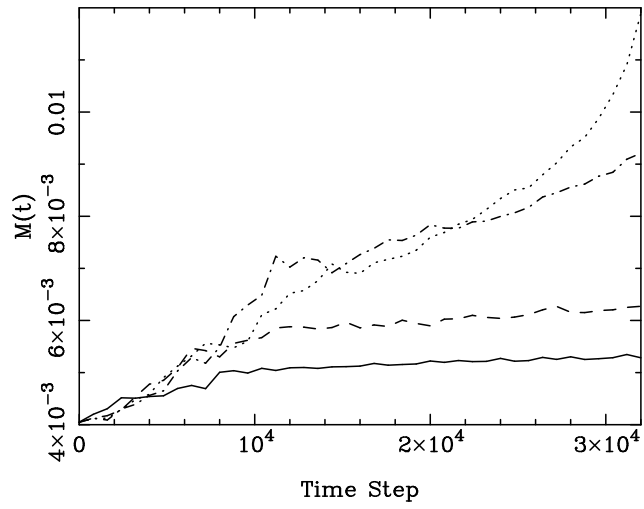}
}
\caption{Enclosed mass within central r=3.5 of the N-body disk
for runs with different number N of active disk
particles, and a constant $a_{soft}={0.1}$.
Solid: N=1 million particles. Dashed: N=10 million particles.
Dotted: N=20 million particles. Dash-Dotted: N=40 million particles.}
\label{Figure4}
\end{figure}

In Figure \ref{Figure4}, we show a set of simulations with 
changing particle numbers while holding the softening parameter 
$a_{soft}=0.1$ (the smallest softening choice that will be
explored in the current paper).  For the small particle-number runs,
the rapid mass inflow is seen to saturate at an earlier time step,
which corresponds to the time (as revealed from the morphological
plots) when the spiral activity is damped by the excess
heating due to the insufficient number of particles,
especially in the outer disk region where the surface density
is low, the grid size is large, and the particle
numbers are small.  For large particle number runs, the rapid mass
inflow is seen to remain at a constant rate all the way until
the end of the run, which corresponds to 25 galactic rotations
at the reference radius r=20.  This indicates that the increased
particle number from Z96 (20 million/100,000=200) is adequate in offsetting 
the reduction in softening (f=1.5/0.1=15), conforming to the $f^2$
scaling law mentioned in the previous paragraph.  

Furthermore, we note that the maximum mass inflow rate observed here
(from the near constant slope of the large particle-number runs with
$a_{soft}=0.1$) corresponds to about 60\% increase in enclosed mass 
within r=3.5 (the bulge region) over 10
rotation periods, or roughly 1/5 of a Hubble type.  This
level of mass accretion is more than sufficient to transform
the Hubble type of a galaxy by several stages in a Hubble time,
consistent with the level of mass flow rates derived for
physical galaxies (Zhang \& Buta 2007, 2015).

\begin{figure}
\vspace{170pt}
\centerline{
\includegraphics{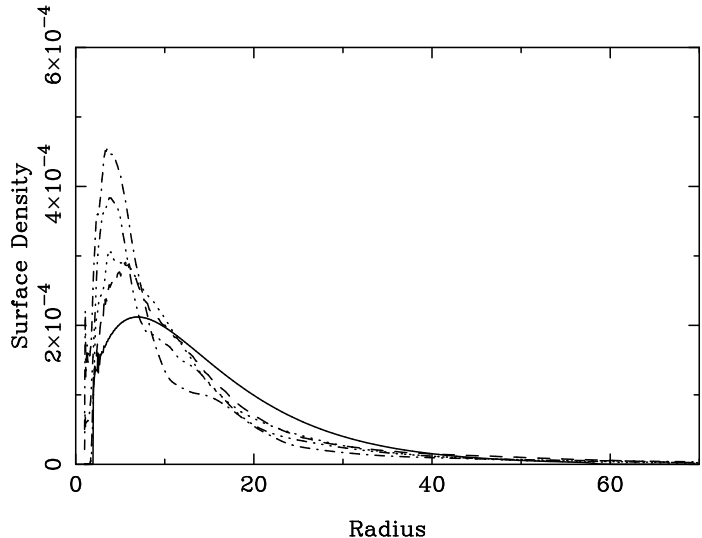}
}
\caption{Basic state surface densities at the beginning (solid), and the
end of four N-Body runs with different number of particles.
Dashed: end surface density for 1 million particles.
Dotted: end surface density for 10 million particles. 
Dash-and-Single-Dotted: end surface density for 20 million particles.
Dash-and-Triple-Dotted: end surface density for 40 million particles.
$a_{soft}=0.1$}
\label{Figure5}
\end{figure}

In Figure \ref{Figure5}, we show 
the disk surface density at the beginning of the run,
and at the end of the four runs with different number of
particles (as presented in Figure \ref{Figure4}, but here
with different line styles from Figure \ref{Figure4} for each case
due to the need to represent the ``before'' surface density).
It can be seen that the large mass inflow of the 
large-particle-number runs indeed builds
a more substantial bulge.  However, the 20 million
particle run in fact produced more substantial bulge-building
than the 40 million particle run.  This is consistent
with the trend observed in Figure \ref{Figure4}, where it is seen that the
40 million particle run has somewhat reduced mass inflow rate
near the end of the run.  This saturation of the mass flow
rate at a specific level of particle number matched to the
decrease of softening length is reassuring since we expect the
collective effects to be relatively independent of particle number
once the excess relaxation effect is held under control by
the use of sufficient number of simulation particles. 

\section{FOUR SETS OF VARIABLE-SOFTENING N-BODY RUNS}

The initial tests of the last section show that by significantly
decreasing the amount of particle softening, and simultaneously 
increasing both the number of simulation particles used as well as
the degree of grid resolution, we can achieve increased 
radial mass flow rates to levels comparable to that needed for a
significant change of galaxy morphological types over
a Hubble time, as indicated in observational studies
(Zhang \& Buta 2007, 2015).

A natural question arrives as to whether this observed increase 
in radial mass flow rate is a true physical effect, or else is 
a numerical artifact due to noise brought about by the small
softening and the corresponding increased collisional relaxation.  
To answer this question, in this section we present
detailed correlations of density wave modal characteristics
and the observed mass flow behavior in the N-body simulations.  Softening
will be shown to change the equilibrium amplitude of the modes
formed, but it hardly changes the {\em level of agreement} between 
the mass flow rate {\em predicted} using the modal parameters found 
in the simulations, and the corresponding mass flow rate {\em measured} 
in the same simulations (i.e.  smaller equilibrium wave amplitudes
are found to correspond to smaller measured mass flow rates, 
given precisely by the theoretical predictions
using the corresponding wave amplitudes).

In what follows, we will first establish the modal nature of 
the density wave patterns formed from originally featureless 
(axisymmetric) basic
state of the galactic disk.  After presenting the modal
characteristics, we will go on to demonstrate the presence
of the potential-density phase shift distribution, which
leads to the secular torque interaction between the basic
state mass distribution and the wave-mode potential field.
This torque interaction leads to the angular momentum exchange
between the basic state and the density wave, which further
leads to the radial mass flow behavior.  We demonstrate
that both the qualitative and quantitative mass flow behavior
observed in these simulations are consistent solely with
the mode/basic-state interaction picture, and are inconsistent
with either the transient-wave- or noise-induced mass flow behavior.

Four sets of 2D N-body simulations are performed
utilizing the same polar grid (220 radial grid cells, 
256 azimuthal grid cells), the same number of particles (20 million), 
the same duration of simulation run (32768 steps, corresponding to roughly 
25 galactic rotation periods at radius 20), and the same time resolution of 
1256 time steps per galaxy rotation period time at r=20, but with 
the particle softening parameter having the values of $a_{soft}=1.5, 
0.75, 0.25, 0.1$, respectively, in the Plummer softening scheme 
(equation \ref{eq:soften}).  

All the runs have the same basic state 
specification as that of Z98 (see also Appendix B of the current
paper), which was also the basic state 
used in the previous section (\S2).  This basic state specification allows 
the setup of a galactic resonant cavity between the inner bulge region 
and the corotation radius in the mid-disk, and thus the formation of 
unstable density wave mode.  The resulting corotation radius ranges
from $r_{co} \approx 28$ (for $a_{soft}=1.5$) to $r_{co} \approx 23$
(for $a_{soft}=0.1$) at the initial emergence stage of the mode,
and the corotation radius tends to decrease in the late stages as
the secular mass inflow leads to a more centrally concentrated
mass distribution.

\subsection{Morphological Evolution}

\begin{figure}
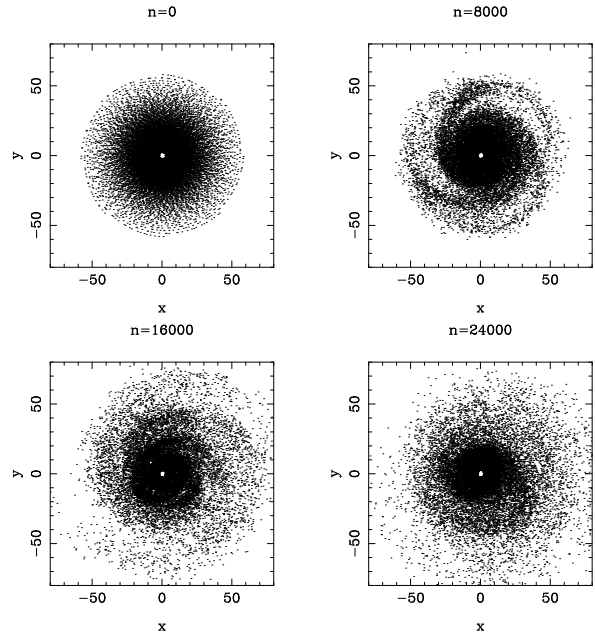

\vspace{210pt}
\includegraphics{Figure6a.ps}
\includegraphics{Figure6b.ps}
\includegraphics{Figure6c.ps}
\includegraphics{Figure6d.ps}
\caption{N-body morphology of a spiral/bar mode at different
time steps. The rotation period at r=20 is about 1256 steps.
The softening parameter is $a_{soft}=1.5$.}
\label{Figure6}
\end{figure}

In Figure \ref{Figure6} to Figure \ref{Figure9}, 
we present the morphological evolution of the 
density wave pattern with basic state and numerical grid specifications
given in Appendices B,C, and with particle softening 
parameter $a_{soft}=1.5, 0.75, 0.25$ and 0.1, respectively.  
The initial disk mass assignment is an axisymmetric modified
exponential disk, and the rotation curve is nearly constant at a value
$v_c \approx 0.1$ in the normalized unit.
The initial velocity dispersion assignment corresponds to the instability 
parameter $Q_T \approx 1$ across the disk.  

\begin{figure}
\vspace{210pt}
\includegraphics{Figure7a.ps}
\includegraphics{Figure7b.ps}
\includegraphics{Figure7c.ps}
\includegraphics{Figure7d.ps}
\caption{N-body morphology of a spiral/bar mode at different
time steps. The rotation period at r=20 is about 1256 steps.
The softening parameter is $a_{soft}=0.75$.
}
\label{Figure7}
\end{figure}

\begin{figure}
\vspace{210pt}
\includegraphics{Figure8a.ps}
\includegraphics{Figure8b.ps}
\includegraphics{Figure8c.ps}
\includegraphics{Figure8d.ps}
\caption{N-body morphology of a spiral/bar mode at different
time steps. The rotation period at r=20 is about 1256 steps.
The softening parameter is $a_{soft}=0.25$.
}
\label{Figure8}
\end{figure}

\begin{figure}
\vspace{210pt}
\includegraphics{Figure9a.ps}
\includegraphics{Figure9b.ps}
\includegraphics{Figure9c.ps}
\includegraphics{Figure9d.ps}
\caption{N-body morphology of a spiral/bar mode at different
time steps. The rotation period at r=20 is about 1256 steps.
The softening parameter is $a_{soft}=0.1$.
}
\label{Figure9}
\end{figure}

A coherent spiral-bar pattern spontaneously emerges from the originally
featureless disk in each case.  The morphology of the m=2 (two-armed)
dominant mode will be shown later in Figure \ref{Figure14}, during its
initial emergence phase.  The mode in fact emerges at progressively earlier 
time steps as softening decreases, this will be shown more clearly later
during detailed analyses.  This is mainly due to the larger seed of 
noise present for modal amplification in the smaller softening cases.

The patterns in various cases contain other contaminants
besides the m=2 mode, for example the $a_{soft}=1.5$ case has some
m=3 component, and the two small-softening cases ($a_{soft}=0.25,
a_{soft}=0.1$) are seen to have local instability
clumps formed in the outer region of the simulation disk.
The small softening cases also tend to have higher heating effect
which submerges the nonlinear pattern within noise at the
later stages of the simulation, though the underlying m=2 mode
can be shown to be persistently present in every case, for the entire
duration of the simulation run (corresponding to
25 rotation periods at radius 20.  See also Appendix E).

In the small softening case of $a_{soft}=0.1$, the pattern is seen 
to have evolved from a spiral-like morphology in the early stage
of the simulation to a bar-like morphology in the later stage, partly 
due to the rapid mass inflow in this small softening case.
The increased central concentration of mass then
favors a bar mode.

\subsection{Phase Evolution}

\begin{figure}
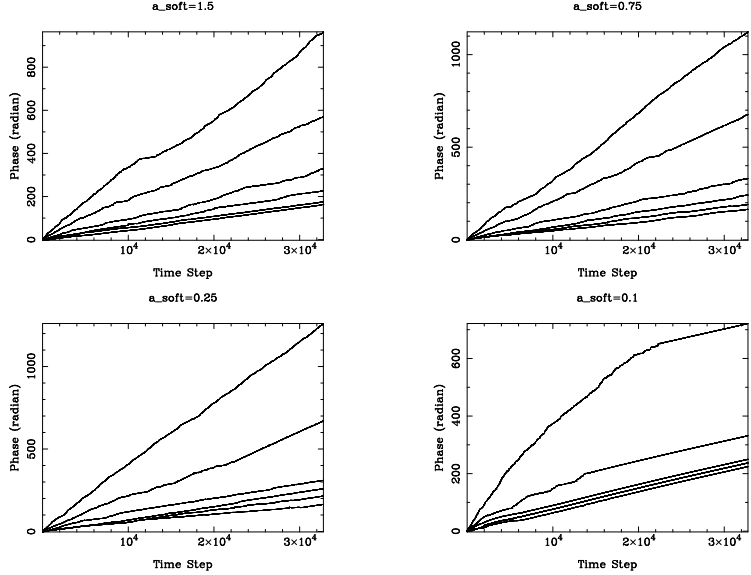

\vspace{180pt}
\includegraphics{Figure10a.ps}
\includegraphics{Figure10b.ps}
 \includegraphics{Figure10c.ps}
 \includegraphics{Figure10d.ps}
\caption{Phase evolution of the m=2 potential at radial
locations (from top to bottom) r=5,10,15,20,25,30,
respectively, using four different
softening parameters.}
\label{Figure10}
\end{figure}

In Figure \ref{Figure10}, we plot the time evolution of the 
azimuthal phase of the m=2 (two-armed) perturbation-potential component 
from the above four runs, at a series of radii across the simulation disk. 

The first thing we notice from this plot is the coherent, approximately
linear, evolution of the phase at the various radial locations, for
all softening choices, indicating that the patterns involved are
rotating at nearly constant angular speed at each radius (though
the pattern speed may change across the different radii, indicating the winding
up of the pattern, especially for inner galaxy disk, and for large
softening cases).  This shows that the patterns that spontaneously
emerged are likely to be unstable {\em modes} of the disk rather
than random transient waves, with the latter not expected to have
a coherent phase evolution over the lifetime of a galaxy.

Furthermore, we observe that while the phase evolution curves for 
the large softening runs diverge at the different radial locations, 
indicating a change of pattern speed for the different portions of 
the spiral arms (or the wrapping up of the pattern with time), the
run with the smallest softening length ($a_{soft}=0.1$ case) 
shows an extremely constant pattern speed between r=10 and r=30 (i.e. between
the inner Lindblad resonance and slightly beyond corotation radius).  
As a matter of fact, two of the 6 curves nearly overlap
(that is why there appear to be only 5 curves).  

Note also that in all cases the more rapid 
phase advance between r=5 and r=10 locations is likely to be
due to a nuclear pattern with a separate pattern speed, since
the outer main pattern has its inner Lindblad resonance at around r=10.

\subsection{Pitch Angles of the Patterns}

\begin{figure}
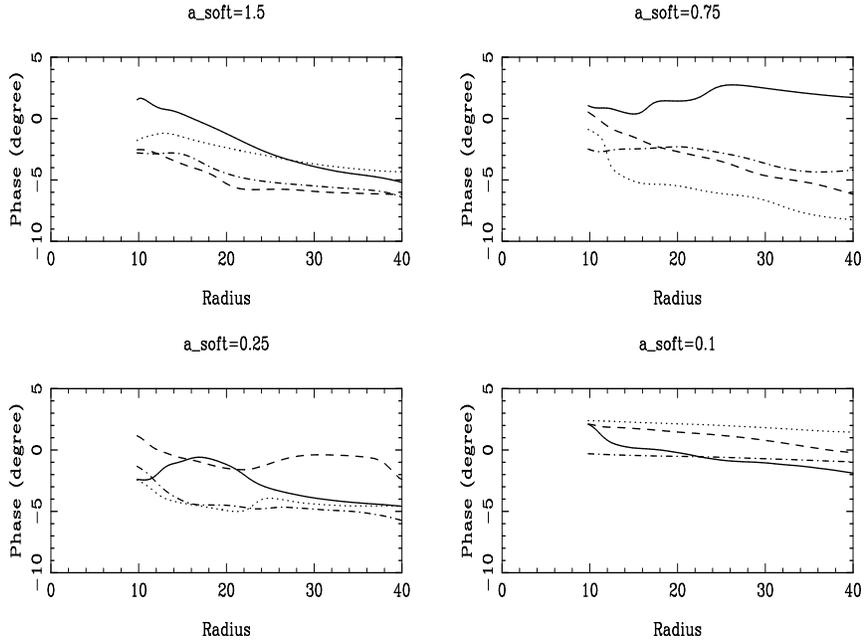

\vspace{220pt}
\includegraphics{Figure11a.ps}
 \includegraphics{Figure11b.ps}
 \includegraphics{Figure11c.ps}
 \includegraphics{Figure11d.ps}
\caption{Phase versus radius of the m=2 potential
at time steps 8000 (solid), 16000
(dashed), 24000 (dotted), 32000 (dash-dotted),
for runs with four different softening length choices.}
\label{Figure11}
\end{figure}

In Figure \ref{Figure11} we plot the m=2 potential phase-versus-radius
at different time steps for the four runs, in order to study the variation
of pitch angle of the spiral-bar pattern versus radius.

For most plots, it can be seen that the slopes of the m=2 phase-versus-radius
starting from r=10 (which is close to the inner Lindblad resonance location of
the dominant mode) to r=40 (which is close to the outer Lindblad resonance,
or the outer limit of the mode) are within a narrow range, especially if we
exclude the step n=8000 (solid) lines, which may be contaminated by the
noisy features that arose during the mode emergence phase.
This near constancy of the slopes of the phase-versus-r curves indicates the 
presence of logarithmic spirals, or else skewed-bars, 
with nearly constant pitch angle.  Some of the regional bumps on the curves
could be due to contamination from local instability clumps or noise. 

Of particular interest is the plot for $a_{soft}=0.1$,
which, as we had previously shown, harbors a pronounced bar pattern in the
later stage of the run.  The evidence of the spiral-to-bar
morphological change can be seen here from the phase-versus-radius plot.   
At the earlier stages of the run (step 8000 and earlier) there is a more
steep gradient for the phase versus radius curves, indicating
the presence of a (skewed) spiral pattern, whereas for the
later stage of the run the phase
versus radius curves flatten out, reflecting the reduced
skewness of the bar.  However, a comparison with the $a_{soft}=0.1$ 
phase-versus-time plot (Figure \ref{Figure10}) shows
that between r=10 and r=30 the phase-versus-time
curves have nearly constant slope throughout the simulation run, indicating 
a constant pattern speed throughout both the spiral and bar
phases for this evolving mode.  This constant pattern speed
during the modal morphological evolution
is also supported by the time-segmented power spectrum analysis of 
Figure \ref{Figure13}, which we will present next.  
These results thus corroborating one another to show that the spiral and bar
modes are from the same modal family (i.e. the
spiral evolves smoothly to a bar, rather than spiral disappearing
and a bar of a different pattern speed later appearing).
Therefore, even with the modal shape evolving from spiral to bar, 
we are still led to the quasi-steady modal picture,
rather than a transient and recurrent pattern picture\footnote{In
essentially all the numerical simulations of spontaneously-formed
density waves in disk galaxies,
the unstable bar modes were obtained from initially
going through a spiral phase (see, for example, Sparke
\& Sellwood 1987).  This is because the formation of
density wave modes requires the removal of angular
momentum from across the inner disk (within corotation) and the deposit
of angular momentum to the outer disk region outside corotation,
and this removal and deposition of angular momentum
in turn depend on the presence of potential-density
phase shift of the pattern involved (Z98), and the
phase shift value is smaller for bars than for spirals
(since bars generally are less skewed than spirals).
Therefore a bar mode employs a spiral modal shape during 
its youthful developmental stage to remove angular
momentum needed for its growth, until it reaches
the adequate amplitude and modal shape to become
its mature self.  This process incidentally lends support to
many of the so-called ``super-fast bars'' found in the 
Zhang \& Buta (2007) and Buta \& Zhang (2009) samples, which
have the bar corotation radius located intermediate in the bar rather than
at the bar end as dictated by conventional wisdom from passive orbit
analysis.  If all bars are formed through an initial
spiral phase, and if the corotation radius 
during the spiral phase is located midway in
the spiral arms, then we should not be surprised to find
a fraction of bars to have this property as well, since
the transition stage between spirals and bars can appear
as either an open spiral or a skewed bar.}.

\subsection{Power Spectra}

\begin{figure}
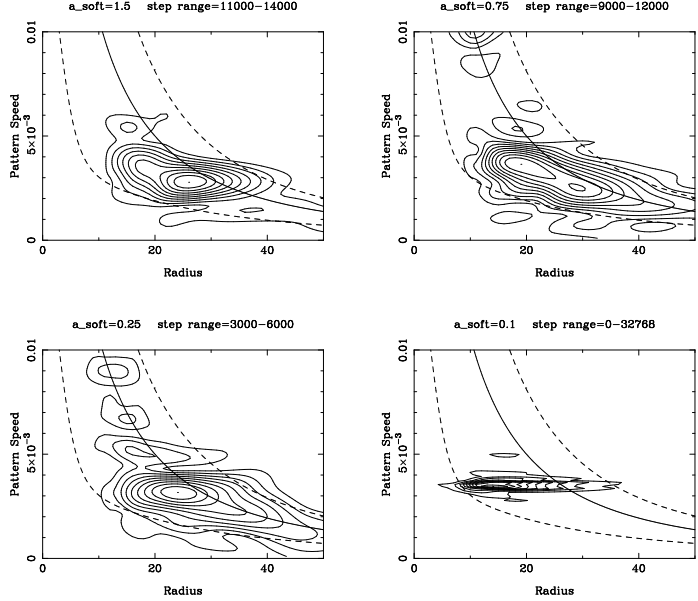

\vspace{180pt}
\includegraphics{Figure12a.ps}
 \includegraphics{Figure12b.ps}
 \includegraphics{Figure12c.ps}
 \includegraphics{Figure12d.ps}
\caption{Contour plot showing comparison of the
power spectra from m=2 potentials at different radii during
indicated time intervals, for the four different
softening parameters and with 20 million particles.
The solid line indicates
galactic rotation speed $\Omega$, the two dashed lines
are $\Omega \pm \kappa/2$, where $\kappa$ is the epicycle
frequency.}
\label{Figure12}
\end{figure}

In Figure \ref{Figure12} we present the contour plots of
the power spectra of the m=2 potential perturbations, 
from the above four runs with differing softening
parameters during selected time intervals, and over the
galactic radial range where the dominant mode is expected
to have significant amplitude.  The solid line in each figure
indicates the galactic rotation curve in angular speed,
the dashed curves delineate the range where modal
amplification within the galactic resonant cavity
is most likely to occur (Bertin et al. 1989a,b). 

Note that for the larger softening runs ($a_{soft}=1.5, 0.75, 0.25$) we have
only plotted during the time interval where the first dominant
mode emerges and reaches quasi-steady state.  This mode will further
evolve and change pattern speed during the remainder of the simulation
run\footnote{This is the reason we have only plotted the mode
for limited time range so as to avoid pattern-speed smearing.  The
end of the above plotting range does NOT indicate the end of the modal
activity.  In Appendix F, we will present a complete set of
power spectra evolution plots for the $a_{soft}=0.25$ case, throughout
the entire duration of the simulation run to demonstrate the longevity of
the mode.}.  Later in the text (Figure \ref{Figure33} - \ref{Figure36})
we will present further evidence of modal evolution with time.
Note also that the pattern speed for the $a_{soft}=0.75$
case varies slightly with radius (i.e. the power spectrum has a
nonzero slope), indicating the tendency for the
winding up of the pattern with time for this softening run.

Cross-comparison with the phase-versus-time/phase-versus-radius 
plots of the previous two sub-sections shows that changes in 
pattern speed and spiral winding angle appear {\em continuously} both
in time as well as in space throughout the simulation run, 
rather than appear as random fluctuations that a transient spiral
pattern would produce. 

\begin{figure}
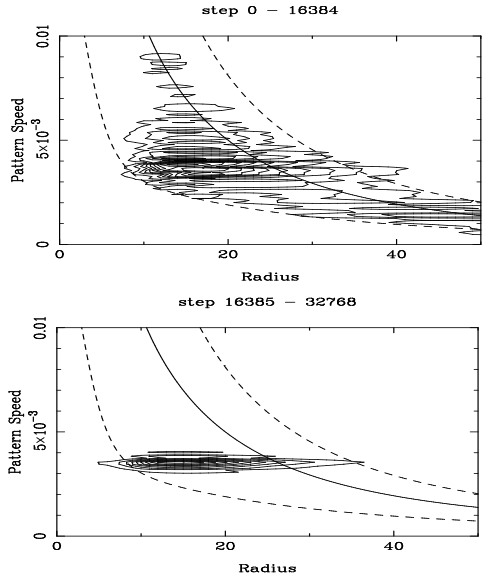

\vspace{160pt}
\centerline{
\includegraphics{Figure13a.ps}
\includegraphics{Figure13b.ps}
}
\caption{Power spectra from m=2 potential at different radii
for the 20 million particle and $a_{soft}=0.1$ run.
Top: step = 0 - 16384, or the spiral-dominated phase.  
Bottom: step = 16385 - 32768, or the bar-dominated phase.
The solid line indicates
galactic rotation speed $\Omega$, the two dashed lines
are $\Omega \pm \kappa/2$.}
\label{Figure13}
\end{figure}

For the small softening length $a=0.1$ run,
despite of the transition from spiral-shaped to bar-shaped morphology,
the pattern speed of the mode is seen to remain constant.
This can be seen especially clearly if we
break up the time duration of the power spectrum
calculation into two segments, as shown in Figure \ref{Figure13}. 
It can be seen here that the main
power spectrum's peak stayed at nearly the same vertical
scale (indicating the same pattern speed) for both time segments.
The conclusion of the constancy of the pattern speed
throughout the spiral-to-bar morphological evolution
during the $a_{soft}=0.1$ run is further confirmed when we re-examine
the m=2 phase-versus-time plot (Figure \ref{Figure10}).
This shows that with the choice of small particle softening
(and in physical galaxies, the {\em zero} particle softening in 
the Newtonian force law), and with the
thickness of the disk here partly represented by grid softening, allow
self-organization and global-self-consistency of the galaxy
density wave mode to be accomplished and maintained during both
the spiral and bar evolution phases.  

\subsection{Potential-Density Phase Shift and Radial Mass Flow}

Having established the modal characteristics for the
simulated density wave patterns, we now look into the issue
of radial mass flow in the parent disk galaxies
containing these patterns.

\begin{figure}
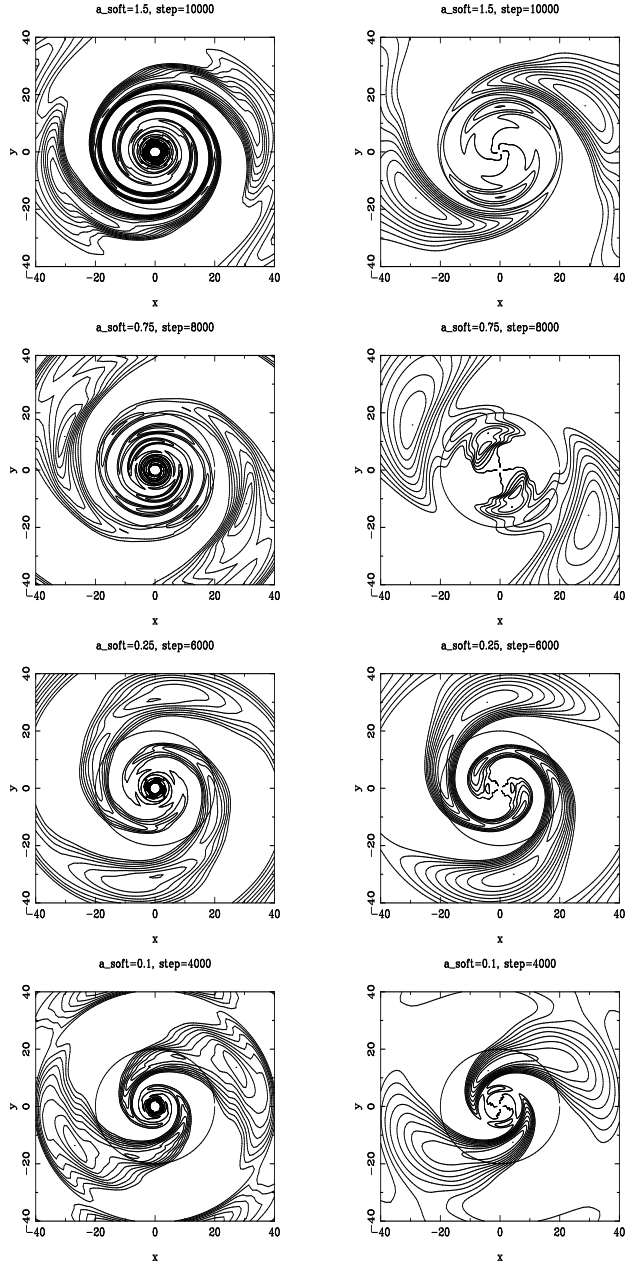

\vspace{440pt}
\includegraphics{Figure14al.ps}
\includegraphics{Figure14ar.ps}
\includegraphics{Figure14bl.ps}
\includegraphics{Figure14br.ps}
\includegraphics{Figure14cl.ps}
\includegraphics{Figure14cr.ps}
\includegraphics{Figure14dl.ps}
\includegraphics{Figure14dr.ps}
\caption{Morphology of the density (left frames)
and (negative) potential (right frames) of the m=2 
modal component for the four runs at the respective
time steps (10000, 8000, 6000, 4000) as indicated in the figure. 
Each frame is individually scaled to have 10 contours.
Only positive contours are plotted.  The circle
in each case corresponds to the approximate location
of the corotation resonance for that particular choice
of the softening.}
\label{Figure14}
\end{figure}

In Figure \ref{Figure14}, we present the contour plots of 
the morphologies of the m=2 modal density and potential 
(with the potential negated so as to ease the comparison with density) for 
the four runs at a progressively decreasing time step for each case, 
corresponding to the initial emergence period of the mode for the
particular softening-parameter choice.
We can see that a spiral-like morphology is present
in all cases (this morphology will later evolve into
a more bar-like morphology for the $a_{soft}=0.1$ run).
Furthermore, we see that the potential patterns display
a more open appearance than their corresponding density
patterns, which are more tightly-wound.
{\em This difference in the winding of the density and
potential patterns reveals the radial distribution of the
azimuthal potential-density phase shift},
which we had plotted schematically in Figure \ref{Figure2},
and which we will analyze quantitatively below.
Formal definition of the equivalent azimuthal phase shift
between the perturbation potential and density patterns in
disk galaxies is given in Appendix A of the current paper
(equation \ref{eq:ps}).  

The existence of the quasi-steady azimuthal phase shift distribution,
with potential lagging density within the corotation radius, 
and potential leading density outside corotation, means that 
there is a secular torque applied by the perturbation potential
on the basic state mass distribution, thus a resulting secular 
angular momentum exchange between the wave and the basic state
at the quasi-steady state of the wave mode.   
This in turn leads to the inflow of basic state mass inside corotation, 
and the outflow of matter outside corotation.  

\begin{figure}
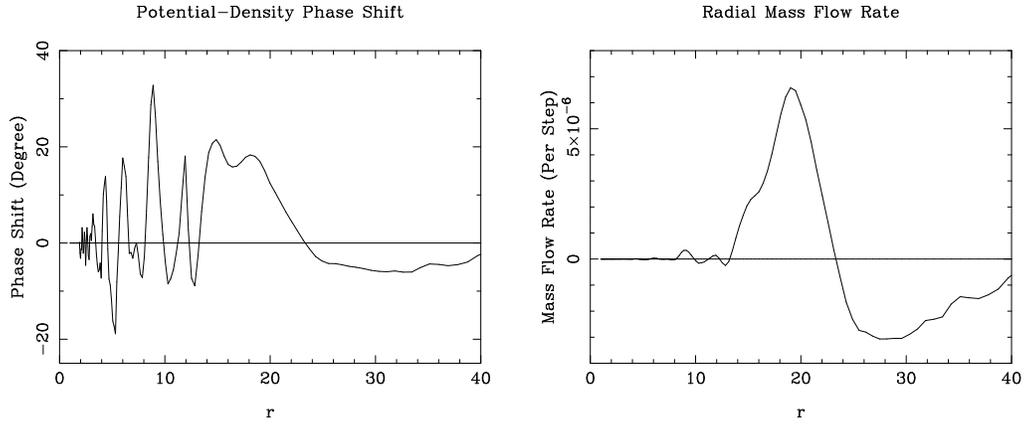

\vspace{160pt}
\includegraphics{Figure15a.ps}
\includegraphics{Figure15b.ps}
\caption{Potential-density phase shift (left frame) and mass flow rate
(right frame) versus r at time step 12800 for $a_{soft}=1.5$ run. These
two figures have identical zero crossings though different amplitude
distributions.}
\label{Figure15}
\end{figure}

In Figures \ref{Figure15}, \ref{Figure16},
\ref{Figure17}, and \ref{Figure18},
we plot the results of the radial dependence of the
azimuthal potential-density phase shift, as well as the
corresponding instantaneous mass flow rate,
calculated using the perturbation density $\Sigma_1$
and perturbation potential ${\cal{V}}_1$ at each radius
according to the expressions given in Appendix A
(equations \ref{eq:ps}, \ref{eq:accre}), at a progressively
decreasing time step for each of the four 
softening choices of 1.5, 0.75, 0.25, and 0.1, respectively.
The time step for plotting is chosen near the peak of the 
first emerged density wave mode in the respective run.

Note that if the density-wave modal surface-density and kinematic
distributions are truly globally self-consistent and
quasi-stationary, and if the galaxy admits only one
dominant mode, the phase shift (as well as mass flow rate) distribution will
experience zero-crossing exactly at one corotation radius.
The radial distribution of the phase shift would be a positive
hump followed by a negative hump, with the transition
between the two humps occurring at the unique corotation radius.
In our simulations, we observe secondary oscillations in the phase shift curve
in the central region of the disk as well as in
the outer region.  These have various causes:
(1) Noise in the perturbation density and potential
used, especially near the region where (and when) the
perturbation quantities themselves are small;
(2) Outer disk truncation effect in a finite-domain
simulation (Z98);
(3) Non-steady nature of the modes formed;
(4) the formation of nested patterns inside the 
primary pattern.  

\begin{figure}
\vspace{160pt}
\includegraphics{Figure16a.ps}
\includegraphics{Figure16b.ps}
\caption{Potential-density phase shift (left frame)
and mass flow rate (right frame)
versus r at time step 10400 for
$a_{soft}=0.75$ run.  These two figures have identical
zero crossings though different amplitude distributions.}
\label{Figure16}
\end{figure}

\begin{figure}
\vspace{160pt}
\includegraphics{Figure17a.ps}
\includegraphics{Figure17b.ps}
\caption{Potential-density phase shift (left frame)
and mass flow rate (right frame)
versus r at time step 5600 for
$a_{soft}=0.25$ run.  These two figures have identical zero
crossings though different amplitude distributions.}
\label{Figure17}
\end{figure}

\begin{figure}
\vspace{160pt}
\includegraphics{Figure18a.ps}
\includegraphics{Figure18b.ps}
\caption{Potential-density phase shift (left frame)
and mass flow rate (right frame)
versus r at time step 8000 for
$a_{soft}=0.1$ run.  These two figures have identical
zero crossings though different amplitude distributions.}
\label{Figure18}
\end{figure}

From the derivation of expressions given in Appendix A,
we see that the mass flow pattern 
{\em has exactly the same zero crossings}
as the phase shift pattern, yet different radial amplitude profile, 
due to the fact that for the mass flow rate expression the local surface 
density, the fractional density-wave amplitude,
and the galactic rotation curve also play a role, 
in addition to the phase-shift or normalized torque-integral contribution.
Thus the appearance of large-value,
multiply-oscillating phase shift features near
the galactic central (or outer) region does not necessarily signal
their importance if the wave amplitude there is small,
and the phase shift oscillations are mainly due to noise.
This point is made self-evident from comparison of the 
left and right frames in Figures \ref{Figure15} - \ref{Figure18}.

\subsection{Comparison of Analytical and
Numerical Accretion Rates}

The evidence we have presented so far supports the view that the 
spiral and bar patterns obtained in these N-body simulations were intrinsic
modes of the basic state of the disk, and they emerge spontaneously 
out of originally featureless (axisymmetric) disks beyond the threshold 
for nonaxisymmetric instabilities.  These instabilities display long-range 
correlation and are self-organized.  That natural laws
implicitly allow such organization to spontaneously appear is part
of the reason the universe displays the observed complexity and hierarchical
organization, starting from the very simple set of fundamental
laws and nearly homogeneous initial conditions of the Big Bang.

In what follows, we present another set of analyses of N-body results,
specifically to check whether the radial mass flow rates obtained in these 
simulations (responsible for bulge building as well as for the spreading out
of the outer disk) agree with the analytical prediction 
(equation \ref{eq:accre} in Appendix A), using the
parameters of the global density wave modes in the same
simulations.  An agreement between the two lends support both 
to the analytical formalism, as well as to the relevance of 
using N-body simulations to characterize the self-organized 
density wave modes. 

\begin{figure}
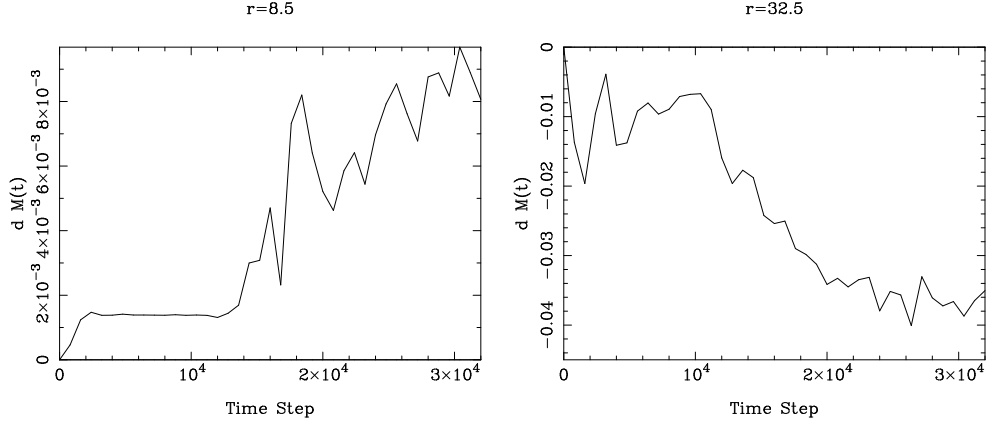

\vspace{145pt}
\includegraphics{Figure19a.ps}
\includegraphics{Figure19b.ps}
\caption{Time evolution of the enclosed mass for a typical 
radius inside corotation, and a typical radius outside
corotation, relative to the initial time step n=0 mass at 
the same respective radius, for the softening length $a_{soft}=1.5$ run}
\label{Figure19}
\end{figure}

In Figure \ref{Figure19} we plot the time evolution of enclosed
mass at two different radii (one inside and one outside corotation), 
after subtracting their respective values at the beginning of the simulation
for each time step represented, for the previous simulation run 
using softening length $a_{soft}=1.5$.  
We observe a general trend of mass inflow inside corotation resonance 
(CR) radius, and mass outflow outside CR.  Note that the CR radius
for this large softening choice is between 25-30, with larger value 
towards the earlier part of the simulation run.  We therefore
choose to plot the outside-CR mass flow curve at radius 32.5.
For progressive smaller softening choices, the CR radius
reduces accordingly.  So for later plots ($a_{soft}=0.75,0.25,0.1$)
we will choose to plot the outside-CR mass flow curves at radius 27.5.

\begin{figure}
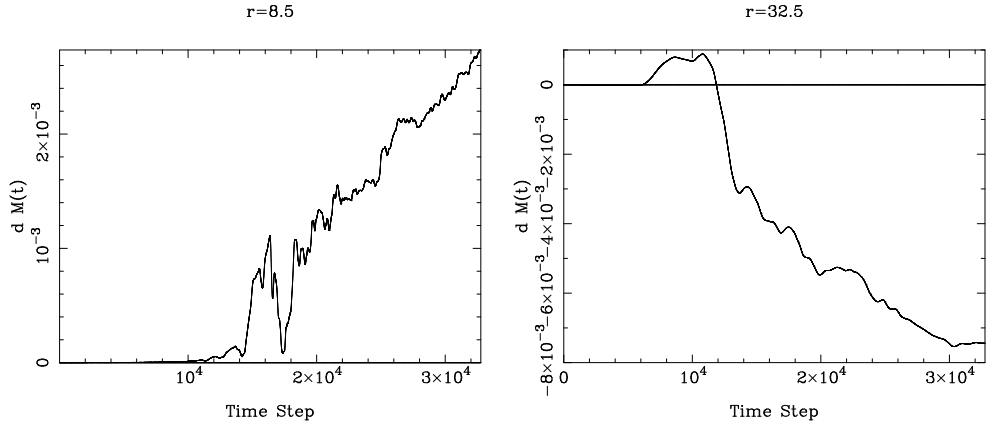

\vspace{145pt}
\includegraphics{Figure20a.ps}
\includegraphics{Figure20b.ps}
\caption{Predicted integrated mass evolution according to
equation (\ref{eq:accre}) within the two different radii, 
for softening length
$a_{soft}=1.5$ run}
\label{Figure20}
\end{figure}

In Figure \ref{Figure20}, we plot the {\em expected} mass increase
within the above two radii by {\em integrating in time} of the instantaneous 
mass inflow rate predicted by equation (\ref{eq:accre}), using the 
density-wave modal characteristics (i.e., the perturbation potential 
and density distributions) as well as galaxy parameters obtained 
in the same simulation run.

Note that the time resolutions for data points in Figures \ref{Figure19} and 
\ref{Figure20}, as well as for subsequent similar pairs of figures, 
are slightly different.  Figure \ref{Figure20} 
and similar figures below are obtained from instantaneous mass flow rates
calculated at every step of the simulation, 
and these are only recorded at selected radii.  Figure \ref{Figure19} and 
similar figures, on the other hand, are obtained using the whole disk
surface mass maps written out at selected time intervals ($\Delta t= 800$ 
time steps in this case), so the time resolution is coarser.  Nonetheless 
the gross features of these two kinds of plots can be compared 
without losing crucial information.

We observe from the comparison of Figure \ref{Figure19} and Figure 
\ref{Figure20} that at the location inside CR, r=8.5, the observed mass 
increase rate is about 3 times as large as the analytically predicted rate,
though the trends of the mass increase are similar between the two.  
This large quantitative discrepancy is particular to the $a_{soft}=1.5$ run,
likely related to the prohibition of the growth of the dominant m=2
intrinsic instability by the large softening.  With the suppression of the 
growth rate of the dominant m=2 mode, some 3-armed as well as other
noisy patterns are seen to emerge.  These non-modal patterns can produce 
radial mass flow yet may not enter into the analytical mass flow prediction.

The mass flow behavior outside corotation likewise shows disagreement
between the two plots:  The actually 
measured mass increase in Figure \ref{Figure19} shows a consistent
outflow pattern as befitting its location outside corotation, 
whereas the expected mass flow calculated using the torque equation 
showed initially a mass inflow, and only afterward changed to outflow.  
This shows that the nonlinear modal organization does not exactly
reproduce the quiescent linear modal picture of the smooth
launching of the mass flow activity.  Also, once again the magnitude
of the actually measured outflow rate is significantly higher than
predicted for this case.
 
\begin{figure}
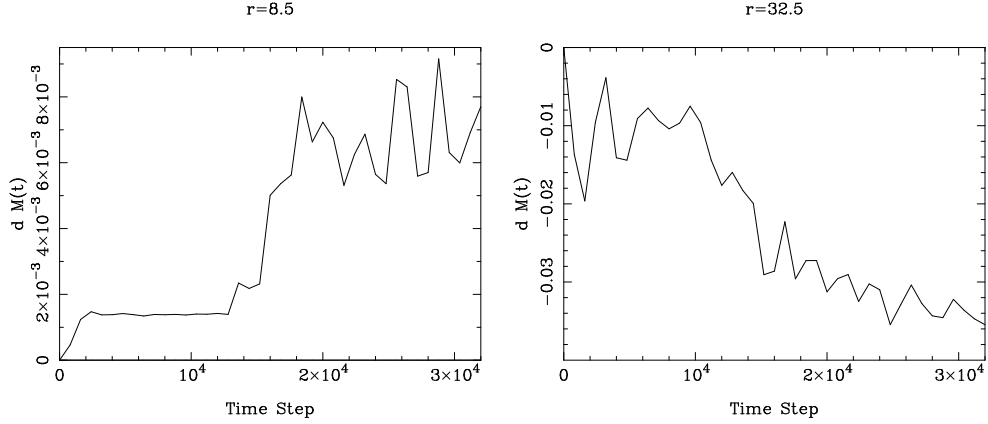

\vspace{145pt}
\includegraphics{Figure21a.ps}
\includegraphics{Figure21b.ps}
\caption{Time evolution of the enclosed mass for a typical
location inside corotation, and a typical location outside
corotation,
relative to the initial n=0 mass at the same respective radius,
for the softening length $a_{soft}=1.5$ run,
using only the even harmonics in potential calculation throughout
the N-body simulation}
\label{Figure21}
\end{figure}

\begin{figure}
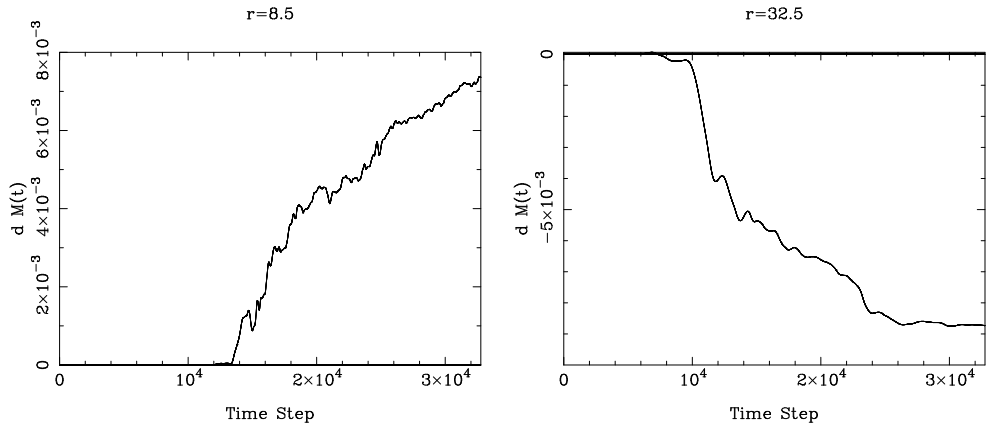

\vspace{145pt}
\includegraphics{Figure22a.ps}
\includegraphics{Figure22b.ps}
\caption{Predicted integrated mass evolution according to
equation (\ref{eq:accre}) within the two different radii,
for softening length
$a_{soft}=1.5$ run,
using only the even harmonics in potential calculation throughout
the N-body simulation}
\label{Figure22}
\end{figure}

We had performed another simulation run with
$a_{soft}=1.5$, but with {\em the potential calculation keeping
only the even azimuthal harmonics} during the Fourier transform
process to obtain the grid potential, to reflect the fact that
observed spiral and bar patterns show predominant bi-symmetric
patterns.  The results for the observed and predicted
mass accumulations inside and outside CR are shown in Figures
\ref{Figure21} and \ref{Figure22}.  Here the predicted and measured 
mass flow rates within CR are seen to be in better agreement,
apparently due to the fact that the enforcement of even-harmonics-only 
coerced a much more effective growth of the collective instability which has
bi-symmetry.  It is now clear that as long as the global
instability has accomplished its growth potential implicit
in the mode/basic-state correspondence, through
either the tinkering of symmetry (as shown here) or through
the reduction of softening (as shown below), the
agreement between the theoretical prediction and the actual
measurement in N-body simulations will tend to be good.

The even harmonics enforcement is most effective
in encouraging larger wave amplitude for the $a_{soft}=1.5$
choice.  For the other smaller softening choices we
will present next ($a_{soft}=0.75-0.1$), 
the even harmonics enforcement in fact hampers the wave
amplitude growth in the {\em later stages} of the calculation
because there is the natural tendency in the
small $a_{soft}$ runs to form local instability clumps 
in the outer disk, and through the enforcement of bi-symmetry these 
would-be local clumps that were seen to morph into diffused noise which tend to 
heat the outer disk and inhibit larger wave amplitude to be maintained.  
Still, tests show that even for $a_{soft}=0.75$ the
enforcement of even harmonics encouraged a more vigorous
modal growth during the first 1/3 of the simulation
run, and as a result the mass flow rate for that duration of time 
has better agreement between the predicted and measured values.

\begin{figure}
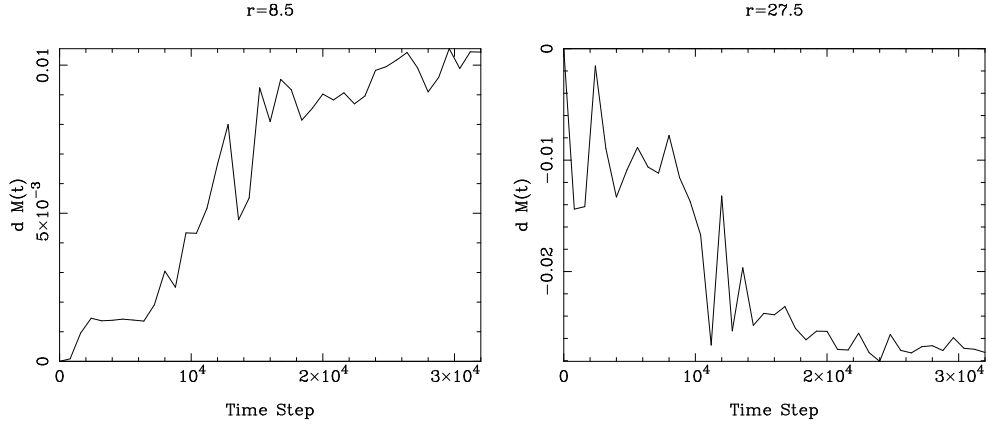

\vspace{145pt}
\includegraphics{Figure23a.ps}
\includegraphics{Figure23b.ps}
\caption{Time evolution of the enclosed mass for a typical 
location inside corotation, and a typical location outside corotation,
relative to the initial n=0 mass at the same respective radius, 
for the softening length $a_{soft}=0.75$ run.}
\label{Figure23}
\smallskip
\end{figure}

\begin{figure}
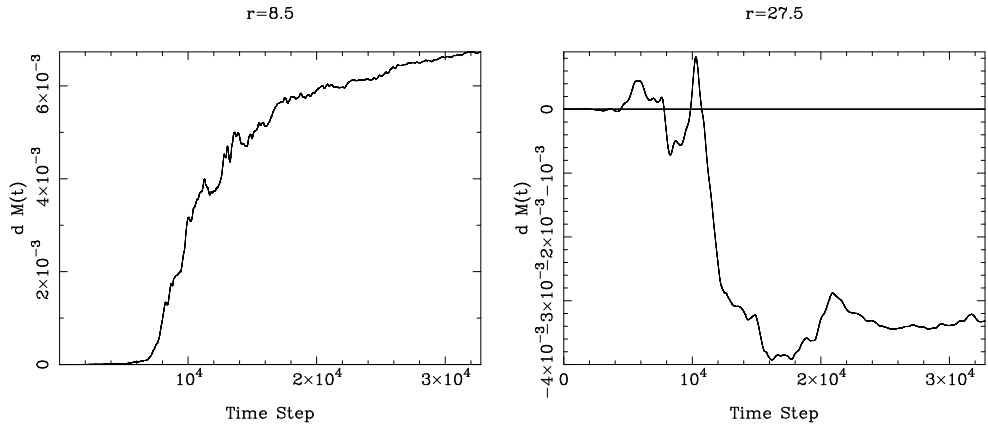

\vspace{145pt}
\includegraphics{Figure24a.ps}
\includegraphics{Figure24b.ps}
\caption{Predicted integrated mass evolution according to
equation (\ref{eq:accre}) within the two different radii, 
for softening length $a_{soft}=0.75$ run.  The second
frame (for r=27.5) used only even harmonics in the mass flow calculation.}
\label{Figure24}
\smallskip
\end{figure}

In Figure \ref{Figure23} we plot the actually measured enclosed mass 
evolution compared to its value at the beginning of the simulation, 
for $a_{soft}=0.75$ run, with the left frame shows the typical mass flow trend
inside CR, and the right frame is typical of the behavior outside CR.
In Figure \ref{Figure24}, we plot the expected accumulation of mass 
within the two radii by integrating the instantaneous mass inflow rate 
according to equation \ref{eq:accre}.

We see that at r=8.5, the trend of mass accumulation with time
between these two figures is extremely similar.  Furthermore, the value of
analytically inferred rates is only about 30-40\% smaller than
the actual observed rates.  This is a much better agreement than 
the case we had shown before for the $a_{soft}=1.5$ run using all harmonics. 
This shows that decreasing softening does make a significant difference 
in our ability to correctly model the collective effects.
As we have commented before, enforcing by-symmetry in the potential
calculation will lead to further improved agreement between the
predicted and measured mass flow rates, for the initial 1/3 of the
simulation duration, before the heating effect sets in.

In the $a_{soft}=0.75$ run, the predicted mass accretion
for r=27.5 has some large fluctuations due to the
contamination of multi-armed patterns.  We therefore chose to
use only {\em even harmonics in the accretion calculation
for the r=27.5 location (even though the N-body simulation
itself was run with the full harmonics set)}.  This 
procedure in the analysis helps
to filter out effect of the transient features
and leads to better agreement between the measured
and predicted mass accumulation rate at r=27.5, though
some residual fluctuations still exist.
Once again the magnitude of the outflow is smaller for
the predicted than the measured for this run, as for
$a_{soft}=1.5$.

\begin{figure}
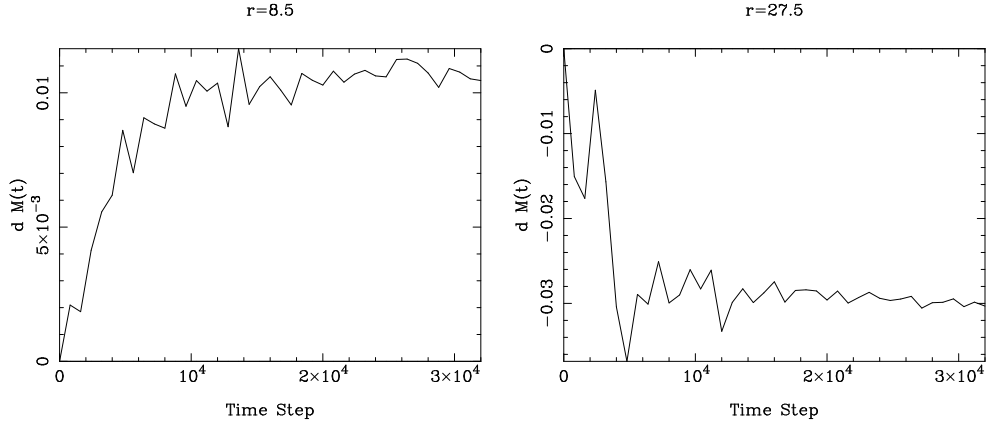

\vspace{145pt}
\includegraphics{Figure25a.ps}
\includegraphics{Figure25b.ps}
\caption{Time evolution of the enclosed mass for a typical 
location inside corotation, and a typical location outside
corotation,
relative to the initial n=0 mass at the same respective radius, 
for the softening length $a_{soft}=0.25$ run}
\label{Figure25}
\smallskip
\end{figure}

In Figure \ref{Figure25}
we plot the actually measured enclosed mass evolution
at the two different radii compared to its value at the
beginning of the simulation, for the $a_{soft}=0.25$ run.
In Figure \ref{Figure26}, we plot the predicted
accumulation of mass within the two radii by integrating the
theoretical instantaneous mass inflow rate equation
\ref{eq:accre}.

For this run, for the location r=8.5 the measured enclosed
mass evolution curve has similar profile as that predicted.
Furthermore, we see that the numerical values of analytically
inferred and actually-measured mass accumulation rates
are {\em nearly identical}.  This shows that reducing the 
softening length to $a_{soft}=0.25$ allows the inter-particle 
interactions to fully represent the collective effect implicit 
in the global modal characteristics.

The predicted mass accumulation curve outside
CR (r=27.5) was more noisy than the inner region behavior due to
the formation of several local instability
clumps at the outer region of the computation domain.
So we have once again {\em used only the even
harmonics for the mass accretion calculation for the
r=27.5 location} (not for the N-body simulation itself,
which was performed using full harmonics), which 
filtered out the contamination
of the local transients.  Here we witness that the numerical
values for the predicted and measured outflow rates are in
better agreement than for the previous two larger-softening
cases.

In Figure \ref{Figure27} we plot the actually measured enclosed-mass 
evolution at the two different radii, compared to their values at the
beginning of the simulation, for the softening choice of $a_{soft}=0.1$, 
as the most extreme case of softening parameter tests.

\begin{figure}
\vspace{145pt}
\includegraphics{Figure26a.ps}
\includegraphics{Figure26b.ps}
\caption{Predicted integrated mass evolution according to
equation (\ref{eq:accre}) within two different radii, 
for softening length $a_{soft}=0.25$ run.  The
second frame (for r=27.5) used only even harmonics 
for mass flow calculation.}
\label{Figure26}
\smallskip
\end{figure}

\begin{figure}
\vspace{145pt}
\includegraphics{Figure27a.ps}
\includegraphics{Figure27b.ps}
\caption{Time evolution of the enclosed mass for a typical 
location inside corotation, and a typical location outside
corotation,
relative to the initial n=0 mass at the same respective radius, 
for the softening length $a_{soft}=0.1$ run}
\label{Figure27}
\smallskip
\end{figure}

\begin{figure}
\vspace{145pt}
\includegraphics{Figure28a.ps}
\includegraphics{Figure28b.ps}
\caption{Predicted integrated mass evolution according to
equation (\ref{eq:accre}) within the two
different radii, for softening length $a_{soft}=0.1$ run}
\label{Figure28}
\smallskip
\end{figure}

Note that the inflection of the curves near time step 5000 for
the $a_{soft}=0.1$ run, inside CR, is linked to the launch of spiral to
bar modal shape transition.  The spiral phase is seen to lead to higher
rate of mass inflow compared to the bar phase, due to the 
more significant skewness and thus the resulting larger potential-density
phase shift of the spiral pattern as compared to the bar pattern.
This higher mass flow rate for the spirals as compared to bars
was also confirmed in the physical galaxy samples
analyzed in Zhang \& Buta (2007, 2015).

In Figure \ref{Figure28}, we plot the expected mass accumulation
of mass within the two different radii for the $a_{soft}=0.1$ run
by integrating the instantaneous mass inflow rate equation \ref{eq:accre}.
We see that the observed and predicted mass flow curves within
the r=8.5 region are quite comparable, with the predicted
rates smaller than the actual rates by about 30\%.  Here the difference 
is obviously not due to the suppression of collective effects,
because the softening length used is even smaller than in the 
$a_{soft}=0.25$ case.  The likely cause for this residual discrepancy 
may be attributed to the viscosity-induced accretion effect
(or dynamical friction) not accounted for by the density-wave torque 
calculation.

The mass flow curves outside CR for $a_{soft}=0.1$ have similar trends,
though the predicted rate once again is smaller in magnitude.

Reexamine all of the outside-CR curves for the previous runs,
we see that they share the same feature of a more pronounced
drop in (measured) enclosed mass at the beginning of the run, signaling
a rapid mass outflow episode as the self-organization of the mode
begins to take place, that was not reflected in the corresponding predicted
mass outflow (this large drop in measured enclosed mass was seen
to be the main reason of the difference in magnitudes of the
measured and predicted outflows in most cases, because after this
initial drop the slopes of the outflows are seen to be quite
comparable between the measured and predicted).
This common discrepancy may be partly a result of
the fact that the outer disk has very low surface density,
and the density wave is poorly represented there, so the 
self-organization process may depart more significantly
from the smooth modal-emergence picture. 

We also note from the above runs that the net amount of mass 
{\em increase} within r=8.5 does not have as substantial a 
difference between the different softening runs (apart from 
the extreme case of $a_{soft}=0.1$), compared to that given 
in Figure \ref{Figure3} and Figure \ref{Figure4} for r=3.5.
This difference is partly accounted for by the shape of the
modified exponential surface density for the choice of basic state,
and as the secular evolution proceeds the entire curve squeezed
inwardly, as opposed to a surface density increase {\em at every radius}
as a result of mass inflow (see below the surface density
evolution in Figure \ref{Figure30}).

We choose this larger radius r=8.5 to confirm the analytical mass
flow equation because near the very central region the presence of
nested resonances complicates the analysis.  Even
r=8.5 itself is officially within the inner Lindblad resonance
of the outer mode\footnote{The resonance itself is shielded from being
active to the wave by the rapid increase in velocity dispersion in this 
region, which forms the so-called ``Q-barrier'' that allows the inwardly
propagating (short trailing) density wave train to be reflected to become
outward propagating (long trailing) wave train towards corotation region
to complete the feedback loop and to form growing modes (Mark 1976).}.  
A location further out in radius turned out to exhibit even earlier
saturation of the accretion behavior because the disk surface
density is depleted as a result of the accretion process
(see Figure \ref{Figure30} below), thus becomes unable to support 
the full amount of mass inflow required by the density wave mode.  

That r=8.5 location can still be a good representative of
the accretion flow due to the outer mode is seen also in the
m=2 morphology plot we presented earlier (Figure \ref{Figure14}). 
There we see that the dominant mode does penetrate to within
the central r=8 region (the nuclear nested mode is well within
the central r=5 region), so the mass inflow predictions
calculated for r=8.5 are still those for the outer dominant mode.
This dilemma of the simulation can be avoided in physical galaxies
through the vertical accretion of cold gas onto the galaxy disk,
followed by its inward channeling to replenish the depleted
surface density of the disk, and to support continued mass 
accretion throughout the lifetime of a galaxy.

\begin{figure}
\vspace{180pt}
\includegraphics{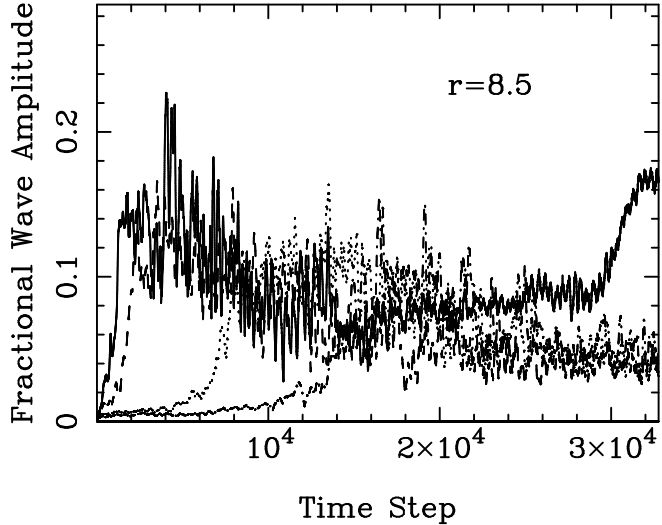}
\caption{Evolution of the fractional wave amplitude
for the four softening choices, at radius r=8.5.
Solid: $a_{soft}=0.1$, 
Dashed: $a_{soft}=0.25$, 
Dotted: $a_{soft}=0.75$, 
Dash-Dotted: $a_{soft}=1.5$, 
}
\label{Figure29}
\end{figure}

In Figure \ref{Figure29}, we plot the evolution of the
fractional wave amplitude (defined as the {\em geometric
mean} of the density- and potential-wave fractional amplitudes)
during the simulation run, for the four softening choices
and at the radius r=8.5.  Note the progressive delayed
emergence of the wave mode, as well as the decrease of 
equilibrium amplitude, as the softening parameter is increased.  
Note also that for $a_{soft}=0.1$ case, the amplitude is re-invigorated 
towards the end of the simulation run, due to the strengthening 
of the central bar as a result of previous mass inflow -- 
this fact can also be discerned from Figures
\ref{Figure27} and \ref{Figure28} as the steepening 
of the enclosed mass slope in the last quarter of the run,
as compared to the corresponding figures for $a_{soft}=0.75$
[Figures \ref{Figure23}, \ref{Figure24}], and for $a_{soft}=0.25$
[Figures \ref{Figure25}, \ref{Figure26}], where the enclosed
mass flattens out towards the end of the simulation, indicating
the slowing down of accretion behavior (which is matched by their
respective decreased wave amplitudes as shown in Figure \ref{Figure29}).  
Of course, the mass flow rate is determined not only by the fractional wave
amplitude, but also by the potential-density phase shift,
as well as by other wave and basic state parameters (Appendix A, equation
\ref{eq:fac}), all of them working in concert to produce
the final radial mass flow rate.  

\begin{figure}
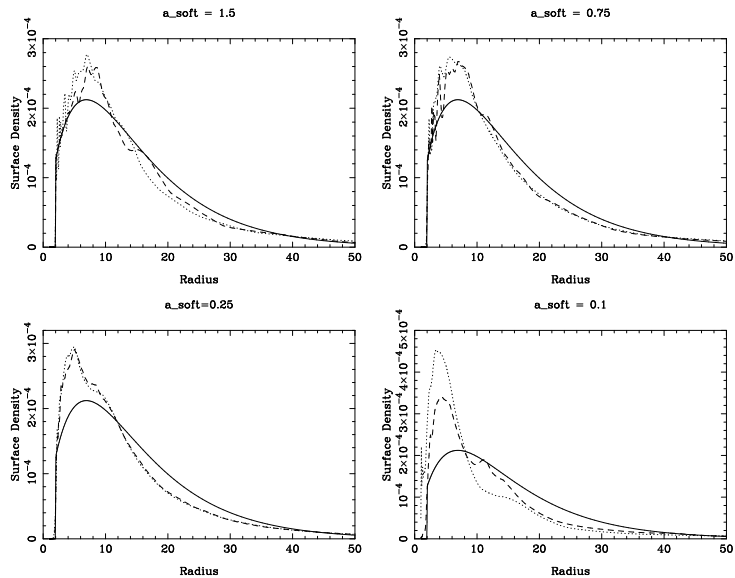

\vspace{190pt}
\includegraphics{Figure30a.ps}
\includegraphics{Figure30b.ps}
\includegraphics{Figure30c.ps}
 \includegraphics{Figure30d.ps}
\caption{Comparison of the surface density evolution of
the N-body runs using four different
softening parameters, and all with 20 million particles.
The different curves in each figure are for time steps
of 0 (solid), 16384 (dashed), and 32768 (dotted), respectively.
Note that the last frame (for $a_{soft}=0.1$) has a different
vertical scale than the other three frames.}
\label{Figure30}
\end{figure}

In Figure \ref{Figure30}, we plot the surface density
evolution of the above four runs.  This plot corroborates
what we had found before from the plots of mass inflow
within the central r=3.5 and r=8.5 radius, as well
as outside the corotation radius.  The $a_{soft}=0.1$ choice
is seen to be the most effective in producing enhanced
mass flow rates, due to its larger wave amplitude and
most persistent density wave activity.

\section{ANALYSES AND DISCUSSION}

\subsection{On the Modal and Quasi-Steady State Hypotheses
of Density Waves in Physical and Simulated Disk Galaxies}

\subsubsection{Previous Work}

The mass flow rate equation used in the last section is exact 
when the density wave pattern involved is a {\em spontaneously formed mode}
that has achieved {\em quasi-steady state} (Z98).  
When this state is achieved, the rate of global wave amplification
through over-reflection at corotation is balanced by the local
dissipation at the spiral-arm collisionless shock.  The wave
amplitude as well as its shape is unchanged (at least on the timescale
of local dynamical time, or galaxy rotation period), and
the only measurable secular change is the slow redistribution
of disk matter both inside and outside corotation, as well as
the heating of disk stars and the energy injection into interstellar
medium (ISM).  The quasi-stationary spiral structure (QSSS) hypothesis
had been used in the past few decades by many workers to explain the
grand-design spiral structures in galaxies (Bertin et al. 1989a,b and the
references therein), though in these previous analyses the mechanisms 
for wave damping and for secular evolution of the basic state were 
not properly identified.  

How good are the modal and QSSS hypotheses?
Are there any objective ways we can judge these?  Earlier
workers (for example, Elmegreen \& Elmegreen [1983,1989]) employed
statistical arguments to support the QSSS hypothesis, and N-body
simulations provided support to the modal and QSSS hypotheses to 
varying degrees (Donner \& Thomasson 1994; Z98).  
Zhang \& Buta (2007, 2015), Buta \& Zhang (2009) used near- and 
mid-infrared images of galaxies to show that for nearby grand-design 
galaxies, the level of quasi-steady state for the density wave
patterns involved can be judged from the coherence of
the potential-density phase shift curves, and from the
{\em level of agreement} between the phase-shift positive-to-negative
zero-crossing predictions for the corotation resonance (CR) radii and the
actual resonance features in galaxies.  Independent studies
have confirmed the accuracy of the potential-density
phase shift method for determining the CRs in grand design galaxies
(i.e., Haan et al. [2009], who compared several methods for CR
determination previously published in the literature, and stated 
``For our galaxies the phase-shift method appears to be
the most precise method with uncertainties of (5-10)\% ...";
as well as Martinez-Garcia et al. [2009,2011] who used a color gradient
method to determine CR, and among their sample galaxies
which overlapped with that of Zhang \& Buta [2007] or Buta
\& Zhang [2009], there is good agreement in the CR locations determined
by these two independent approaches).

\subsubsection{Analogy with Fully-Developed Turbulence}

In what follows, we demonstrate another effective approach for 
determining the quasi-steady state (QSS) of the modes\footnote{We 
use QSS in the following instead of the original QSSS abbreviation 
because we are considering the quasi-steady state of both 
the spiral and bar modes. Note that in order for the phase
shift and torque relation to be applicable, the quasi-steady
state for the wave mode only needs to be maintained on the order
of local dynamical time scale (i.e. galaxy rotation period),
so the secular changes in the modal shape due to the secular
changes of the basic state do not hamper the applicability
of the phase-shift/volume-torque calculation approach.},
in addition to our previous approach of comparing the phase shift
zero crossings with the resonant features on galaxy images.
This time, instead of using just the radial distribution
of the potential-density phase shift, we will actually 
make use of the mass-flow-rate curve, even though the
two curves have the same zero crossings, because as it turns
out the magnitude of the mass flow rate curve provides crucial
information about the quasi-steady modal status as well.

\begin{figure}
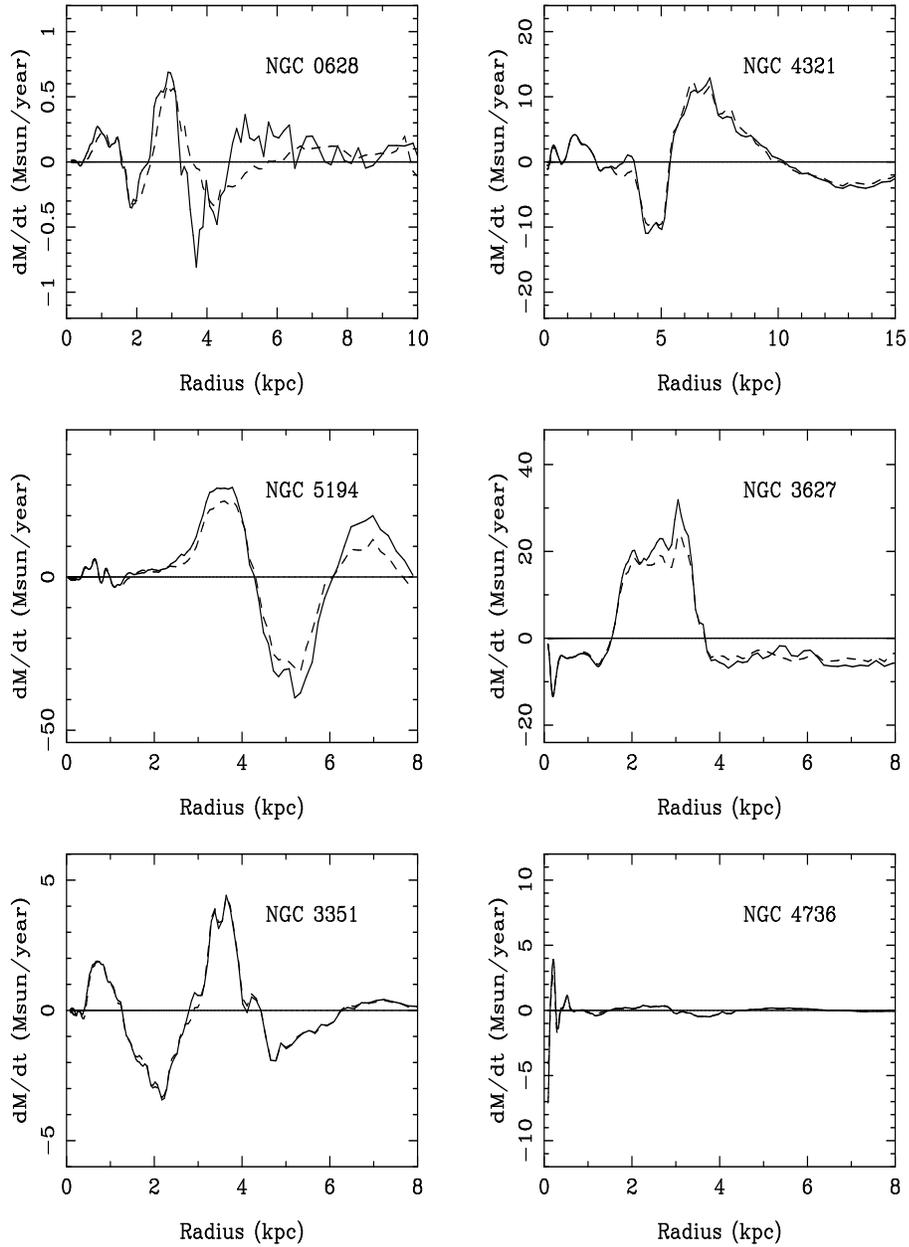

\vspace{430pt}
\includegraphics{Figure31a.ps}
\includegraphics{Figure31b.ps}
\includegraphics{Figure31c.ps}
\includegraphics{Figure31d.ps}
\includegraphics{Figure31e.ps}
\includegraphics{Figure31f.ps}
\caption{Radial mass flow rates for the six sample galaxies
of Zhang \& Buta (2015).  The solid lines are calculated
using the full Fourier spectra of density and potential,
and the dashed lines using only the m=2 components.}
\label{Figure31}
\end{figure}

In Figure \ref{Figure31}, we present the comparison of radial mass 
flow rates of six galaxies calculated using the torque equation 
(\ref{eq:accre}), first using only the m=2 Fourier components (dashed curves) 
for the perturbation density and potential, and subsequently
using the full set of Fourier components (solid curves).
The solid curves had previously been published in Zhang \& Buta (2015).
We see immediately that apart from minor differences,
the m=2 calculation and the m=full (short-hand
for using all Fourier components) calculation agree to
a high degree.  In particular, slightly larger disagreements
were obtained for the two galaxies (NGC 5194, or M51; and NGC 3627,
which is part of the Leo Triplet interacting galaxy group) that
are known to be undergoing tidal interactions.  We also notice that along the
trend of Hubble type evolution from late to early (left to right,
top to bottom for these six galaxies in Figure \ref{Figure31}),
the agreement between the m=2 and m=full mass flow rates
gets progressively better.

\begin{figure}
\vspace{170pt}
\centerline{
\includegraphics{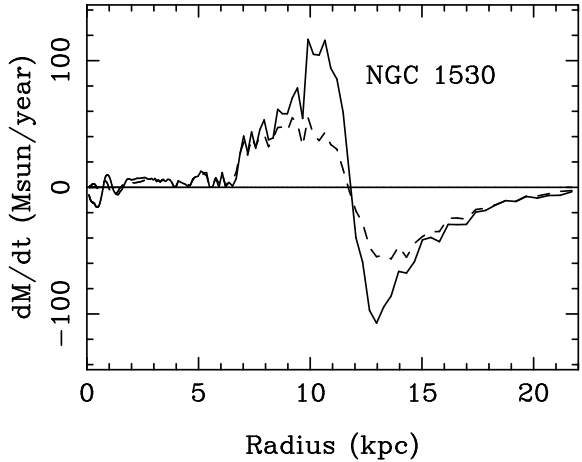}
}
\caption{Radial mass flow rate for NGC 1530 first analyzed in
Zhang \& Buta (2007).  The solid line is calculated
using the full Fourier spectra of density and potential,
and the dashed line using only the m=2 components.}
\label{Figure32}
\end{figure}

In Figure \ref{Figure32}, we present similar comparison
of m=2 and m=full mass flow rates for galaxy NGC 1530,
first analyzed in Zhang \& Buta (2007).  This galaxy has exceptionally 
high mass flow rate even though it is in a relatively isolated
environment.  We observe that the m=full mass flow rate was significantly
higher than the m=2 mass flow rate (though the zero-crossings
of the two curves are similar), indicating that the
amplitude of the mode is yet to settle into its quasi-steady
value even though the modal density profile appears to be
stabilized: The amplitude of the mode is obviously on the wane.  

In Z98, we have demonstrated that the growth rate of the 
mode $\gamma_g$ due to global amplification and feedback cycle is 
positively correlated with the {\em fraction} of m=2 sinusoidal modal 
component in the density wave composition (since only the m=2 modal 
component has the correct phase relation between the over-reflected 
wave train from corotation and the incoming wave train, in order to form 
growing mode; whereas the higher harmonic components cancel
out due to their lack of correct phase relations).  On the other hand, 
the effective dissipation rate $\gamma_d$ of the mode, once the wave 
has acquired sufficient nonlinear amplitude, is approximately independent 
of the exact nonlinear distortions of the wave profile (since $\gamma_d$ is 
determined mainly by the potential density phase shift pattern for a
wave with a given amplitude, and the phase shift was shown to be relatively
independent of the degree of nonlinearity in the azimuthal
density profile of the mode [Figure 5 and Figure 6 of Z98]).  
Therefore, for high-amplitude, very nonlinear density wave mode 
such as present in NGC 1530, the mode is able to manipulate the degree of 
nonlinearity of its azimuthal profile to boost up the damping rate
as compared to the growth rate, in order to evolve towards
a quasi-steady equilibrium amplitude.

The results shown above, that at quasi-steady state of the wave mode 
the m=2 and m=full contribution to radial mass flow are approximately equal,
might appear at first sight to be surprising, i.e.  how could something 
that is supposed to be only ``a part'' of the whole be equal to the whole?
Here we need to realize the specialness of a self-organized density
wave mode that has achieved quasi-steady state.  The mode had gone through 
many cycles of fine adjustments among its various physical processes so that
the global self-consistency condition is achieved.  Among these
processes, one of them is the randomization of the orbital energy of 
the basic state matter during the energy and angular momentum exchange 
between the wave mode and the basic state, through the mediation of the 
collisionless shock at spiral arms.  This randomization process happens both 
for the stellar and for the ISM components of the galaxy 
disk mass, and there are in fact two aspects to it (Z99): 
one aspect is the randomization of the orbital energy corresponding
to $\Omega_p \times L_{wave}$, where $\Omega_p$ is the
pattern speed of the wave and $L_{wave}$ is the angular
momentum density of the wave, this contributes to the damping
of the wave and the secular decay of the mean stellar orbit; 
the second aspect is the randomization of the residual 
orbital energy corresponding to $(\Omega - \Omega_p) \times L_{wave}$ 
(for matter inside corotation.  There is a corresponding sign change 
for matter outside corotation).  This second aspect results in the 
secular heating of the disk stars (which produces the age-velocity dispersion 
relation of the solar neighborhood stars [Z99]), 
as well as the energy injection into the interstellar
medium to fuel a cascade process akin to the driven
turbulence in fluid dynamics (which produces the Larson-Law
size-line-width correlation of the interstellar clouds.  See
Larson [1981]; Zhang et al. [2001]; Zhang [2002]; Falceta-Goncalves et al.
[2015]).

Stars appear to accomplish the randomization of the orbital energy 
also through a kind of cascade/randomization process
from the large to the progressively smaller scales.
The smaller scale nonlinear exchange is nothing more than
the cascaded-down version of the large-scale
exchange, and does not create a new term in the energy
balance equation.  Another way to look at this matter
is that at the QSS, the growth rate of the m=2 mode is
equal to the growth rate of the full nonlinear mode,
since the nonlinear mode does not have an independent global
amplification mechanism other than the cascading of energy
from the m=2 component.  Therefore, at the QSS the m=2 mode dissipation
rate (or the m=2 torque integral) is {\em formally}
equal to the m=2 growth rate\footnote{We say
formally here because the torque integral acquires its meaning
as corresponding to the dissipative angular momentum
exchange between the wave and basic-state matter {\em only for the
overall full nonlinear set of components at the QSS}.  For each m
component individually or for waves not at the QSS the torque
integral does not uniquely predict anything, since it is
a spatial average of time-dependent quantities.  In other words,
the torque integral, when used to predict the wave/basic-state
angular momentum exchange, is a form of {\em closure relation} enforced
through the consideration of global energy balance,
at the QSS of the wave mode.},
with the m $\neq$ 2 components formally have zero growth
rate as well as zero dissipation rate, even though
in reality the m=2 mode cannot accomplish energy dissipation
without going through the cascade into nonlinear components.

This aspect of local interactions in density waves is in essence 
the same as what Kolmogorov had hypothesized for the energy cascade process 
of fully-developed fluid turbulence in the inertial range (Frisch 1995).
The fact that galaxies can randomize the originally coherent orbital 
velocity/energy is due to the global-self-consistency constraint of 
the spontaneously-formed density wave mode at the quasi-steady state, 
which produced both the local instability/collisionless shock condition 
at the spiral arms, as well as the detailed correlation between kinematic 
and positional distributions of disk matter that enable both wave growth
and damping, accompanied by dissipative basic state evolution. 

\begin{figure}
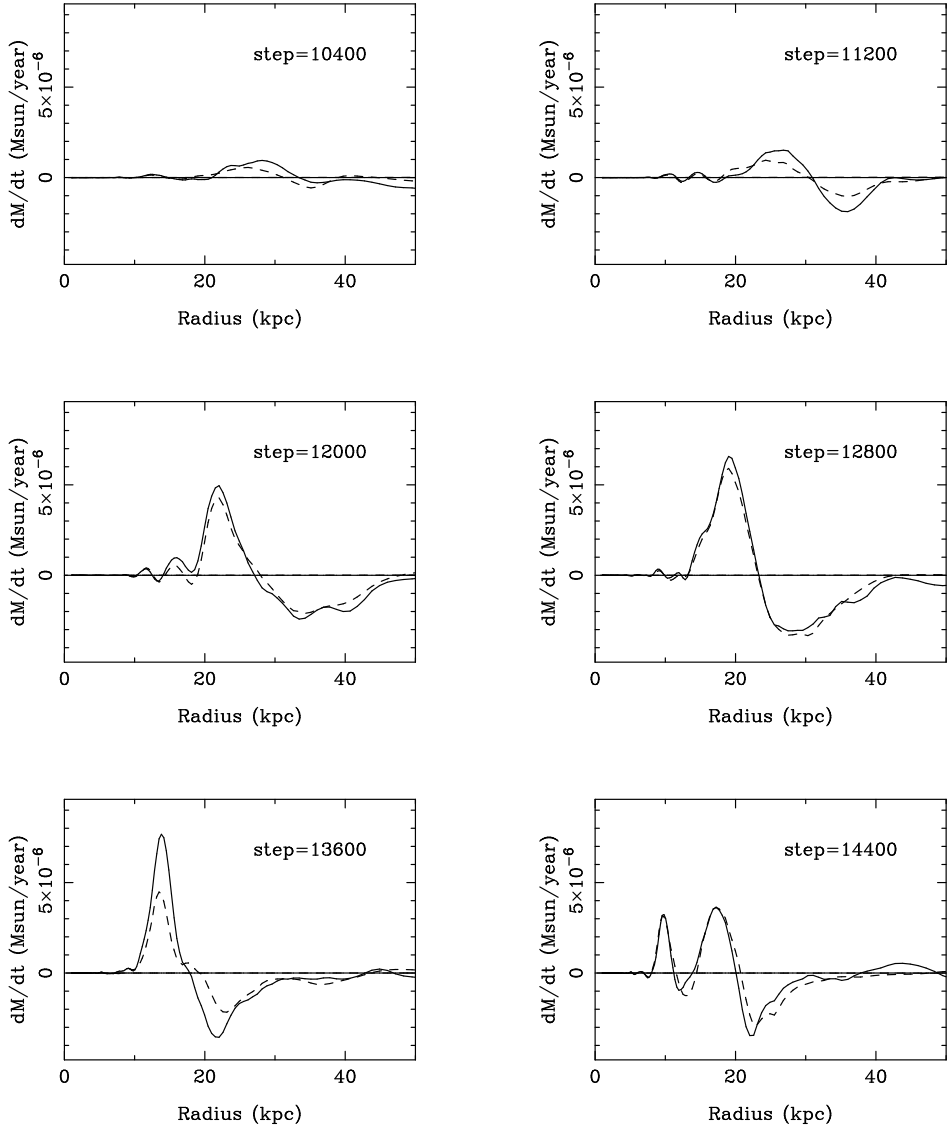

\vspace{410pt}
\includegraphics{Figure33al.ps}
\includegraphics{Figure33ar.ps}
\includegraphics{Figure33bl.ps}
\includegraphics{Figure33br.ps}
\includegraphics{Figure33cl.ps}
\includegraphics{Figure33cr.ps}
\caption{Radial mass flow rates for the N-body spiral
mode with $a_{soft}=1.5$, at six different time steps,
using the full (solid lines) or m=2 (dashed lines) Fourier components.}  
\label{Figure33}
\end{figure}

\begin{figure}
\vspace{410pt}
\includegraphics{Figure34al.ps}
\includegraphics{Figure34ar.ps}
\includegraphics{Figure34bl.ps}
\includegraphics{Figure34br.ps}
\includegraphics{Figure34cl.ps}
\includegraphics{Figure34cr.ps}
\caption{Radial mass flow rates for the N-body spiral
mode with $a_{soft}=0.75$, at six different time steps,
using the full (solid lines) or m=2 (dashed lines) Fourier components.}
\label{Figure34}
\end{figure}

\begin{figure}
\vspace{410pt}
\includegraphics{Figure35al.ps}
\includegraphics{Figure35ar.ps}
\includegraphics{Figure35bl.ps}
\includegraphics{Figure35br.ps}
\includegraphics{Figure35cl.ps}
\includegraphics{Figure35cr.ps}
\caption{Radial mass flow rates for the N-body spiral
mode with $a_{soft}=0.25$, at six different time steps,
using the full (solid lines) or m=2 (dashed lines) Fourier components.}
\label{Figure35}
\end{figure}

\begin{figure}
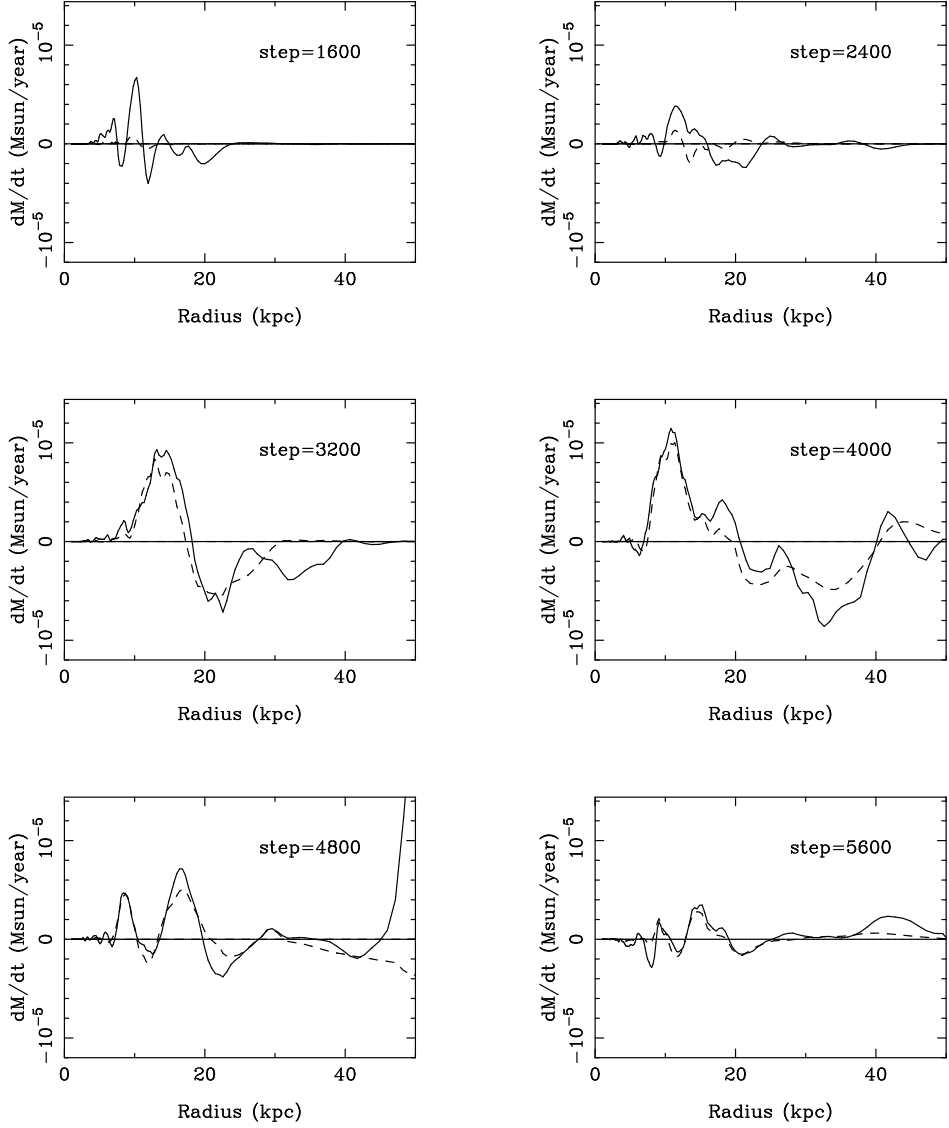

\vspace{410pt}
\includegraphics{Figure36al.ps}
\includegraphics{Figure36ar.ps}
\includegraphics{Figure36bl.ps}
\includegraphics{Figure36br.ps}
\includegraphics{Figure36cl.ps}
\includegraphics{Figure36cr.ps}
\caption{Radial mass flow rates for the N-body spiral
mode with $a_{soft}=0.1$, at six different time steps,
using the full (solid lines) or m=2 (dashed lines) Fourier components.  
Note the contribution
of a local instability clump in the last two frames, which have
poor m=2 and m=full agreement, as to be expected of this non-modal
feature.}
\label{Figure36}
\end{figure}

In what follows we present similar m=2 versus m=full mass flow
rate comparison calculated using the surface density 
and potential obtained in our previous set of four N-body 
simulations.  In Figures \ref{Figure33}, \ref{Figure34},
\ref{Figure35}, and \ref{Figure36}, we show the respectively
calculated m=2 and m=full mass flow rates for the four runs,
at six different time steps, and for a duration of 3000 steps
which correspond to about two and a half galactic rotation
periods at radius 20, during the emergence and stabilization
of the first dominant mode in each case.  

From these four figures, we can tell that for every softening
choice {\em when} the modal shape and kinematics have achieved 
quasi-steady state, the agreement between the m=2 and m=full 
mass flow rates is very good, just as what we had observed 
for physical galaxies.  Note that the QSS can be short-lived
even after its initial acquirement, because the disk surface
density and modal shape continue to evolve throughout
the simulation run for each case.  Still, exploration into
further time period of the modal evolution (not shown here
because of the space constraint, but see Appendix E) shows that a new QSS phase
can once again be achieved (signified by good agreement
between m=2 and m=full mass flow curves) after its temporary
loss as a result of the basic state evolution.  In particular,
the $a_{soft}=0.1$ run showed a most robust correlation
of m=2 and m=full mass flow curves through most of the simulation
duration (25 galactic rotation periods at radius 20), 
likely a result of the fact that small softening
allowed effective inter-particle correlations to
quickly re-establish each new global-self-consistency state 
as the basic state continually evolved.

\subsection{Accuracy and Implications of
the Secular Radial Mass Flow Rates Obtained in N-Body
Simulations}

In Appendix D, we present detailed analyses of the effect of the choice 
of softening parameter on the accuracy of particle-mesh N-body simulations, 
particularly for the polar-grid configuration.  We see there that grid noise
can be of concern for small-softening, large-grid-size simulations 
(see also Efstathiou et al. 1985, Figure 1).  For simulation parameter 
combinations where $a_{soft}$ is small, the grid used is coarse, and 
for outer disk region where the cell size is large, grid noise has a 
significant impact on the accuracy of individual star's orbit -- if the 
accuracy of individual orbit is what we are after in the simulation effort. 

However, as we have shown in this paper, when simulating global density 
wave modes in galactic disks, grid noise does not appear to have a 
detrimental effect on the macroscopic properties of the modes, as well as 
on the {\em correlation} of modal parameters and their associated basic 
state evolution rates.

This positive outcome becomes easier to understand if we recall that in
a disk system possessing collective instabilities, individual simulated 
orbits had long been known to diverge exponentially from true orbits 
in the presence of random noise, especially for simulations dealing with 
collective instabilities (see, e.g. Miller [1971], Pfenniger [1986], 
Romeo [1990], Weinberg [1993]).  This exponential divergence between
the simulated individual orbits and their true counterparts in globally 
unstable systems is present {\em even when large softening length is used}.

On the other hand, the collective behaviors of the modes being modeled
were found not to be impacted by the microscopic inaccuracies of the
simulated individual orbits -- otherwise NONE of the previously
published N-body simulations of galaxies containing unstable density
wave patterns can be trusted!  This fidelity of simulated global pattern 
in the face of drastic (exponential) individual orbit inaccuracy comes 
about because self-organized global structures such as spiral and bar
modes are ``dissipative structures'' in the sense of Prigogine and 
coworkers (Prigogine 1980), or else ``strange attractors'' in the sense 
used in nonlinear dynamical systems (Wiggins 2003 and the references 
therein).  These systems automatically seek their attractor/modal solutions
even if perturbed from corresponding nonequilibrium steady states.  
This property is the well-known ``asymptotic stability'' of the 
dissipative nonequilibrium quasi-steady state.  In the case of galaxies
possessing density wave modes, the effective collision/scattering processes
at the crest of density wave patterns re-establish the correlations needed
for the self-organization and collective dissipation processes, even if
microscopically these orbits are not exactly what they would be for
an infinite-precision calculation.  From this perspective it also
becomes clear why a smaller-softening choice which allows the proper 
establishment of correlations among particles in a simulated galaxy disk 
is more important than the accurate modeling of a large-softening disk which
lacks the proper correlations among particles: the near-collision (or
the scattering) condition allowed by small softening (or efficient 
inter-particle interaction) is what establishes the correlations among
particles to allow the collective instability to successfully operate.

As we have seen, macroscopic properties of the evolution of the global
density wave patterns, as well as the evolution of the basic state,
are not impacted by the inaccuracies in the behavior of
the individual particles' orbits.  Furthermore, if the noise in individual 
stellar orbits had mattered, even in physical galaxies we should not have 
been able to observe the beautifully organized grand-design density wave 
structures, since in these galaxies the stellar orbits are perturbed 
by chance encounters with the Giant Molecular Cloud Complexes, the tidal 
companion galaxies, as well as the formation of disk open and globular 
star clusters.  The stability of the global pattern characteristics against 
random fluctuation is indeed what allowed these patterns to emerge as a 
predominant form of organization in disk galaxies, and to have the modal 
characteristics correlate with that of the basic state\footnote{We
recall that one of the criteria of the Hubble classification scheme 
is the empirically observed correlation of spiral pitch angle with the 
degree of prominence of the bulge.  This correlation has been reproduced
in the modal picture of the galactic density waves (Bertin et al. 1989a,b).}, 
rather than with the chance elements of internal and external random 
perturbations.

One question that may arise from examining the results of this work
is that since the different softening-parameter runs gave different 
mass flow rates, what magnitude of the mass flow should we associate 
with the secular morphological evolution of physical galaxies?  

We note that the aim of these simulations is not to derive an 
{\em absolute} mass flow rate and claim that it corresponds 
to that in {\em all} physical galaxies, but rather to demonstrate the 
{\em correlation} between the N-body simulated mass flow rate and
the rate predicted by the volume-torque/mass-flow-rate equation 
(i.e. equation [\ref{eq:accre}]).  A confirmation of this correlation 
shows that the analytical equation predicts the correct mass flow rates for
{\em all} ranges of the wave and basic state parameters, thus the rate
equation can be used in physical galaxies with the actually measured
parameters for the wave and the basic state.

In the simulation results presented in this section, we had 
indeed shown that once the collective effects are allowed to be 
fully represented either by the use of sufficiently small $a_{soft}$, 
or else by the enforcement of special bi-symmetry condition,
{\em all} of the measured mass flow rates in N-body simulations
at a wide range of softening lengths have the correct correspondence with the 
respective analytical predictions, especially for the radial range 
inside corotation, which is our main interest in the secular evolution 
studies.  Therefore, these pattern morphologies and mass flow rates 
can {\em all} find correspondence with physical galaxies of varying
Hubble types.  In fact, a little contemplation will show that
the larger softening choices correspond to the 
thicker inner disk scale heights and smaller secular mass flow rates
of the earlier Hubble type galaxies, whereas the smaller softening
choices correspond to the thinner disk and the larger secular
mass flow rates of later Hubble type galaxies (Zhang \& Buta
2007, 2015). 

\subsection{Irreversibility and Singularity-Hierarchy in
Natural Systems}

Returning to some of the questions we posed in the Introduction section, 
we note that accompanying the emergence of global patterns,
in an originally featureless system past its instability threshold,
there is also the spontaneous emergence of new meta-laws.  
The torque-integral/mass-flow rate equation we have confirmed
using N-body simulations in this work is an example of emergent
meta-law.  The original derivation of this law (Z98,Z99) required
the assumption of global self-consistency between the
various physical processes in the galaxy (i.e. the compatibility
of the energetics of the wave mode versus the energetics of 
the basic state, which is essentially the requirement for
global energy and angular momentum conservation), as well
as the assumption of quasi-steady state of the wave mode,
which is an empirically well-supported hypothesis.

Given these two very reasonable basic assumptions, the outcome
of the secular evolution of the basic state of the galactic disk
is inevitable, given that the pattern's matter distribution
and kinematics naturally lead to the characteristic potential-density 
phase shift pattern which implies a secular torque action by the 
wave mode on the basic state, and thus secular energy and 
angular momentum exchange between the two components.  The operation 
of this exchange process is necessarily dissipative (as there is the
irreversible conversion of stellar orbital energy into
random motion energy of stars).  Furthermore, the working of the
dissipation mechanism requires the presence of collisionless
shock at the wave crest, or a temporary local gravitational
instability at the spiral arms (Z96).  This further
implies that the location of the spiral arms is an effective
singularity of the underlying differential (fluid) formulation
(the arguments for the inevitability of the formation of 
this singularity or collisionless shock through the azimuthal
steepening of the spiral wave solution using an
Eulerian equation set were given in Z96,
based on the earlier work of Lubow, Balbus, \& Cowie [1986]).
Z98, Z99 showed that the breakdown of the {\em differential}
form of the Poisson equation at the density wave
crest (the site of local instability and collisionless
shock) is a necessary ingredient for the global self-consistency 
of the governing equations to be achieved.  The {\em integral form}
of the Poisson equation is what we use to obtain the global
(phase shifted) perturbation potential from the perturbation density.
This should come as no surprise, since we know
that across any hydrodynamic shock, the differential
formulation invariably breaks down (see, for example, Shu [1992]),
and the properties of the shock-jumps are established by {\em enforcing
global conservation relations}, just as what we have done
in this work by enforcing global self-consistency requirement
between the kinematics and mass distribution of the density
wave mode, and between the energetics of the wave mode and the
basic state of the disk (Z98, Z99). 

Generalizing what we have learned in this work, we realize
that in nature irreversibility may be the rule rather 
than the exception, even though the differential
forms of governing equations that we commonly employ
are all time-reversible.  We now realize that
these differential laws are likely to be idealizations, and 
their origins may be emergent under the general boundary
conditions of the evolving universe.  In this sense
laws and meta-laws (or fundamental laws and emergent
laws) have no clear distinction.  It all depends on the scale 
we conduct our observations and analyses in.

We also realize that effective singularity hierarchies (as reflected,
for example, in the spontaneous emergence of density wave modes, 
and the attainment of new closure relations at the QSS) may be 
the general mechanism nature employs to separate the different 
domains of physical and biological sciences.  Within a
given hierarchy, deductive approach may be used to arrive
at new inferences.  However, when we are crossing the boundary
of a given hierarchy (such as in the study of the emergence
and maintenance of global spiral modes, as well as the
secular evolution of the basic state as a result of the
interaction with these modes), we need to rely on
empirical evidence (such as the QSS hypothesis of the mode), 
meta-principles (such as the second law of thermodynamics and the
theory of dissipative structures), as well as reasonable
global closure relations (such as global energy and angular momentum
balance considerations) in order to establish a new 
axiomatic structure for the subsequent analyses.

With the guidance of these general principles,
we can then explore the detailed workings of
the various facets of the proposed new paradigm (in the
case of spiral galaxies, this involves the discoveries
of the mechanisms of wave mode amplification through over-reflection 
and feedback; wave stabilization through damping by the basic state energy
and angular momentum input; the realization of local
dissipation or energy cascade by the action of collisionless shocks; 
the demonstration of the presence of potential-density phase shift 
in a self-sustained density wave mode by both the integral form of 
the Poisson equation and the equations-of-motion of disk matter,
etc.).

We now also have a tentative answer to the one question
that Feynman raised that we had quoted in the Introduction of this
paper: Schrodinger's equation most likely does not {\em directly} predict
the existence of frogs, musical composers, and morality.
Since the physical universe is organized in a hierarchical
fashion with effective singularities separating the
different hierarchies, and since emergent laws governing
the higher level phenomena cannot be {\em deductively} derived
from the lower level ``fundamental laws'', the claim
that complex phenomena of the universe can be
logically derived from a simple set of fundamental
equations is unfounded.  Instead,
the differential forms of equations governing given
regimes of physics are expected to be idealizations that have 
limited range of applicability, and will break down
at instability/singularity front of the next emerging hierarchy.
In such cases, guided by empirical evidence and physical
intuition, new global-self-consistency or closure relations, and new
meta laws (formulated perhaps as a new set of differential
equations with their new range of applicability) 
need to be obtained to guide the analyses.

The universe since the Big Bang is in a process of
irreversible evolution.  If we take into account
the subtleties of quantum mechanics, no process is
truly reversible (i.e. no physical system can become
truly isolated).  Yet we have learned in this work that
some processes are more irreversible than others,
i.e., those that involve self-organized structures
that drastically increase the rate of entropy production
generate levels of irreversibility exponentially
faster than processes that are passive.  Dissipation
is an important element of the formation of nonequilibrium self-organized
structure because it helps to stabilize the dynamical
equilibrium state (as highlighted by the fluctuation-dissipation 
theorem) by damping out random noise fluctuations.

\section{CONCLUSIONS}

The low radial mass rates found in the past N-body simulations of disk 
galaxies had cast doubts on the effectiveness of secular evolution as an 
important dynamical process for the morphological transformation of galaxies 
along the Hubble sequence.  In this work it is shown that such low numerical 
mass flow rates, even in those simulated disks that admit unstable
density wave modes (which is a necessary condition for obtaining the
coordinated mass flow pattern that could lead to galaxy evolution along the
Hubble sequence), were chiefly the result of the artificial ``softening'' of 
gravity which is a common practice in galaxy simulations to avoid the
rapid relaxation caused by the used of smaller number of particles 
compared to that in physical galaxies.  By decreasing the amount of 
softening and simultaneously increasing the number of simulation particles, 
as well as increasing the grid and time resolution, realistic levels 
of mass inflow comparable to those obtained for physical galaxies are now
achievable in N-body simulations, and these mass flow rates
are shown to agree with theoretical predictions using the
density wave parameters obtained in the same simulations.
The results of this study thus support the importance of 
secular evolution, in particular {\em radial stellar mass accretion}, as 
an extremely relevant dynamical process in transforming the morphologies 
of galaxies during the past Hubble time. 
The exploration of the process of the emergence and 
maintenance of density wave modes as an example
of ``dissipative structures'' also sheds light on more
general questions regarding the emergence and the role of ordered hierarchy
of structures and laws in the evolution of the universe.

\section*{REFERENCES}

\smallskip

Anderson, P.W. 1972, {\em Science}, 177, 393
\smallskip

Antonov, V.A. 1962, {\em Vest. Leningr. Gos. Univ.}, 7, 135
\smallskip

Baba, J., Saitoh, T.R., \& Wada, K. 2013, {\em ApJ}, 763, 46
\smallskip

Balbus, S.A. 1988, {\em ApJ}, 324, 60
\smallskip

Balogh, A., Treumann, R.A. 2013, 
{\em Physics of Collision Shocks} (New York: Springer)
\smallskip

Bertin, G., Lin, C. C., Lowe, S. A., \& Thurstans, 
R. P. 1989a, {\em ApJ}, 338, 78; 1989b,
{\em ApJ}, 338, 104
\smallskip

Binney, J., \& Tremaine, S. 2008, {\em Galactic Dynamics},
(Princeton: PUP)
\smallskip

Buta, R., \& Zhang, X. 2009, {\em ApJS}, 182, 559
\smallskip

Carlberg, R.G., \& Sellwood, J.A. 1985, {\em ApJ}, 292, 79
\smallskip

D'Onghia E., Vogelsberger, M., \& Hernquist, L. 2013, 
{\em ApJ}, 766, 34
\smallskip

D'Onghia E. 2015, {\em ApJ}, 808, 8
\smallskip

Donner, K.J., \& Thomasson, M. 1994, 
{\em A\&A} 290, 785 (DT94)
\smallskip

Efstathiou, G., David, M., Frenk, C.S., \&
White, S.D.M. 1985, {\em ApJS}, 57, 241
\smallskip

Elmegreen, B.G., \& Elmegreen, D.M. 1983, {\em ApJ}, 267, 31
\smallskip

Elmegreen, B.G., \& Elmegreen, D.M. 1989, in
{\em Evolutionary Phenomena in Galaxies},
eds. J.E. Beckman \& B.E.J. Pagel (Cambridge: CUP), 83
\smallskip

Feynman, R.P., Leighton, R.B., \&
Sands, M. 1963, {\em The Feynman Lectures on Physics} vol. I, 7-7
\smallskip

Feynman, R.P., Leighton, R.B., \&
Sands, M. 1964, {\em The Feynman Lectures on Physics} vol. II, 41-12
\smallskip

Falceta-Goncalves, D., Bonnell, I.,
Kowal, G., Lepine, J.R.D., \& Braga, C.A.S. 2015,
{\em MNRAS}, 446, 973
\smallskip

Frisch, U. 1995, {\em Turbulence: The Legacy of
A.N. Kolmogorov} (Cambridge: CUP)
\smallskip

Haan, S. et al. 2009, {\em ApJ}, 692, 1623
\smallskip

Hockney, R.W., \& Eastwood, J.W. 1988, 
{\em Computer Simulation Using Particles}
(New York, CRC)
\smallskip

Kalnajs, A.J. 1965, {\em Ph.D. thesis}, Harvard Univ.
\smallskip

Kalnajs, A. J. 1972, {\em Astrop. Lett.}, 11, 41
\smallskip

Kormendy, J. 1979, {\em ApJ} 227, 714
\smallskip

Kormendy, J., \& Kennicutt, R. 2004, {\em ARA\&A}, 42, 603
\smallskip

Kreuzer, H.J. 1981, {\em Nonequilibrium
Thermodynamics and its Statistical Foundations}, (Oxford: OUP)
\smallskip

Larson, R.B. 1981, {\em MNRAS}, 194, 809
\smallskip

Lin, C. C., \& Lau, Y. Y. 1979, {\em Stu. Appl. Math.}, 60, 
p. 97
\smallskip

Lin, C. C., \& Shu, F. H. 1964, {\em ApJ}, 140, 646
\smallskip

Lindblad, B. 1963, {\em Stockholms Obs. Ann.}, 22, 5 
\smallskip

Lubow, S.H., Balbus, S.A., \& Cowie, L.L. 1986, 
{\em ApJ}, 309, 496
\smallskip

Lynden-Bell, D., \& Kalnajs, A. 1972,
{\em MNRAS}, 157, 1
\smallskip

Lynden-Bell, D., \& Wood, R. 1968,
{\em MNRAS}, 138, 495
\smallskip

Mark, J.W.-K. 1976, {\em ApJ}, 205, 363
\smallskip

Martinez-Garcia, E.E., Gonzalez-Lopezlira, R.A.,
\& Bruzuel, A.G. 2009, {\em ApJ}, 694, 512
\smallskip

Martinez-Garcia, E.E., Gonzalez-Lopezlira, R.A.,
\& Bruzuel, A.G. 2011, {\em ApJ}, 734, 122
\smallskip

Miller, R.H. 1971, {\em J. Comp. Phys.} 73, 90
\smallskip

Miller, R.H. 1976, {\em J. Comp. Phys.} 21, 400
\smallskip

Pettitt, A.R., Tasker, E.J., \& Wadsley, J.W. 2016,
{\em MNRAS}, 458, 3990
\smallskip

fenniger, D. 1986, {\em A\&A} 165, 74
\smallskip

Prigogine, I. 1980, {\em From Being to Becoming} 
(New York: Freeman) 
\smallskip

Roberts, W.W. 1969, {\em ApJ}, 158, 123
\smallskip

Roberts, W.W., \& Shu, F.H. 1972, 
{\em Astrophys. Lett.}, 12, 49
\smallskip

Romeo, A. 1990, {\em Ph.D. Thesis}, ISAS 
\smallskip

Romeo, A. 1994a, {\em A\&A} 286, 799 
\smallskip

Romeo, A. 1994b, {\em Ap\&SS} 216, 353 
\smallskip

Romeo, A. 1997, {\em A\&A} 324, 523 
\smallskip

Romeo, A. 1998, {\em A\&A} 335, 922 
\smallskip

Rybicki, G.B. 1971, {\em Ap\&SS} 14, 15
\smallskip

Sellwood, J.A. 1987, {\em ARA\&A} 25, 151
\smallskip

Sellwood, J.A. 2014, {\em RvMP}, 86, 1
\smallskip

Sellwood, J.A., \& Binney, J.J. 2002, {\em MNRAS}, 
336, 785
\smallskip

Shu, F.S. 1992, {\em The Physics of Astrophysics},
vol. 2 (Mill Valley: Univ. Sci. Books)
\smallskip

Sparke, L.S., \& Sellwood, J.A. 1987,
{\em MNRAS}, 225, 653
\smallskip

Thomasson, M. 1989, {\em Research Report}, No. 162, 
Chalmers Univ.
of Techn., Goteborg
\smallskip

Thomasson, M., Donner, K.J., \& Elmegreen, B.G. 1991,
{\em A\&A}, 250, 316
\smallskip

Toomre. A. 1981, in {\em Structure and Dynamics
of Normal Galaxies}, eds. S.M. Fall \& D. Lynden-Bell (Cambridge: CUP), 111
\smallskip

Weinberg, M. 1993, {\em ApJ} 410, 543
\smallskip

White, R.L. 1988, {\em ApJ} 330, 26
\smallskip

Wiggins, S. 2003, {\em Introduction to Applied
Nonlinear Dynamical Systems and Chaos}, 2nd ed. (New York: Springer)
\smallskip

Zhang, X. 1992, {\em PhD Dissertation}, Univ. of
California, Berkeley
\smallskip

Zhang, X. 1996, {\em ApJ} 457, 125 (Z96)
\smallskip

Zhang, X. 1998, {\em ApJ} 499, 93 (Z98)
\smallskip

Zhang, X. 1999, {\em ApJ} 518, 613 (Z99)
\smallskip

Zhang, X. 2002, {\em Ap\&SS} 281, 281
\smallskip

Zhang, X. 2003, {\em Journal of the Korean Astronomical Society}, 36, 223
\smallskip

Zhang, X. 2008, {\em PASP}, 120, 121
\smallskip

Zhang, X. \& Buta, R. 2007, {\em AJ} 133, 2584
\smallskip

Zhang, X. \& Buta, R. 2015, {\em NewA}, 34, 65 
\smallskip

X., Lee, Y., Bolatto, A., \& Stark, A. A. 2001,
{\em ApJ}, 553, 274
\smallskip

\section*{APPENDIX A:  POTENTIAL-DENSITY PHASE SHIFTS AND
RADIAL MASS FLOW RATES}

The original derivation of the collective dissipation mechanism 
responsible for the secular redistribution of galaxy disk mass
can be found in Z96, Z98, Z99. Here we give a brief
summary of the analytical results used in the main text of the 
current paper.

For a disk galaxy with non-axisymmetric density
and potential perturbations of $\Sigma_1$ and ${\cal V}_1$, respectively,
the (z-component) torque applied by the total potential on the
material at a unit-width annular ring at a galactic distance R
(which, at the quasi-steady state of the density wave mode,
is equal to the averaged rate of angular momentum flow from the
wave to the disk material) is
 
\begin{equation}
T(R)
= - R
\int_0^{2  \pi}
\Sigma_1(R,\phi)
{ {\partial {{{\cal V}_1}(R,\phi)}} \over {\partial \phi}}
d \phi
= 2 \pi R 
 {\overline{ {{dL} \over {dt}}} (R)}
,
\label{eq:torque}
\end{equation}
where $L$ is the angular momentum density of the disk material.
It can be shown that the only way for this torque and angular momentum
exchange expression to be non-zero is to have the potential perturbation
phase-shifted in azimuth from the density perturbation,
which is naturally satisfied by a spontaneously formed, skewed
density wave mode (Z96).  For such a mode the
phase shift distribution versus galactic radius is of a
characteristic two-humped shape, positive inside corotation
and negative outside, with the zero-crossing of phase shift
versus galactic radius curve happen at corotation radius.

Based on the above torque expression,
the equivalent phase shift $\phi_0$ between two nonlinear wave forms is
defined as (Z98),
\begin{equation}
\phi_0 = {1 \over m} \sin^{-1}
\left ( {1 \over m}
{{\int_0^{2  \pi}
\Sigma_1
{ {\partial {{{\cal{V}}_1}}} \over {\partial \phi}}
d \phi}
\over
{ \sqrt{\int_0^{2  \pi}
{\cal{V}}_1^2
d \phi}}
\sqrt{\int_0^{2  \pi}
{\Sigma_1}^2
d \phi}  }
\right ),
\label{eq:ps}
\end{equation}
where m is the number of spiral arms, and the sign of the
phase shift is defined such that the phase shift is positive
when the potential lags the density in the direction of galactic rotation.
The equivalent phase shift is the amount of phase shift
which would be present between two sinusoidal wave forms if each is
endowed with the same energy as the corresponding nonlinear wave form,
and which would lead to the same value for
the torque integral as would the nonlinear waveforms.
Note that in the above expression the perturbation waveforms must
have their azimuthal mean values subtracted.

The (inward) radial mass accretion rate at a galactic radius 
R is related to the mean orbital decay rate $-dR/dt$ of an 
average star through
\begin{equation}
{dM(R) \over dt} = - {dR \over dt} 2 \pi R \Sigma_0(R)
\end{equation}
where $\Sigma_0(R)$ is the mean surface density of the basic state
of the disk at radius R.

We also know that the mean orbital increase rate of a 
single star $dR/dt$ is related to its
angular momentum gain rate $dL^*/dt$ through
\begin{equation}
{dL^* \over dt} = V_c M_* {dR \over dt}
\end{equation} 
where $V_c$ is the mean circular velocity at radius R, 
and $M_*$ the mass of the relevant star.

Now we have also
\begin{equation}
{dL^* \over dt} = {\overline{ {{dL} \over {dt}}}} (R)
{M_*  \over \Sigma_0}
\end{equation}
where ${\overline{ {{dL} \over {dt}}} (R)}$
is the angular momentum gain rate of the basic
state disk matter per unit area at radius R.

Since, as shown in equation (\ref{eq:torque}),
\begin{equation}
{\overline{ {{dL} \over {dt}}}} (R)
= - { 1 \over {2 \pi}} \int_0^{2 \pi}
\Sigma_1 { {\partial {\cal V}_1} \over {\partial \phi}} d \phi
\label{eq:eq4}
\end{equation} 
(Z96), we have finally

\begin{equation}
{dM (R) \over dt} = 
{R \over {V_c}} \int_0^{2 \pi} 
\Sigma_1  {{\partial {\cal V}_1} \over {\partial \phi}} d \phi
\label{eq:accre}
\end{equation}
where the subscript 1 denotes the perturbation quantities.
This result, even though derived through the stellar orbital decay rate,
is in fact general, and can be applied to the mass accretion rate
of both stars and gas, as long as the relevant perturbation surface
density is used.  The perturbation potential field used in the calculations
needs to be the total potential (i.e. stars plus gas), however.  
We see also from the expressions of mass flow (equation \ref{eq:accre})
and phase shift (equation \ref{eq:ps}) that they share {\em the same zero 
crossings} (that of the zero crossings in the torque integral expression), 
but have different radial amplitude modulations.

From the above equation, it can also be shown that for waves with
moderate pitch angles, the mass flow rate can be further
written in terms of the parameters of the wave and basic state
(Z98 \S5.1):

\begin{equation}
{dM (R) \over dt} =
\pi F^2 R V_c \tan (i) \sin (m \phi_0) \Sigma_0
,
\label{eq:fac}
\end{equation}
where $F$ is the fractional amplitude of the wave (defined as the
{\em geometric mean} of the fractional potential wave amplitude
and fractional density wave amplitude), $i$ is the pitch
angle of the wave, $m$ is the number of arms (usually taken to be 2),
$\phi_0$ is the azimuthal potential-density phase shift,
and $\Sigma_0$ is the local surface density of the basic state
mass distribution.

For the phase shift distribution of a spontaneously formed spiral
or bar mode, the direction of torque (or the sign of phase shift)
leads to mass inflow inside corotation and outflow outside corotation 
(Z96, Z98).  The existence of the phase shift distribution
across the galactic radius for a spontaneously formed density
wave mode thus serves as the natural engine for the secular evolution
of galaxies.  The role of the density wave pattern is functional
as well as structural, and the most important function of it
is to accelerate the entropy evolution of the parent system, as
is the case for all dissipative structures (Prigogine 1980).

\section*{APPENDIX B CHOICE OF BASIC STATE PARAMETERS}

We have chosen to use normalized units in the N-body calculations 
for this work, similar to that used in Donner \& Thomasson (1994).
In these units, the total mass of the disk, including
the active disk mass $m_d$, the inert halo mass $m_h$, and
the inert bulge mass $m_b$, sums to 1, i.e.

\begin{equation}
m_{total}=m_d+m_h+m_b=1.
\end{equation}
Z96 used the set of mass ratios $m_d:m_h:m_b=0.5:0.4:0.1$,
the same as used in Donner \& Thomasson (1994), whereas Z98, Z99 
used an alternative set of $m_d:m_h:m_b=0.4:0.5:0.1$,
which is what is used for the basic state in the current paper.
This slightly differing choice in the second set allows a more
long lasting dominant mode to be present during the simulation run.

The disk surface density used in the simulations is in the form of a 
modified exponential (which is an exponential with a central
hole, accounting for the fact that the active disk mass decreases
in the center, and the mass in the central region is dominated
by the inert bulge mass)

\begin{equation}
\Sigma_d (r) = \Sigma_{d0} (e^{-r/R_d} - e^{-2r/R_d})
,
\end{equation}
where we have chosen $R_d=10$ for the disk scale length
in the normalized unit, and $\Sigma_{d0}$ is a constant chosen to make 
the total disk mass of the correct amount $m_d$ in the normalized unit.  
The curve for this basic state mass distribution
can be found in Figure \ref{Figure5}
and Figure \ref{Figure30} of the current paper
as the ``initial'' surface density
distribution (solid line).

An inactive bulge and a rigid halo (i.e., used purely for their gravitational
effect on disk particles, while their own mass distributions are not
recomputed as a response to the disk evolution) are also used, which are 
assumed to be of the regular exponential shape: 

\begin{equation}
\Sigma_h (r) = \Sigma_{h0} e^{-r/R_h}
,
\end{equation}

\begin{equation}
\Sigma_b (r) = \Sigma_{b0} e^{-r/R_b}
,
\end{equation}
The scale lengths are $R_h=5$ and $R_b=1$ for the rigid halo and bulge
component, respectively.  These choices of the three mass components
result in a nearly constant rotation curve, with velocity $v_c \sim 0.1$ 
in the normalized unit for the simulation parameter choice of the main body 
of the paper (and double that for the grid softening test presented
in Appendix D3). 

The choice for the initial radial velocity dispersion for
the galaxy simulations in this work leads to an initial instability
parameter Q of 1.0 throughout the disk.  

This particular choice of the basic state allows a resonant
cavity to be set up between the corotation region (around
r $\sim$ 30 in the normalized unit for the largest softening run,
and r $\sim$ 20 for the smallest softening run) and the inner bulge region
(which serves as a Q-barrier to reflect the inwardly-propagating
wave train to complete the feedback cycle), so a normal
mode of the galactic disk can spontaneously emerge.

\section*{APPENDIX C CHOICE OF SIMULATION PARAMETERS AND
DESCRIPTION OF NUMERICAL PROCEDURES}

The simulation grid of this work is a polar grid similar to the original
one used by Miller (1976), with radial grid rings distributed in an 
exponential fashion, and the azimuthal grid spokes distributed in a 
uniform fashion, i.e.

\begin{equation}
r=L e^{\alpha u}
,
\end{equation}
\begin{equation}
\phi=\alpha v,
\end{equation}
where $u$ has integer values that ranges from 0 to $n_{u} - 1$,
and $v$ has integer values that range from 1 to $n_{v}$.
$L$ is the constant length scale factor, usually chosen to be 1 in the
normalized unit, and $\alpha = 2 \pi / {n_v}$.
$n_{v}$ is usually selected to have a power of two so as to allow the 
use of the FFT algorithm for force calculation
in the azimuthal direction.  The exponentially 
spaced radial grid and uniformly spaced azimuthal grid with compatible 
choice of $n_u$ and $n_v$ ensure that the grid shape remains roughly 
square throughout the radial range of a galaxy disk (one of the main reasons 
for adopting an exponentially spaced radial grid).  This choice of grid 
spacings also naturally matches the mass density distribution in disk 
galaxies, which has higher density and more finely-resolved density wave 
patterns in the inner disk.  

Donner \& Thomasson (1994) selected a grid of $n_{u}$ = 
50 radial grid cells, and $n_{v}$ = 64 for the azimuthal grid cells,
together with 50,000 active disk particles.  These parameters have been 
slightly changed in the simulations of Z96 to
(110, 128), and in Z96, Z98 to (55, 64).   
The simulations in the main text of
this paper used a simulation grid of (220, 256), and
that in Appendix D3 used a grid of size (110, 128).

The active disk particles were initially assigned on rings
with constant particle number per ring ($n_{ring}=2 \times [n_v /2 +1])$, 
but variable ring spacings to represent the changing disk surface density
with galactic radius.
The cloud-in-cell (CIC) interpolation approach (see, e.g.,
Hockney \& Eastwood 1988, Chapter 5) is employed for grid mass
assignment from the disk particle mass distribution. 
Fast Fourier Transform (FFT) approach was employed for grid potential 
calculation based on grid mass, and forces on grid points
were calculated using finite difference of the grid potential.
The forces on disk particles are then calculated once more
through the CIC method from grid force, and these disk particles
are subsequently moved around using a time-centered leap-frog scheme.  
Sufficient time step resolution is chosen to ensure stability and
accuracy.

The length of the time step is chosen through a parameter called
the number of time steps per crossing time $S_{stpcrt}$ at a given 
radius (usually chosen to be $R_d=20$, which is within or
close to the corotation radius).  This number is defined as the time steps
it takes for a particle to cross the specified disk radius $R_d$,
i.e.

\begin{equation}
\Delta t \cdot S_{stpcrt} = {{R_d} \over {V_c(R_d)}}
,
\end{equation}
where $\Delta t =1$ is the normalized unit of time step, and $V_c$ is
the rotation curve value.  In the original Donner \& Thomasson (1994)
simulation, $S_{stpcrt}$ was chosen to be 50, which corresponds
to 314 time steps per rotation period at radius 20. The same value
was chosen also in Z98, Z99, though Z96 chose
$S_{stpcrt}=100$ to correspond to the doubling of the time resolution
for that particular set of simulation.  We have chosen $S_{stpcrt}
= 200$ for the simulations in the main body of the text (and this
corresponds to 1256 steps per rotation period), and $S_{stpcrt}
=100$ for the simulation in Appendix D3, all at radius r=20.
These $S_{stpcrt}$ choices are made in 
order guarantee the numerical stability of the computation algorithm.

After $S_{stpcrt}$ is determined, a renormalized gravitational
constant $G_{new}$ is computed and is used to replace the $G_{old}$
in the usual Newtonian equations.  The new and old gravitational
constants are related through

\begin{equation}
G_{new} = {{{R_d^2} \over {{\Delta t}^2 \cdot {S_{stpcrt}^2}}} \over
{m_d {{V_d^2(R_d)} \over {G_{old}}} +
m_h {{V_h^2(R_d)} \over {G_{old}}} +
m_b {{V_b^2(R_b)} \over {G_{old}}} }}
,
\end{equation}
where $V_d$, $V_h$, $V_b$ are the rotation velocity contribution
of the different mass components to the total rotation curve of
the galaxy, i.e.

\begin{equation}
m_d V_d^2 + m_h V_h^2 + m_b V_b^2 = m_{total} V_c^2 = V_c^2,
\end{equation}
since $m_{total}=1$ in the normalized units.
With these choices of $\Delta t$ and $G_{new}$
the desired $S_{stpcrt}$ will be achieved.

Further details of the numerical procedure can be found in
Miller (1976), Thomasson (1991), Donner \& Thomasson (1994),
as well as in Z96, Z98, Z99.

\section*{APPENDIX D. GRID NOISE ASSOCIATED WITH THE
USE OF SMALL PARTICLE SOFTENING PARAMETER}

In this appendix we address the issue of whether the quality
of the simulation results presented in this paper is negatively
affected by the use of small particle-softening parameter
$a_{soft}$.  We will show that the {\em effective} disk thickness
in these 2D simulations is maintained mainly by the finite
size of the grid, whereas the effective inter-particle interaction
which allows the operation of collective instability is enhanced
by the choice of small particle softening parameter $a_{soft}$.
The combined effects of the two, coupled with sufficient number
of simulation particles and finer grid, allow the {\em macroscopic}
features of the modes to be faithfully reproduced despite
the increased noise in individual particle's orbit.

\subsection*{D1. The Differing Roles of Grid and Particle Softening}

In the simulations described in the main body of this paper, while the
particle softening parameter $a_{soft}$ had been drastically reduced in
some cases, the corresponding grid softening had not been reduced to 
a similar extent (the newer grid is only about a factor of 4 finer 
in linear resolution compared to that used in Z98, whereas the 
particle softening is reduced by a factor of 15 for the extreme
choice of $a_{soft}=0.1$).  It thus appears that particle
softening is the main inhibitor to obtaining higher mass flow rates,
the grid softening has only marginal effect: the new simulation
maintaining $a_{soft}=1.5$ but with 4 times the linear resolution 
compared to Z98 did not seem to affect the mass flow rate by much,
whereas decreasing $a_{soft}$ clearly allowed the radial mass flow rate 
to increase (even starting with a factor of 2 $a_{soft}$ reduction,
as seen in Figure \ref{Figure3}), especially for the very 
central region where enough
mass supply (both from the local mass surface density as well as from the
mass inflow from the outer disk) was available to support the 
continued mass inflow and bulge building.

Finite amount of grid softening, on the other hand, helps to maintain the
desired finite disk thickness effect in these 2D simulations.  The
{\em effective} disk thickness, or effective softening length, is proportional
to the root mean square of the particle and grid softening lengths 
(Romeo 1994b).  The finite disk thickness in our simulations
manifests both in the particle-number dependence of the noise performance 
(Figure \ref{Figure5}) -- since relaxation in a razor-thin disk should
have shown no particle-number dependence (Rybicki 1971); as well as
in the longevity of the spiral-bar patterns formed -- since a razor-thin 
disk is known to have relaxation time scale on the order of a mean orbital
period (Rybicki 1971) whereas the patterns in our simulations (especially
the underlying m=2 component) using sufficient number of particles were 
shown to last more than 25 orbital periods.

In fact, Rybicki (1971) already pointed out the role of finite grid size on
mimicking a finite-thickness disk.  Basing his analyses on the cumulative
effect of short and long-range interactions, Rybicki (1971)
derived that for a numerical disk which has grid size of $h$, 
the relaxation time $t_R$ can be expressed as 

\begin{equation}
t_R = {{\lambda^3 N h t_M} \over {2R}}
\end{equation}
where $\lambda$ is the ratio of random velocity
$\sigma$ to circular velocity $V$,
i.e. $\sigma=\lambda V, \lambda \le 1$, and $t_M$ is the orbital
crossing time, i.e. $t_M=R/V$ where R is the disk radius.

The above expression of Rybicki's is nearly
equivalent to another expression derived in White (1988)

\begin{equation}
t_R = {{\sigma^3 h} \over {5G^2 \mu m}},
\end{equation}
where $\mu$ is the surface density of the disk, and G is
the gravitational constant, and $m$ is the mass of the
particle.  White's expression can be
shown to be equivalent to Rybicki's if we replace the factor
of 5 in the denominator of White's expression by a factor
of $2 \pi$ in Rybicki's (or $\pi$ if the crossing time
is defined using the diameter instead of the radius of the system).
Rybicki (1971) stated that the factor $h$ is {\em either}
the particle softening length $a_{soft}$ {\em or} the grid size
$h$, whichever is greater. Romeo (1994b) and Donner \& Thomasson (1994), 
on the other hand, argued
that the effective softening length is proportional to the root mean square of
particle and grid softening, i.e. $h=\sqrt{s^2+a_{soft}^2}$
where $s$ is the local grid size.

Our current work showed that finite grid size can indeed simulate a 
finite-thickness disk in terms of inhibiting the unwanted fast relaxation,
as Rybicki had suggested, yet it does not seem to significantly diminish
the desired collective dissipation effect as the choice of large
particle softening length does.  This is likely due to the fact
that grid softening is implemented in an {\em anisotropic} fashion, i.e.,
the mass and force assignments are through the {\em same} Cloud-in-Cell,
or CIC, interpolation scheme, which apparently helped to retain the 
correlated angular inhomogeneities of the particle positions and 
kinematics in addition to conserving momentum, in effect achieving
sub-grid resolution.  On the other hand, particle softening is an 
{\em isotropic} scheme that uniformly evens out the fluctuations in 
the force from all directions, thus has a more detrimental effect on 
suppressing the {\em correlated} collective interactions.

\begin{figure}
\vspace{180pt}
\centerline{
\includegraphics{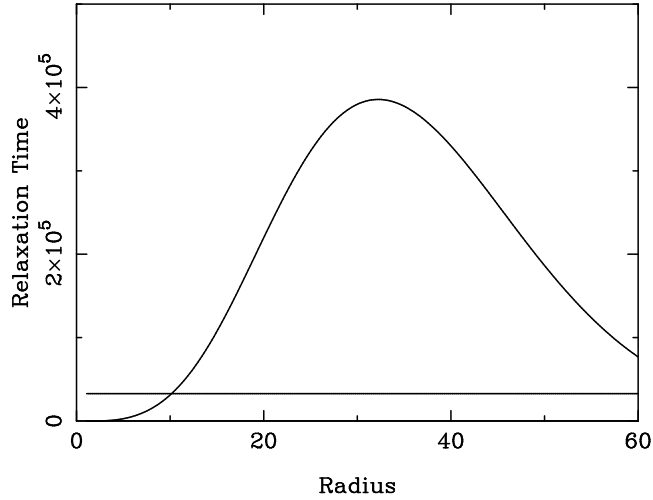}
}
\caption{Relaxation time in the $a_{soft}=0.1$,
20 million particle simulation.
The straight line in the lower part of the figure indicates
the maximum calculation duration used in the simulations here.}
\label{Figure37}
\end{figure}

Figure \ref{Figure37} gives the relaxation time for our previous
20 million particle, $a_{soft}=0.1$ simulation calculated using the
White (1988) expression above, with $h$ given by the root mean square
of the particle and grid softening (times $\sqrt{2}$).  
The straight line in the figure shows the
maximum time step duration (32768) used in the majority of the simulations 
presented in this paper.  It can be seen that over most of the radial range 
of the simulation disk (except for the very central region) the relaxation 
time is much longer than the duration of the simulation, so two-body 
relaxation effect should be minimum, as was also borne out by the 
consistently vigorous spiral/bar activity before the noise due to
collective effect swamped the nonlinear patterns.

\subsection*{D2 Grid Noise in Particle Mesh Simulations}

By choosing to use grid size larger than the particle softening parameter,
we do pay a price: that of increased grid noise.
This manifests as fluctuations of the interpolated forces 
on individual particles as we traverse grid boundaries, which can also be
attributed to aliasing effect in the Fourier transform domain
(Hockney \& Eastwood 1988, \S 5-6-3 and \S 5-6-4).
Grid noise is more pronounced for the smallest softening choices
($a_{soft}=0.1$), and it is also more pronounced in the outer disk
region than in the inner disk for the polar grid used here, 
since the grid spacings in the outer disk are exponentially larger
than in the inner disk (therefore, for all the past polar-grid
simulations presented in the literature, grid noise issue was already
present since even when the particle softening parameter
was chosen to be of the order of the grid size in the inner
disk, for the outer disk region the grid size is much
larger than the particle softening parameter
due to the exponential growth in grid size in polar grid design).

\begin{figure}
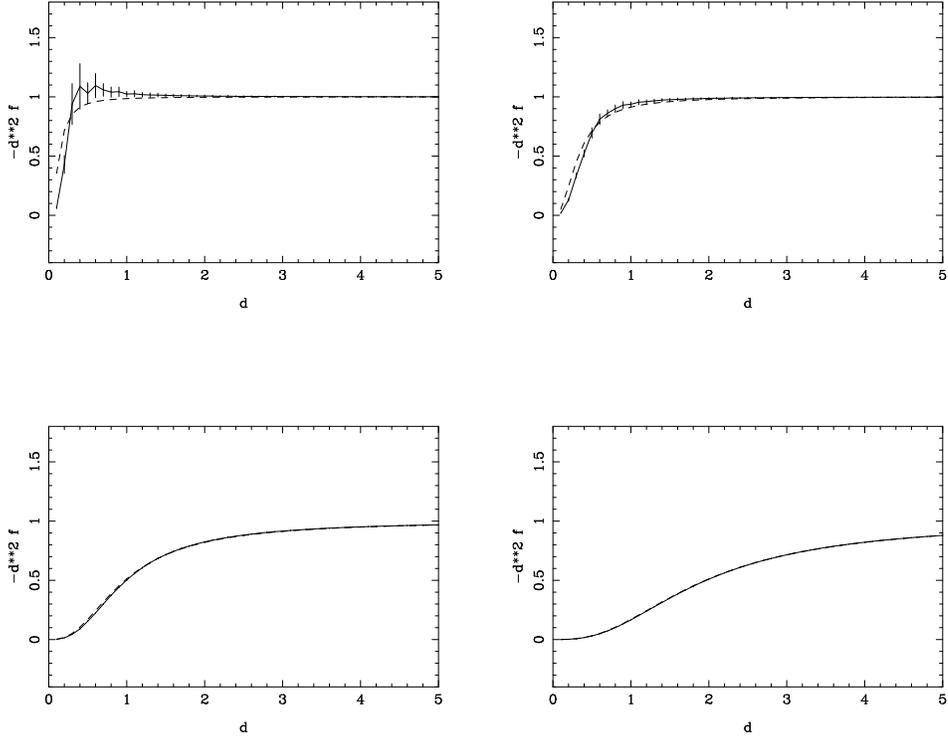

\vspace{245pt}
\includegraphics{Figure38a.ps}
\includegraphics{Figure38b.ps}
\includegraphics{Figure38c.ps}
\includegraphics{Figure38d.ps}
\caption{Grid noise estimates for N-body particle-mesh
simulations using Plummer sphere softening with $a_{soft}=0.1$ 
(top left), $a_{soft}=0.25$ (top right), $a_{soft}=0.75$ 
(bottom left), and $a_{soft}=1.5$ (bottom right),
respectively.  All the calculations start with a point-mass 
source particle at r=10, and with 100 test particles
at an average distance d (varying from near zero to
about 5 in the normalized unit of the disk) from the 
source particle randomly placed within the computational 
grid of 220 exponentially spaced radial sections and 
256 equally spaced azimuthal sections.
Solid lines are expected N-body performance with 1 $\sigma$ rms
fluctuations, and dashed lines are the
corresponding theoretical softened forces.}
\label{Figure38}
\end{figure}

In what follows the results of Monte Carlo calculations are presented
to characterize the magnitude of grid noise for the simulation
parameters we have used in this paper.
A source particle is placed at a select radial location in 
the disk (here we choose two radius values roughly corresponding to the
inner Lindblad resonance and the corotation resonance of the mode,
which give approximately the best-case and the worst-case error bounds
for a modal resonant cavity between the inner disk and the corotation
radius), and 100 test particles are placed at EACH of a range of distances 
from the source particles.  Analytical results (from the Newtonian
equation with softening) are compared with that obtained in the N-body
(interpolated) grid force calculation scheme, averaged for the 100 test 
particles for each distance, at a particular choice of $a_{soft}$.   
In Figures \ref{Figure38}, \ref{Figure39},
and \ref{Figure40}, we plot the noise performance of the Plummer softening 
scheme for the choices of softening parameters and grid sizes used in this 
study.  Note that even though the larger softening case ($a_{soft} =1.5$)
has its CR closer to 30, this large softening case has relatively 
small grid noise so the extent of the noise performance in that case 
(plotted only up to radius of 20 but can be easily extrapolated up to 30) 
is no worse than the worse case scenario 
(for $a_{soft} = 0.1$) represented in these figures.  The trend of 
increasing grid noise with decreasing softening, with increasing galactic 
radius (and thus grid size), and with overall grid-coarseness, is evident 
in these plots.  

\begin{figure}
\vspace{245pt}
\includegraphics{Figure39a.ps}
\includegraphics{Figure39b.ps}
\includegraphics{Figure39c.ps}
\includegraphics{Figure39d.ps}
\caption{Grid noise estimates for N-body particle-mesh
simulations using Plummer-sphere softening with $a_{soft}=0.1$ 
(top left), $a_{soft}=0.25$ (top right), $a_{soft}=0.75$ 
(bottom left), and $a_{soft}=1.5$ (bottom right),
respectively.  All the calculations start with a point-mass 
source particle at r=20, and with 100 test particles
at an average distance d (varying from near zero to
about 5 in the normalized unit of the disk) from the 
source particle randomly placed within the computational 
grid of 220 exponentially spaced radial sections and 256 
equally spaced azimuthal sections.
Solid lines are expected N-body performance with 1 $\sigma$ rms
fluctuations, and dashed lines are the
corresponding theoretical softened forces.}
\label{Figure39}
\end{figure}

\begin{figure}
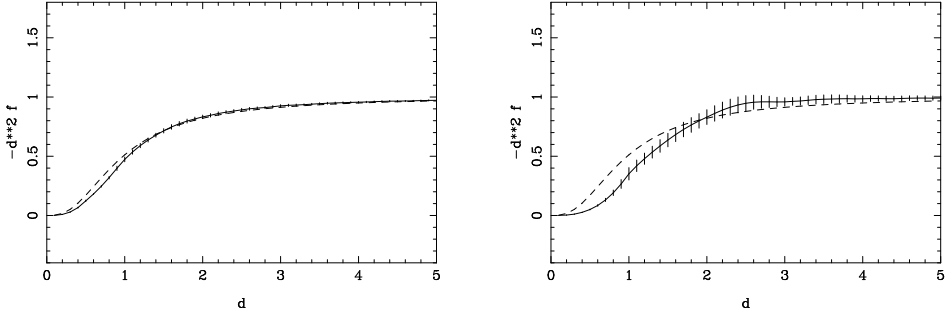

\vspace{125pt}
\includegraphics{Figure40a.ps}
\includegraphics{Figure40b.ps}
\caption{Grid noise estimates for N-body particle-mesh
simulations of Appendix D3
using Plummer-sphere softening with $a_{soft}=0.75$.
All the calculations start with a point-mass source particle 
at r=10 (left) or r=20 (right), and with 100 test particles
at an average distance d (varying from near zero to
about 5 in the normalized unit of the disk) from the source particle,
randomly placed with the computational grid of nr=110 exponentially
spaced radial
sections and 128 equally spaced azimuthal sections.  Solid lines are N-body 
performance with 1 $\sigma$ rms fluctuations, and dashed lines are the
corresponding theoretical softened forces.}
\label{Figure40}
\end{figure}

In the main text, we have shown that despite the fluctuations
of the forces caused by the increased grid noise for small
softening runs, the azimuthally averaged torques and mass flow rates converge
to theoretical expectations for the simulation of the self-organized density
wave modes.  Besides the arguments presented in the main
text based on the ``asymptotic stability'' of the patterns
amidst noise, we can also regard the mass assignment onto
grid as a kind of ``cluster clump'' formation in real galaxies.
Even though the forces are of error with respect to the
smooth disk model we started with, the forces are in fact
{\em exact} with respect to the clustered mass distribution.
And we know, in physical galaxies star cluster formation had no
fundamental impact on a galaxy's ability to form grand
design density wave patterns.  Thus the errors caused by
grid mass assignment should not impact modal pattern in 
simulated galaxies either (apart from the heating
effect it caused which impacts the longevity of the mode), as we have
confirmed from the simulation results in this paper.

\subsection*{D3. Effects of Grid Size and Grid Noise
to Mass Flow Rates In N-Body Simulations}

In this section we demonstrate further that the increased mass flow 
rates in less-softened N-body runs are due to the collective instabilities 
brought about by the decreased softening, and not due to increased 
grid noise brought about by the same softening reduction. 

The first hint for this conclusion can already be found in the previous
simulation result shown in Figure \ref{Figure4}.  It was seen there 
that when an insufficient number of particles are used for a given 
$a_{soft}$ (which generally {\em increases} the few-particle relaxation 
effect, or increases the noise), the mass inflow tapers off at an 
earlier time step.  Whereas for very large particle numbers (which 
generally {\em decreases} the few-particle relaxation effect, or decreases 
the noise), both the spiral/bar activity and the mass inflow continue 
until the end of the run.  This positive correlation between the number 
of simulation particles and the longevity of efficient mass inflow 
shows that noise is unlikely to be the {\em cause} of the increased mass 
flow rate.  Instead, the survivability of the density wave modes 
appears to be the key to the continued mass inflow.  

\begin{figure}
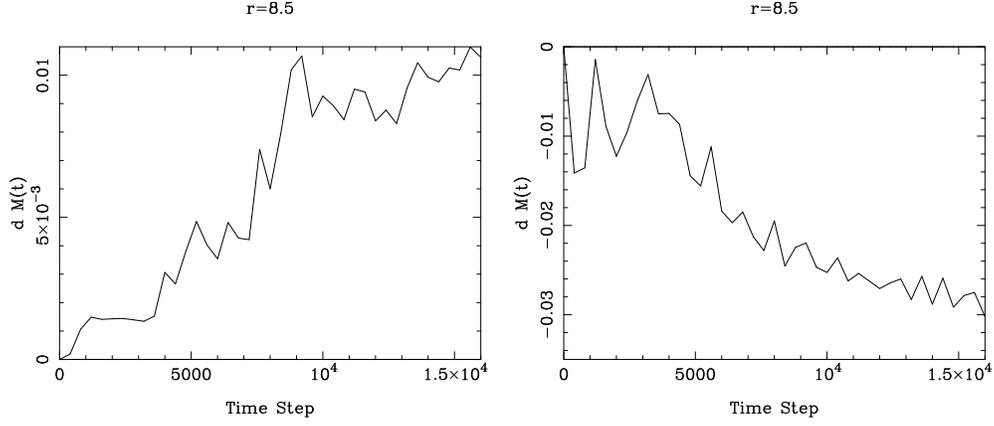

\vspace{145pt}
\includegraphics{Figure41a.ps}
\includegraphics{Figure41b.ps}
\caption{Time evolution of the enclosed mass for a typical 
location inside corotation, and a typical location outside corotation,
relative to the initial n=0 mass at the same respective radius, 
for the softening length $a_{soft}=0.75$ run, with a coarser
mesh (110 radial sections, 128
azimuthal sections) and twice as coarse a time resolution
as before.}
\label{Figure41}
\smallskip
\end{figure}

\begin{figure}
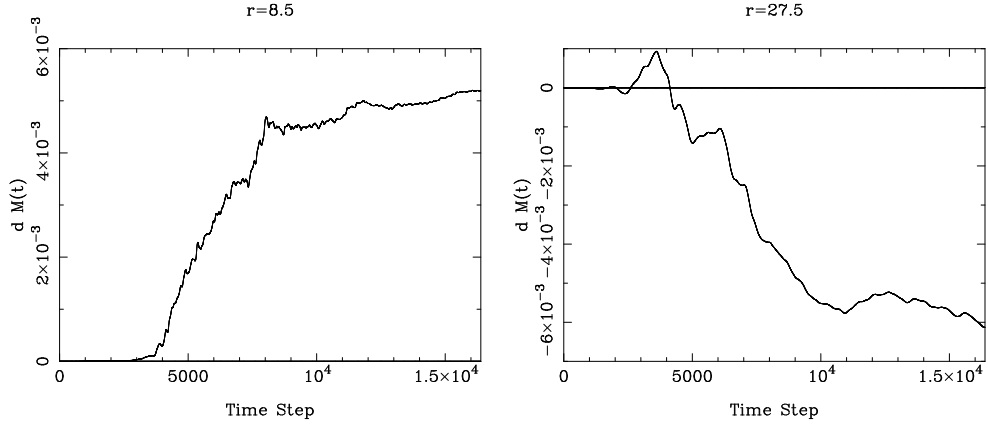

\vspace{145pt}
\includegraphics{Figure42a.ps}
\includegraphics{Figure42b.ps}
\caption{Predicted integrated mass evolution according to
equation (\ref{eq:accre}) within the two different radii, 
for softening length $a_{soft}=0.75$ run, but with a 
smaller grid (110 radial sections, 128
azimuthal sections) and twice as coarse
a time resolution as before.  The second
frame (for r=27.5) used only even harmonics in the mass flow calculation.}
\label{Figure42}
\smallskip
\end{figure}

In Figures \ref{Figure41} and \ref{Figure42}, we plot the same
mass flow evolution as previously simulated in Figures
\ref{Figure23} and \ref{Figure24} for the softening parameter choice
of $a_{soft}=0.75$, this time reducing the grid linear resolution 
by one half, and time resolution by one half correspondingly (i.e. 
16384 steps here cover the same time duration in terms of pattern 
rotation periods as 32768 steps before).  Decreasing grid resolution 
by one half should significantly {\em increase} the grid noise, from
the result we presented above.  Yet the results here show {\rm comparable} 
mass flow rates compared to that obtained from the higher grid 
resolution ones (the slight {\em decrease} in mass flow rate is due
to the contribution of grid-softening effect to mass flow, which is 
apparently not as prominent as particle softening; yet even this decrease
is in the opposite direction to what one would expect if grid noise
is the main contributor to the {\em increase} in mass flow rate
in simulated galaxies), especially for the first one-half of the
duration of the simulation run, when the heating due to noise has 
not clamped the amplitude of the pattern for this coarser-grid run. 
This shows that we are indeed revealing the role of increased 
collective effects brought about by decreased particle softening,
and the increased mass flow rate is not due to
increased grid noise (if anything, increased grid noise, and the resulting
increased heating of disk particles,
slightly hampered the mass flow in the later half of the simulation
period, as evidenced by the flattening of the mass-increase
curves in both Figures \ref{Figure41} and \ref{Figure42}).

Therefore, we conclude that as long as our interests are in the
comparison of macroscopic characteristics of the mode (equilibrium
amplitude, pitch angle, potential-density phase shift and torque)
in the simulations with that in theoretical predictions, as well as with that in observed galaxies, the grid noise present in
simulations with smaller softening, as in the current paper,
is not a debilitating inconvenience.

\section*{APPENDIX E. A CLOSE-UP LOOK AT THE
MODAL CHARACTERISTICS EVOLUTION\footnote{Appendices E, F, and G
are added at the review stage in response to the comments of
an anonymous referee.}}

\begin{figure}
\vspace{210pt}
\includegraphics{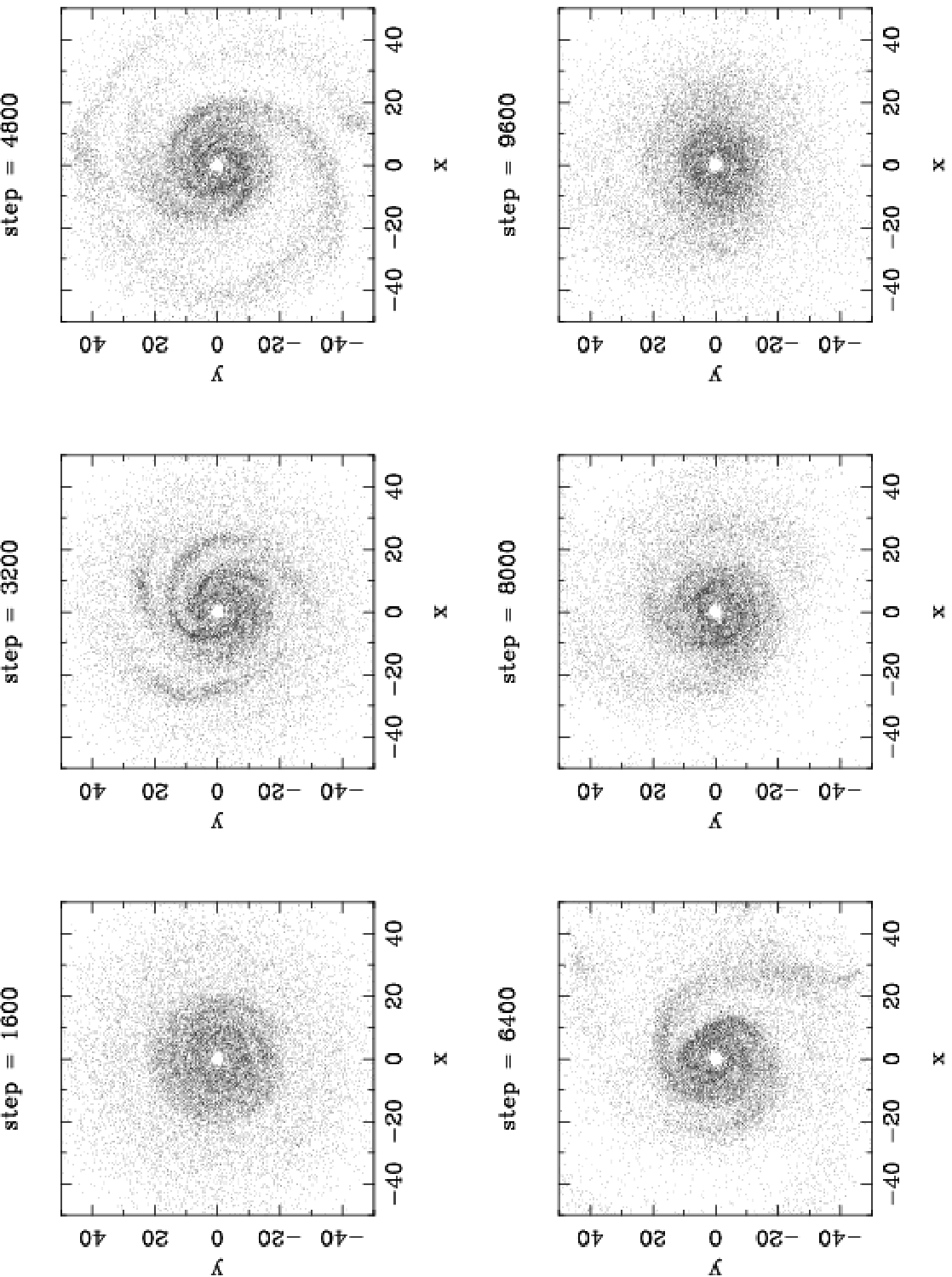}
\caption{Morphological evolution of an N-body spiral/bar mode.
The rotation period at r=20 is about 1256 steps.
The softening parameter is $a_{soft}=0.25$.} 
\label{Figure43}
\end{figure}

\begin{figure}
\vspace{210pt}
\includegraphics{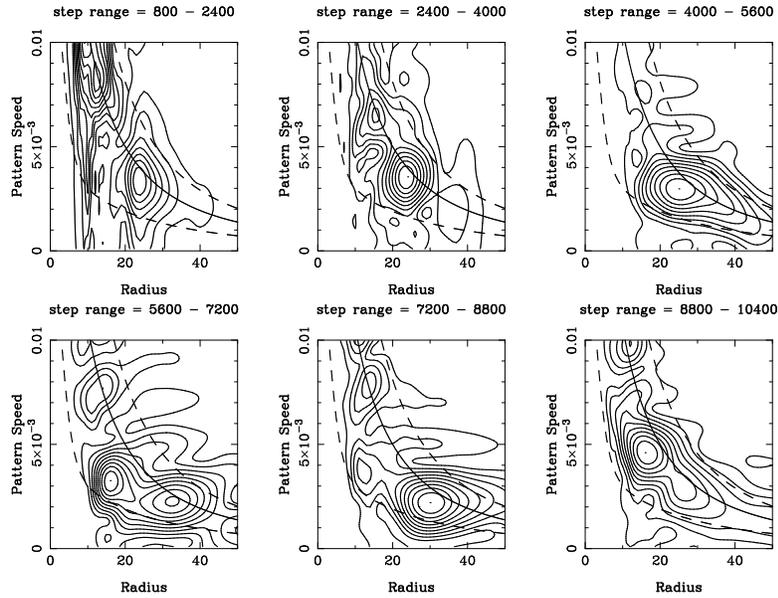}
\caption{Power Spectra of the above mode.  Each frame is calculated
with a central time step identical to that in the corresponding morphology frame,
and with a time step range of 1600.}
\label{Figure44}
\end{figure}

\begin{figure}
\vspace{210pt}
\includegraphics{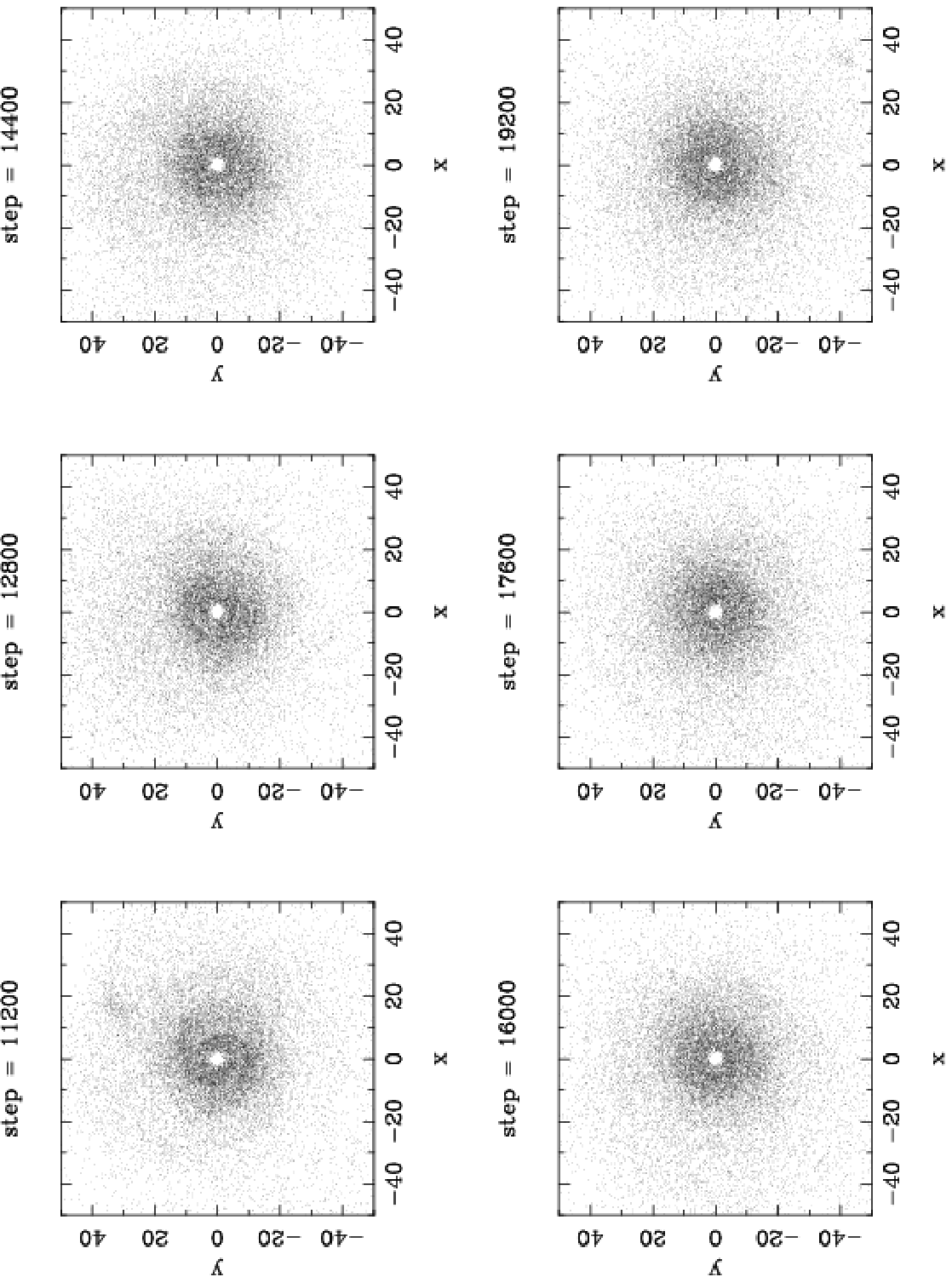}
\caption{Morphological evolution of an N-body spiral/bar mode
(continued from Figure \ref{Figure43}).
The rotation period at r=20 is about 1256 steps.
The softening parameter is $a_{soft}=0.25$.} 
\label{Figure45}
\end{figure}

\begin{figure}
\vspace{210pt}
\includegraphics{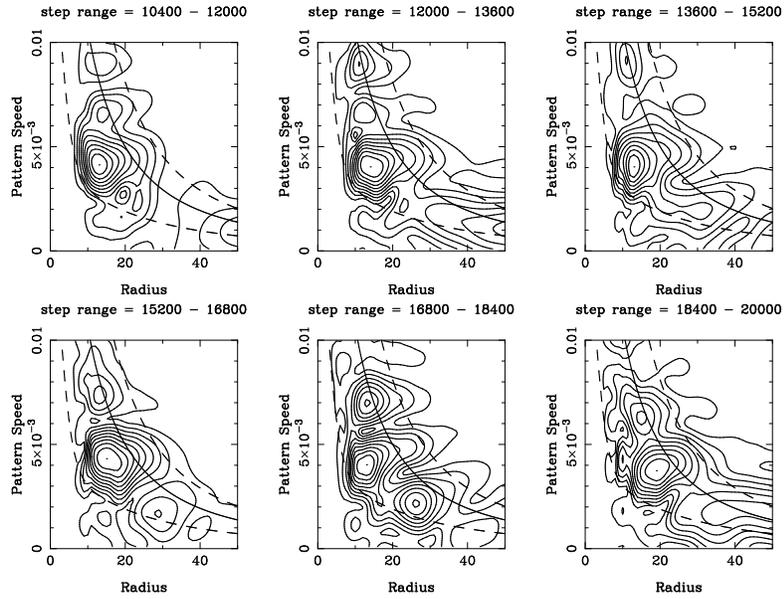}
\caption{Power Spectra of the above mode.  Each frame is calculated
with central time step identical to that in the corresponding morphology frame,
and with a time step range of 1600.}
\label{Figure46}
\end{figure}

\begin{figure}
\vspace{210pt}
\includegraphics{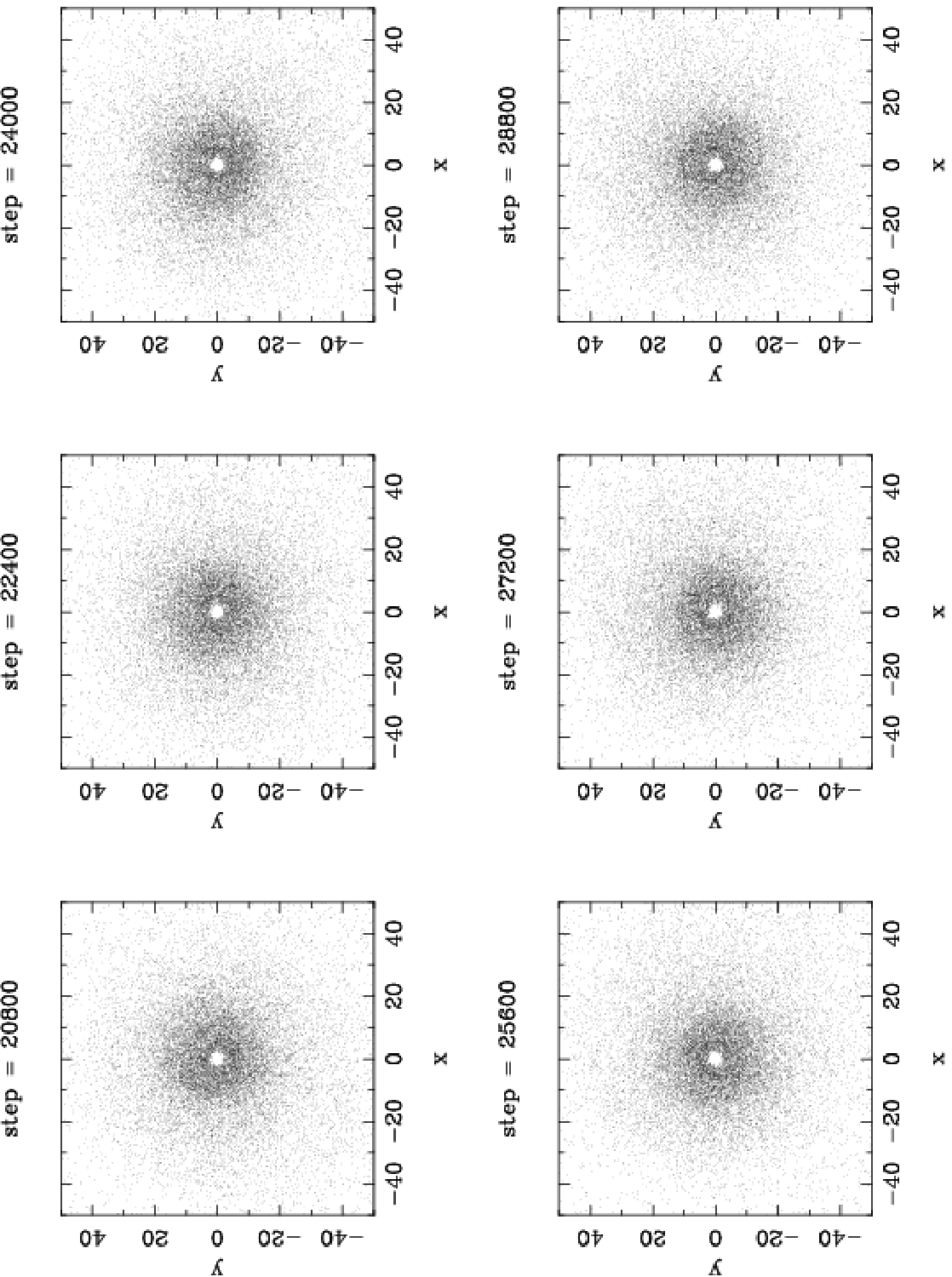}
\caption{Morphological evolution of an N-body spiral/bar mode
(continued from Figure \ref{Figure45}).
The rotation period at r=20 is about 1256 steps. 
The softening parameter is $a_{soft}=0.25$.} 
\label{Figure47}
\end{figure}

\begin{figure}
\vspace{210pt}
\includegraphics{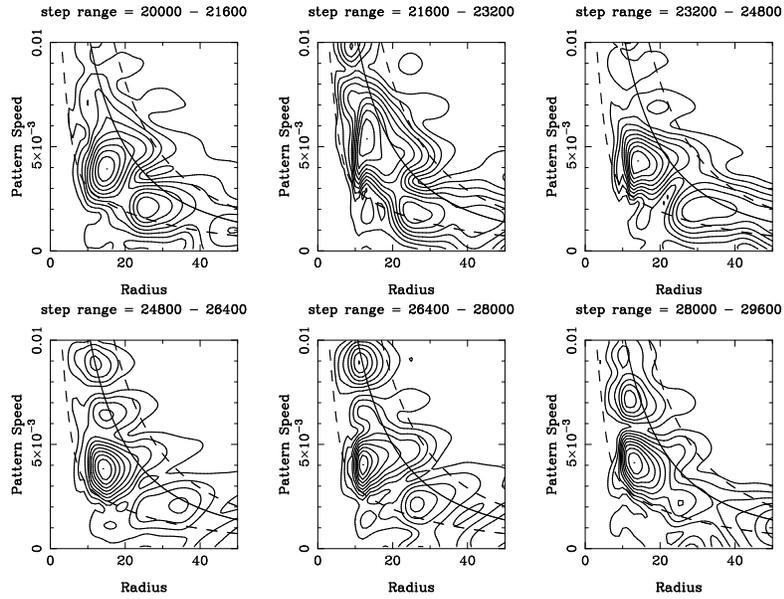}
\caption{Power Spectra of the above mode.  Each frame is calculated
with central time step identical to that in the corresponding morphology frame,
and with a time step range of 1600.}
\label{Figure48}
\end{figure}

In preparation for the discussion to come in Appendix F,
in Figures 43-48 we present the morphological and power
spectra evolution for the spiral/bar mode presented in the main text
using $a_{soft} =0.25$, here with better spatial-temporal
resolution
and covering the entire duration of the simulation run.  

We see from these figures that even though the spiral
pattern involved is not perfect steady, it is a mode 
(or a modal set, since during some time intervals
there are nested modes of different pattern speeds, 
which are present in physical galaxies as well) nonetheless,
rather than a combination of transient wave trains (these
wave trains would have filled up the spectral space with
disorganized noisy components, rather than coherent
and concentrated spectral clumps).  During
certain intervals, the mode transitions to a different
shape in response to the changing basic state, and the power
spectra may reveal additional features.
But after a while the dominant mode (or sometimes the modal set)
always settles back onto its quasi-stable
location on the power-spectra plot.  This is the kind of
``attractor'' or ``asymptotic stability'' behavior we would expect
of a mode.

In Figures 49 and 50, we plot the m=2 density and potential
contours, covering also the duration of the simulation but
with courser time spacing.  We see that until the very
end of the simulation, which is roughly 25 rotation periods at
reference radius of 20, the underlying m=2 mode is still
alive and well, even though the heating of the basic state
of the disk by the modal instability has submerged the
nonlinear spiral pattern (the heating is a result of using
smaller number of particles than in physical galaxies;
of the near-collision condition at the modal collisionless shock;
and of using a small softening length choice).

\begin{figure}
\vspace{200pt}
\includegraphics{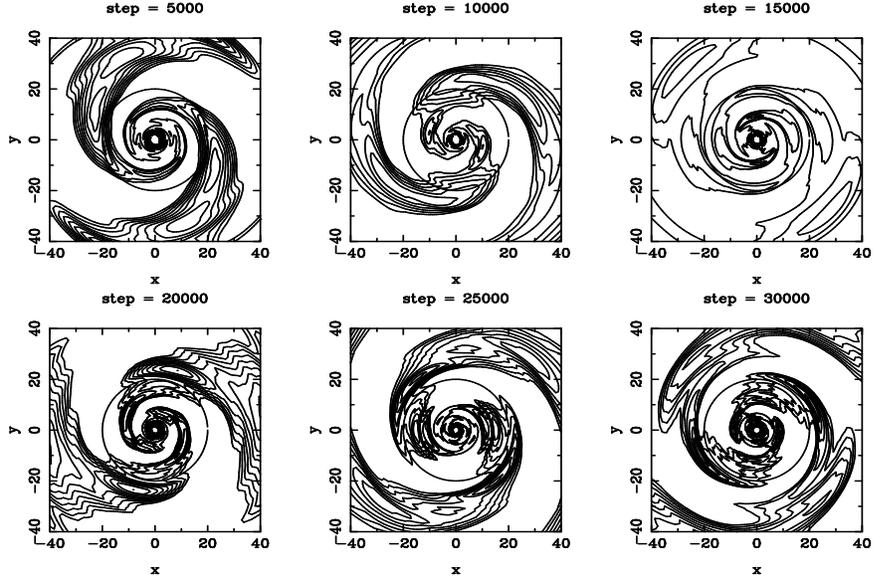}
\caption{Morphological evolution of the m=2 component of the
density of the N-body spiral/bar mode.  
The softening parameter is $a_{soft}=0.25$. 
The circle indicates the reference radius of 20.
Each frame is individually scaled to have 10 contours.
Only positive contours are plotted.} 
\label{Figure49}
\end{figure}

\begin{figure}
\vspace{200pt}
\includegraphics{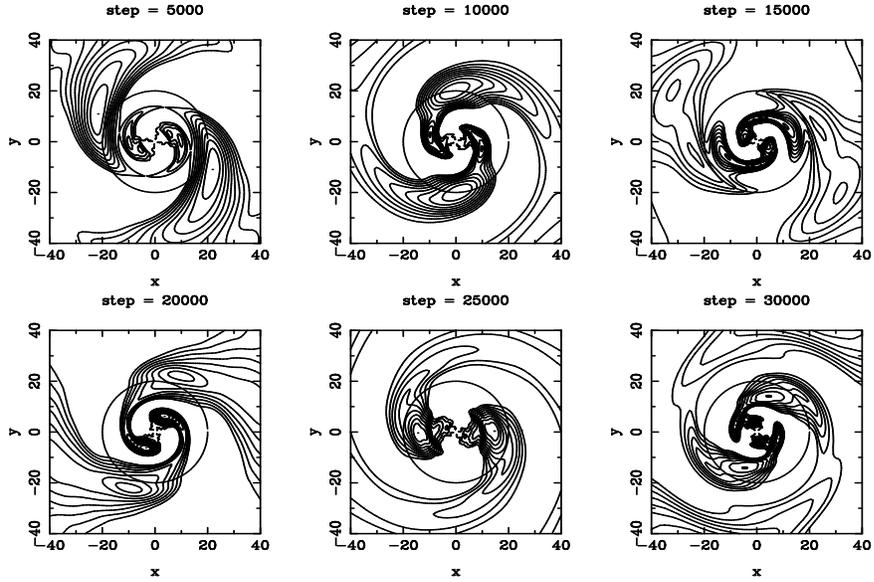}
\caption{Morphological evolution of the m=2 component of the
(negative) potential of the N-body spiral/bar mode.
The softening parameter is $a_{soft}=0.25$. 
The circle indicates the reference radius of 20.
Each frame is individually scaled to have 10 contours.
Only positive contours are plotted} 
\label{Figure50}
\end{figure}

In Figures 51-53, we plot the evolution of the radial mass flow
rates due to the spiral mode, using once again the full density
and potential perturbations (solid lines), as well as only
the m=2 density and potential perturbation components.

A comparison between this last group of figures and that of
the power spectra evolution figures (Figures 44, 46, 48) 
shows several interesting features:

\begin{enumerate}
\item As the evolution advances the matter as well as the spiral
activity becomes concentrated into the central region of the galaxy.
\item A better m=2/m=full agreement (especially for the inner
disk, inside the corotation radius of 20) is associated with 
a more concentrated power-spectrum peak location, indicating that
the former is a good criterion for judging the quasi-steady state
of the mode, as we have argued in the main text (\S 4.1.2).
\item The amplitude of the radial mass flow rate curves
decreases with time (note the reduced vertical scales with advancing time, 
among the three sets of figures, i.e. $10^{-5}, 2 \times 10^{-6},
$ and $5 \times 10^{-7}$, respectively).  Associated with this
decreasing radial mass flow rate towards the
end of the simulation, the agreement between the m=2 and m=full
mass flow rates is better, especially for the central region
where the mass and modal activity are concentrated.
This shows that the slower secular evolution speed
towards the end of the simulation (due to the reduction of
nonlinear spiral amplitude) leads to slower modal evolution,
and thus a better quasi-steady state or global self-consistency.
\end{enumerate}

We conclude that the variation of the spiral pattern (both
its morphology and pattern speed) is to a large extent a result
of the secular evolution of the basic state induced by the
very same density wave mode.  We here recall the saying that 
``one cannot have the cake and eat it too''.  If we seek to obtain 
quasi-steady state of the mode, we cannot have fast secular evolution 
at the same time.  If the secular evolution is fast, the modal 
characteristics evolve rapidly with it, so as to be compatible 
with the evolving boundary condition.  During these transitional times, 
the power spectra may indicate additional peaks other than that 
corresponding to the dominant mode.  This does not mean that 
the mode has now become a transient wave.  The mode is still a mode, 
only it is changing and evolving, just like when a person ages, 
he or she is still the same person (although some people would 
refer this process as the old self dying and a new self being born 
-- which is OK as long as we all know what we are talking about).

\begin{figure}
\vspace{210pt}
\includegraphics{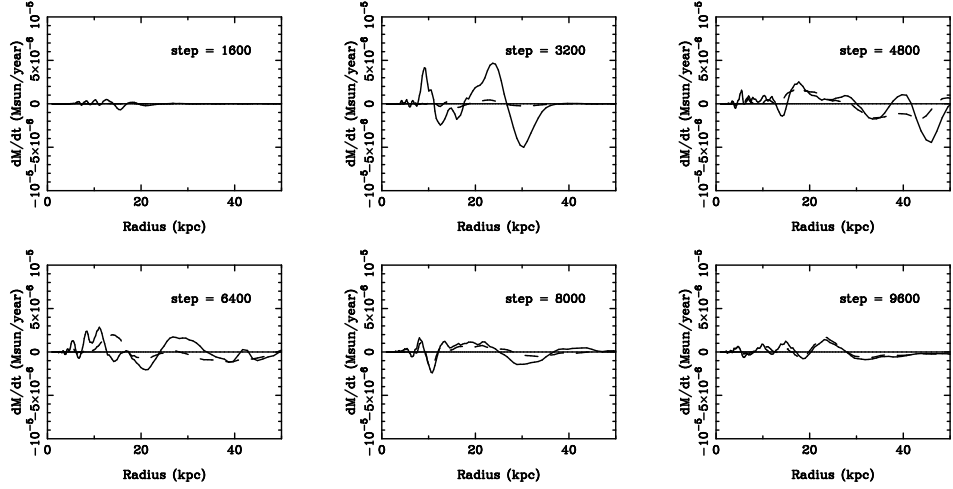}
\caption{Radial mass flow rates for the N-body spiral/bar mode
with $a_{soft}=0.25$, first set of six time frames.
Solid lines: full Fourier components.  Dashed lines: m=2
Fourier components}
\label{Figure51}
\end{figure}

\begin{figure}
\vspace{210pt}
\includegraphics{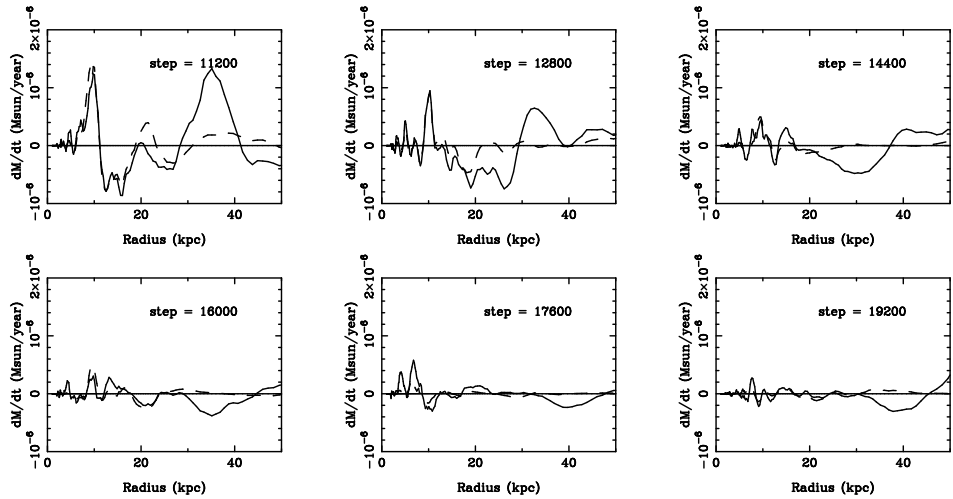}
\caption{Radial mass flow rates for the N-body spiral/bar mode
with $a_{soft}=0.25$, second set of six time frames. Note the
reduction in vertical-axis scale compared to the first set.
Solid lines: full Fourier components.  Dashed lines: m=2
Fourier components}
\label{Figure52}
\end{figure}

\begin{figure}
\vspace{210pt}
\includegraphics{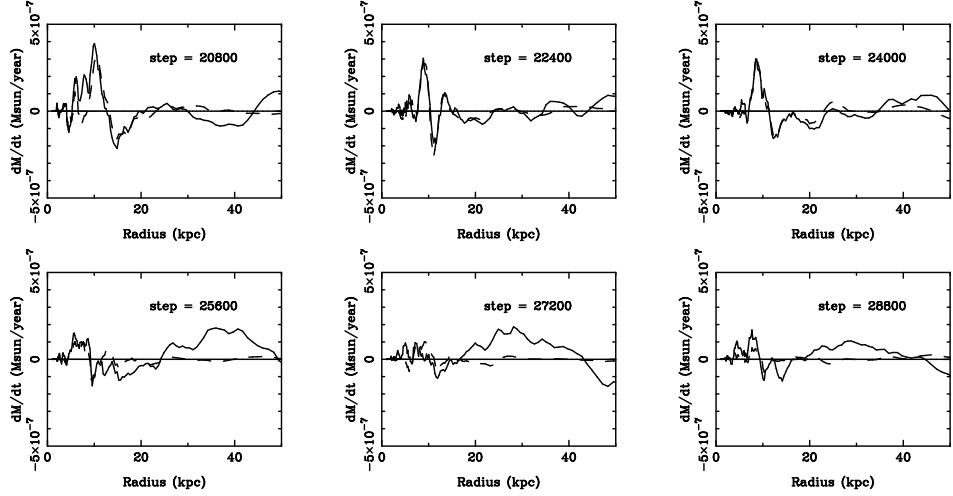}
\caption{Radial mass flow rates for the N-body spiral/bar mode
with $a_{soft}=0.25$, third set of six time frames. Note the
further reduction in vertical-axis scale compared to the
first as well as the second set.
Solid lines: full Fourier components.  Dashed lines: m=2
Fourier components}
\label{Figure53}
\end{figure}

Therefore we should relax our condition for calling a mode
quasi-steady:  As long as it satisfies the global self-consistency
condition (growth rate = damping rate), we can regard it as
quasi-steady even if it constantly evolves to accommodate
the changing basic-state boundary condition as a result of 
the secular evolution of the basic state induced by the very same mode.
For the theoretical results (that of the torque integral
and mass flow rate equations) to hold, the mode only needs
to be quasi-steady on the local dynamical timescale (in this simulation
the local dynamical time at radius 20 is 1296 time steps, slightly
shorter than the time duration between adjacent frames in Figures
43-48), as we have verified in the main body of this paper by 
comparing theoretically-predicted and measured mass flow rates.

We would also like to point out once again that the observed
galaxies, especially the nearby, grand-design, non-interacting,
intermediate to early type galaxies, appear to have even better 
agreement between the m=2 and m=full estimations of torque, than what
we have been able to reproduce in the current generation
of N-body simulations (see Figure 31 of this paper in the main text).  
The agreement between m=2 and m=full torques appears to span several 
sets of nested modes, as in the case of NGC 4321 (M100), NGC 4736, and
NGC 3351.  This agreement between the m=2 and m=full torques indicates
a high degree of global self-consistency for the modes in
observed grand-design galaxies, a result obviously partly
due to a Hubble-time's worth of self-consistent evolution, including
the active participation also of the bulge and halo components
(which are held fixed in our N-body simulations), partly due to
the much higher (by several orders of magnitude) numbers of
degrees of freedom that physical galaxies have, compared to that
available in simulated galaxies, to be used for adjusting the multitude
of correlations to achieve a fine balance between modal growth
and dissipation, between the secular evolution of the basic state
and the transformation of density wave morphology to be compatible
with the renewed basic state.

\section*{APPENDIX F. ROLE OF BASIC STATE SPECIFICATION}

The current paper explores mainly the influence of the choice of simulation 
parameters (in particular the softening parameter) on obtaining a realistic 
LEVEL of radial mass flow rate in N-body simulated disk galaxies.  In this 
appendix, however, we provide additional arguments of why the choice of the 
basic state parameters is of equal importance in obtaining the correct mass 
flow PATTERN that is needed for building up the Hubble sequence through
secular radial mass accretion in the inner disk. Most of the theoretical
results relating to this discussion were previously obtained in Z96,Z98,Z99. 

As we have commented before, the correct choice of the basic state involves 
the use of galaxy disk parameters that allows the global density wave MODES 
to spontaneously emerge and be stabilized (the stabilization is by the same 
collective dissipation process that leads to secular mass redistribution [Z98]).
Why is it so important to choose a basic state specification that allows
unstable modes?  Could not the transient waves on over-stable basic state
also accomplish the objective of secular evolution, as seems to be advocated
in a recent review article of Sellwood (2014), which is titled ``Secular
Evolution in Galaxies''?

First of all, we point out that in none of the transient wave simulations,
carried out by Sellwood and his collaborators over the past three decades, 
as well as by more recent researchers such as D'Onghia, Vogelsberger, 
\& Hernquist (2013); Baba, Saitoh, \& Wada (2013); D'Onghia (2015), etc., 
was there a systematic trend of mass inflow inside corotation, and outflow
outside, such as what we have demonstrated.
The fundamental reason for this is that the transient waves rely on
resonant interactions between the wave trains and the basic state to
exchange angular momentum.  In transient waves, the resonances are broadened,
but the interaction is not coordinated as in the modal case.  For example,
Carlberg \& Sellwood (1985) states that the sense of angular momentum
exchange between the wave and the basic state does not depend on the
winding sense of the spiral at the Lindblad resonances, so a leading and
a trailing wave train can exchange angular momentum with the basic state
in the same way.  Sellwood \& Binney (2002) studied the so-called ``radial
migration'' process of stars due to resonant interaction of waves
and stars near the corotation region, yet they conclude that the end
result of this interaction is such that the entire population of interacting 
stars do not lose or gain angular momentum, so there is no systematic
radial mass flow that leads to bulge building through this so-called
``radial migration'' process studied by them.  These resonant interactions
thus are of entirely different nature than the collective interactions 
between the wave mode and the basic state stars that we studied in this work.
Resonant interaction is top-down, i.e., stars passively respond to an
enforced smooth potential, whereas the collective interaction is sideways,
i.e., individual stars develop correlations in their motion through
the mediation of collective instabilities and collisionless shocks.

For unstable global modes, however, as demonstrated in Z96, Z98, Z99,
their surface density and kinematics are such that they support a 
characteristic distribution of potential-density phase shift between 
the potential spiral and the density spiral of the mode.  This phase shift 
distribution in turn leads to a secular torque by the modal potential 
on the density, and as a result a secular radial mass flow pattern as 
we had mentioned above.  The global self-consistency requirement for 
these modes (since the modes are self-organized and self-sustained) is
what brought on the effective singularity condition at the spiral/bar 
wave crest, which leads to coordinated secular evolution of the basic state 
of the galactic disk, as well as the constant revision of the wave modal 
shape to accommodate the changing basic state.  Thus, in the secular 
evolution scenario the quasi-steady state (QSS) hypothesis should be 
modified to include the coordinated modal shape evolution so as to be 
compatible with the evolving basic state.  To judge whether a given 
wave mode has reached QSS or not, we should use the criterion of
global self-consistency (growth rate = dissipation rate), which can
be further judged by the m=2 versus m=full mass flow rates comparison,
as we had done in the main text of this paper, rather than by the
strict requirement that the modal shape does not change secularly.

For transient wave trains in an over-stable disk, on the other hand,
the global self-consistency condition is not enforced, and it is not
necessary that the spiral arms are the site of effective singularity
and collective dissipation.  The no-global-self-consistency condition means  
no quasi-steady state, thus the transient patterns continuously evolve
(in a haphazard fashion, rather than in a coherent fashion as in the
modal scenario).  The fact that some researchers succeeded in producing a 
continued sequence of wave decay and reemergence within this scenario 
(D'Onghia et al. 2013; Baba et al. 2013) does not mean ALL observed spirals
are such recurrent transients, especially NOT those grand-design 
two-armed spirals which account for about 50\% of disk galaxies 
(Pettitt, Tasker, \& Wadsley 2016), NOR those galaxies that possess 
long-lived resonance features such as nuclear, inner and outer rings.

If our goal is to obtain the coordinated radial mass flow pattern that
is responsible for the secular growth of bulge and galaxy morphological
evolution along the Hubble sequence, then we cannot avoid choosing an
unstable basic state to global modal formation, and employing the collective
interaction of these modes with their parent basic state of the disk,
such as is done in the current paper as well as in its predecessors.
To model galaxies is oftentimes more than to produce a visual appearance
of spirals or bars, but rather to bring out the intrinsic common dynamics
underlying the formation and evolution of a large class of these galaxies.

To study global collective effect thoroughly is inherently more challenging
than to study local effects one at a time.  However, without taking a
globally self-consistent view, we will not be able to arrive at the forced
singularity condition (in some sense, nature itself is FORCED into arriving
at the collective dissipation process in these singularities when given
conflicting requirements that cannot be resolved by a smooth differentiable
solution).  When taking a local view, it is like the fable of the blind
men and elephant, we might mistake a tail, a trunk, a tusk, as the essence
of the elephant.  And even if we take stock of all of these pieces, we still
cannot pile them together to arrive at a LIVE elephant.  Anything that is
alive is the result of co-evolution between the system and the environment.
The co-evolution process is when the correlations are gradually set up.
These multitudes of correlations among the constituent parts are what give
the system its livelihood and its evolutionary dynamics.  

In this context we see that the so-called action-angle approach adopted 
by many previous practitioners of galactic dynamics (Binney \& Tremaine
2008 and the references therein; Sellwood 2014 and the references therein),
which follows a venerable tradition in classical mechanics and
classical kinetic theory, exactly ignored this correlation among the 
constituent particles.  It produces a system that is harmonic by design 
(that is what the angle part represents -- i.e., sinusoidal oscillations; 
and the action part represents conserved integrals).  The validity of
the action-angle approach depends on the assumption that most of the
orbits supporting the galactic potential are made up of so-called regular
orbits that respect the conserved integrals under the smooth average potential;
whereas we learned in the current work that most of the orbits in the
self-organized density wave potential are severely perturbed (by the 
collisionless shock) and thus are almost always chaotic (see also Zhang
\& Buta 2007, in particular Figures 14 and 15), and these chaotic orbits 
are associated with a grainy potential incorporating correlations.  The 
closed-up harmonic system described by the action-angle approach will 
not display secular behavior (apart from orbit resonances, which is a 
local effect), at least not the type involving global collective dissipation
processes as discussed in the current paper, which require inter-particle
correlations, rather than passive response of individual orbits to
an {\em applied} smooth potential.

These said, it is not the current author's assertion that the action-angle
approach has no merits in studying galactic dynamical problems.  There
are at least three scenarios that this approach (or the associated 
phase space distribution function approach governed by collisionless
Boltzmann equation) is in fact applicable:  (1) When the system involved
is in steady state and is not unstable to the growth of global modes.
(2) When the system is unstable to the emergence of modes, but its
behavior is first analyzed in the linear regime.  In that case we are
studying the modal growth regime with no secular evolution of the
basic state.  (3) In non-self-consistent studies of, say, gas response
to an applied stellar potential, then the stellar potential itself
can be studied as a smooth system.  One needs to be cautious about interpreting
the results of this approach when, say, the self-gravity of the gas
becomes important.  In such cases the often-large phase shift
between the stars and gas derived from the non-self-consistent analysis
is not physical (see further the discussion in Appendix G).

To take a global view of galactic dynamics also means the ability to synthesize
many diverse known observations and subject them under the same organizing
principles.  There is a well-known saying: ``Theory destroys facts''.  A
successful theory is not meant to explain disjoint pieces of fact in an
{\em ad hoc} fashion.  It is meant to unify and simplify and weave facts into
a tightly-knit web that possesses inevitability.  Another saying is that
``It is not the complexity of nature that requires explanation.  It is
rather the simplicity of it''.  By simplicity is meant the spontaneous broken
symmetries of nature, that organize nature into a network of patterns.
What is behind a set of simple patterns (such as two-armed grand-design
density wave spirals and bars) is often a profound intrinsic dynamics
that requires our digging deep and connecting wide in order to distill
its essence.  Here ``simplicity'' is not synonymous with ``easy to understand'',
but rather indicates highly ordered patterns unified by a deep organizing
dynamical principle.  ``Complexity'' in this context would mean the noisy
randomness that is controlled more by chance than by necessity: it is
complex by appearance, rather than by the profoundness of its underlying cause.

We are here witnessing a paradigm shift of ``From Being to Becoming'' 
(using the title of Prigogine's 1980 book), the ``Being'' type
of study is the older paradigm of using action-angles and describing
a system that is immutable to change (apart from local resonant interactions),
whereas the ``Becoming'' type of study is one which invokes the inherent
self-organizing power of the many-body systems that drives its own
long-term evolution through the coordinated interaction of its component
parts.  The coordination is not done in a top-down fashion, but is controlled
by the system's intrinsic instability dynamics, which is akin to the 
``invisible hand'' analogy used by economists for understanding the long-range 
order that emerged out of the seeming random interactions of component systems
responding only to short-term and local interests.  This paradigm shift 
is aligned also with the one mentioned by Feynman as quoted in the 
Introduction of the current paper, that of the switch from the study of 
quantitative dynamics to qualitative dynamics: the qualitative dynamics IS
the emergent dynamics of a self-organized dissipative structure arrived 
at in complex systems through spontaneous symmetry breaking and the
establishment of long-range correlation through local interactions.
Methods of our former inheritance, of taking things apart, studying one 
aspect at a time, without worrying about the consequence of the correlations 
among the component parts; or of employing only the linear and deductive 
dynamics, would not be adequate in the studying of the evolutionary 
problems of many degree of freedom systems possessing self-organized 
dissipative structures that can induce secular evolution of the 
parent system through collective effects.

\section*{APPENDIX G. ROLE OF GAS}

The current paper focuses on the role of stars in participating the
collective dissipation process at the density wave crest.  The role of
gas was previously analyzed in Z98, \S 4.  At the request of the
referee of this paper, we give a brief summary below of the role of
gas in supporting the collective dissipation process.

It was the dust lanes at the inner edges of grand-design spirals that 
first impressed many about the potential importance of gaseous shock waves 
of galactic scale on the dynamics of galactic density waves (Roberts 1969).
Many had placed hopes for the damping of the violently unstable stellar
density waves on the dissipative gaseous density waves (Kalnajs 1972; 
Roberts \& Shu 1972; Bertin et al. 1989a,b).  In these early studies, as a
result of the dissipation of gas under the applied stellar potential, an
azimuthal phase shift between the gaseous mass response and the forcing 
stellar potential was observed.  This occurs largely because the studies 
were not done self-consistently (i.e., the gas was modeled as passively 
responding to the forcing stellar potential).  When the self-gravity of
the gas was considered together with the stars, the phase shift between
the gaseous and stellar potential was found to become insignificant
(Lubow, Balbus \& Cowie 1986; Balbus 1988).

Z98 found, in addition, that when the stellar and gaseous media are 
considered together in galactic N-body simulations, the role of gas largely 
parallels that of stars, supporting earlier conclusions.  This is because the 
viscous/dissipative effect of gas by itself is quite inadequate: Taking only 
the microscopic viscosity of gas into account, we obtain a secular mass
accretion timescale many orders of magnitude longer than the Hubble time.
For gas to contribute significantly to secular evolution, the 
mean-free-path of the gas needs to be taken as the mean-free-path of 
cloud-scattering in the star-gas two-fluid medium (Z96, Z98), which is 
on the order of hundreds of parsecs to 1 kpc, therefore stars and gas now 
share similar mean-free-path and similar rate of dissipative interaction.  
They are both particles in a joint ``collisional medium'' with the
collisional/scattering viscous dissipation determined by the 
intermediate-range gravitational interactions at the density wave crest
(and the amount of dissipation is not determined locally, but globally
so that the wave damping exactly offsets wave growth at the quasi-steady
state of the density wave mode).

In the simulations presented in Figures 8, 9 of Z98, it
was found that the curves of potential-density phase shift (which gives
the normalized torque between the density wave and the basic state) for
stars and for gas have similar shapes, with gas phase shift having slightly 
larger values.  This is also confirmed in the observational studies presented
in Zhang \& Buta (2015), where both the stellar and the ISM phase shifts
with respect to the common potential are calculated.
Pettitt et al. (2016) studied the galactic spiral structures excited 
during the tidal interaction with companions, and also found that
the stellar and gaseous spirals have similar shapes, but slightly offset
in azimuth (presumably also offset with respect to their common
spiral potential, since these spiral patterns have skewness, so the
Poisson equation will naturally leads to a potential-density phase shift).

Therefore it is most convenient to think of gas just as another gravitational
mass component, slightly more dissipative than stars, but with only a
quantitative rather than a qualitative difference when participating
in the collective interactions with the density wave potential field.
Note that to properly model the collective dissipation effect the gas
must be modeled as dissipative particles which lose energy during
close encounters/collisions at the density wave crest (as modeled in Z98 based
on the algorithm described in Thomasson 1989), rather than using a
smooth-particle-hydrodynamics (SPH) approach.  In the former the gravitational
viscosity of the gas (as was that of the stars) self-consistently emerges 
as a result of the global collective interaction of the wave and the
basic state matter through the mediation of spiral collisionless shock,
whereas in the latter approach the viscosity is put in "by hand" by the
simulator, thus is not modeled self-consistently.  Incidentally,
modeling ISM clouds as collisional particles was also the preferred
simulation approach suggested in Binney \& Tremaine (2008, pp 520-521).

Through the presence of the potential-density phase shift, allowed by
both the Poisson equation and the equations of motion for a self-organized
unstable density wave mode (which has skewness to support the phase shift),
the star-gas two-fluid mode is thus self-damped: the growth tendency
of the global resonance cavity due to over-reflection at corotation and
feedback at galactic center, is offset by the dissipation tendency as
is revealed and supported by the potential-density phase shift, and mediated
by the collisionless shock that happens partly as a result of the phase shift
(i.e., due to this phase shift, matter rams unexpectedly into the potential
field of the density wave at a supersonic velocity, during its spiral arm
crossing orbital motion.  See Z96, Figure 5).  The phase shift required 
by dissipation is provided by the Poisson equation naturally, due to the 
skewness of the pattern and the nonlocal nature of the Poisson integral.  
Stars and gas on the basic state thus both can have dissipative interaction 
with their common density wave potential.

\end{document}